\documentclass[onecolumn,authoryear]{els-mrw} 

\usepackage{amsmath,amssymb,amsfonts,amsthm,makeidx,graphicx}
\usepackage{txfonts}
\usepackage{helvet}
\usepackage{subcaption}
\usepackage[dvipsnames]{xcolor}

\let\oldcite\cite
\renewcommand{\cite}[1]{[\oldcite{#1}]}

\begin{document}

\chapter{Ultra-peripheral Collisions}\label{chap1}

\author[1]{Jes\'us Guillermo Contreras}%
\author[2]{Spencer Robert Klein}%

\address[1]{\orgname{Czech Technical University in Prague}, \orgdiv{Faculty of Nuclear Sciences and Physical Engineering}, 
\orgaddress{B\v{r}ehov\'a 7, 11519 Prague, Czechia}}
\address[2]{\orgname{Lawrence Berkeley National Laboratory}, \orgdiv{Nuclear Science Division}, \orgaddress{1 Cyclotron Road, Berkeley CA 94720 USA}}

%\date{\today}
%\today

\articletag{UPCs}

\maketitle

\begin{glossary}[Glossary]
\term{Ultra-peripheral collisions} Ultra-peripheral collisions are collisions at impact parameters $b>2R_A$, where nuclei interact electromagnetically but not hadronically.

\term{LHC} Large Hadron Collider at CERN.

\term{RHIC} Relativistic Heavy Ion Collider, at Brookhaven National Laboratory.

\term{pomeron} Quasiparticle carriers of the colorless strong force with the quantum numbers of the vacuum; to lowest order, the pomeron is two gluons.

\term{reggeon} Quasiparticle carriers of the colorless strong force.  In contrast to the pomeron, reggeons accommodate a wider range of quantum numbers; they are thought to be mostly quarks.

\term{BSM} Physics that goes beyond the current standard model.
\end{glossary}

\begin{abstract}[Abstract]
Photons in ultra-peripheral collisions of heavy ions (and protons) 
can be used for a variety of physics purposes - for studies of nuclear structure at low Bjorken$-x$, as probes of beyond-standard-model physics at the highest possible photon energies, and to explore new regimes of quantum mechanics via interferometry between two photon sources that share no common origin.  This review will present the origins of photons in ultra-peripheral collisions and discuss the important physics that is being done with these photons, with particular emphasis on those on the energy frontier for photon physics.  

\end{abstract}

\section{Introduction: UPCs at the energy frontier}
\label{ch:Intro}

Photons are important probes of matter because they couple to all electrically charged particles.  They can be used to study diverse physics topics, both within the standard model, and beyond it. The reach of many of these studies is determined by the maximum photon energies.

Ultra-peripheral collisions (UPCs) are of importance as the energy frontier for photon studies, providing photons at least an order of magnitude more energetic than any other technique. Table~\ref{ta:maxenergy} lists the maximum energies.  At the LHC, thanks to the Lorentz boosts of both ions, photon energies can reach $10^7$ GeV in the target frame.  In the photon--nucleon center-of-mass frame, the maximum energy is 700 GeV for lead targets, and 5 TeV for proton targets. The proton--target energies are an order of magnitude higher than at the HERA $e{\rm p}$ collider, while the photon--ion energies are orders of magnitude higher than those available at fixed target facilities.

These higher energies offer many unique possibilities.  UPC photons can probe the parton content of protons and nuclei down to Bjorken$-x$  of $10^{-6}$ at moderate $Q^2,$ widening the range over which nuclear shadowing is measured, and saturation phenomena probed.   UPC photons also give access to many new phenomena.  Photoproduction of open charm, open bottom, and dijets are all within reach with UPCs at the LHC, for both proton and ion targets.  UPCs also provide opportunities for high-luminosity studies of other physics, such as meson spectroscopy, including exotic mesons. 

The LHC also produces two-photon interactions at unprecedentedly high energies.   With LHC lead--lead collisions, two-photon collisions can reach energies of up to 170 GeV, rising to 840 GeV for proton--lead collisions and 4.2 TeV for ${\rm pp}$ collisions.  This allows  us to search for beyond-standard-model (BSM) physics at unprecedentedly high energies, including via light-by-light scattering.   With heavy-ion beams, two-photon collisions occur at very high rates, allowing for exploration of both higher order quantum electrodynamics and BSM tests, such as the search for an anomalous magnetic moment of the $\tau^\pm$.

Beyond the higher energy, UPCs also exhibit unique quantum phenomena, since the two ions act as a two-source interferometer for vector meson production - like a two-slit interferometer, but with two independent sources.  In coherent photoproduction, it is impossible to determine which nucleus emitted the photon, and which is the target.  The destructive interference between these two possibilities is an example of the Einstein-Podolsky-Rosen paradox.  Moreover, the azimuthal angular dependence of this interference allows for precise measurements of the hadronic radius of nuclei.

Looking ahead, the planned FCC-hh and/or SPPC colliders will reach collision energies an order of magnitude higher than are available at the LHC, allowing parton distribution studies to be extended down to $x\approx10^{-7}$, and BSM searches to probe an order of magnitude higher in energy.

\begin{table}
\caption{List of beam energies and maximum center-of-mass energies for UPC photoproduction ($W_{\gamma \rm p}$) and two-photon interactions ($\sqrt{s_{\gamma\gamma}}$) respectively, for different colliding species at different colliders.}
\begin{center}
\begin{tabular}{|l|l|r|r|r|}
\hline
Facility & System & $\sqrt{s_{NN}}$  & Max. $W_{\gamma \rm p}$ & Max $\sqrt{s_{\gamma\gamma}}$ \\
\hline
RHIC & AuAu & 200 GeV & 25 GeV & 6 GeV \\
     & pAu  & 200 GeV & 52 GeV & 30 GeV \\
     & pp   & 500 GeV & 200 GeV & 150 GeV \\
\hline
LHC 
     & PbPb & 5.1 TeV & 750 GeV & 170 GeV \\
    & pPb  & 8.16 TeV & 1.5 TeV & 840 GeV \\
    & pp   & 14 TeV   & 5.4 TeV & 4.2 TeV \\
\hline
FCC-hh~\cite{FCC:2025lpp}
    & PbPb & 40 TeV & 4.9 TeV & 1.2 TeV \\
 SPPC~\cite{CEPC-SPPCStudyGroup:2015csa}    & pPb  & 57 TeV & 10 TeV & 6.0 TeV \\  
    & pp   & 100 TeV & 39 TeV & 30 TeV  \\
\hline
\end{tabular}
\label{ta:maxenergy}
\end{center}
\end{table}

This review will begin by discussing the photon flux in UPCs, followed by a discussion of photoproduction and then two-photon interactions, before moving on to a discussion of quantum interferometry and then concluding with a discussion of future prospects.  It builds on previous reviews~\cite{Bertulani:1987tz,Bertulani:2005ru,Baltz:2007kq,Contreras:2015dqa,Klein:2017nqo,Klein:2019qfb,Klein:2020fmr,Klein:2020nvu}.

\section{The photon flux from protons and nuclei}
\label{ch:Photonflux}

\begin{figure}[t]
\centering
\includegraphics[width=.4\textwidth]{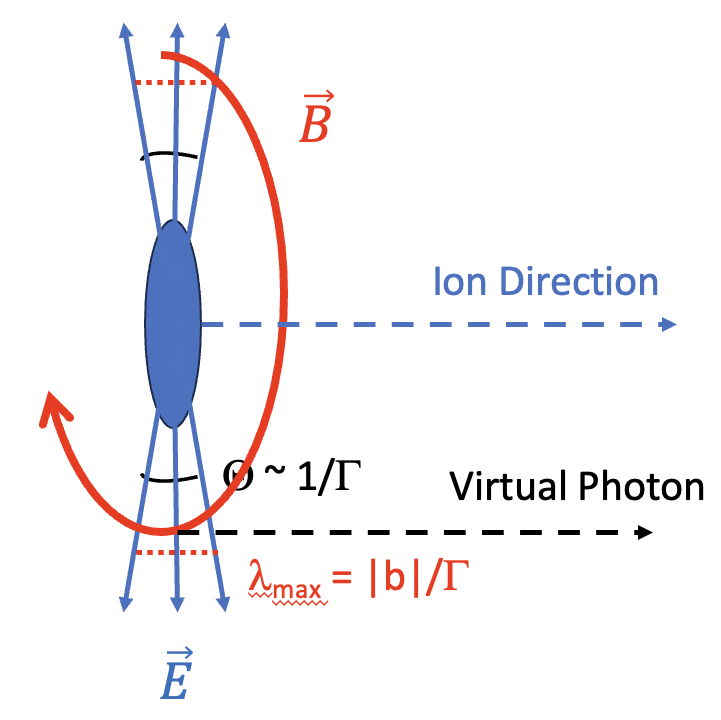}
\includegraphics[width=0.48\textwidth]{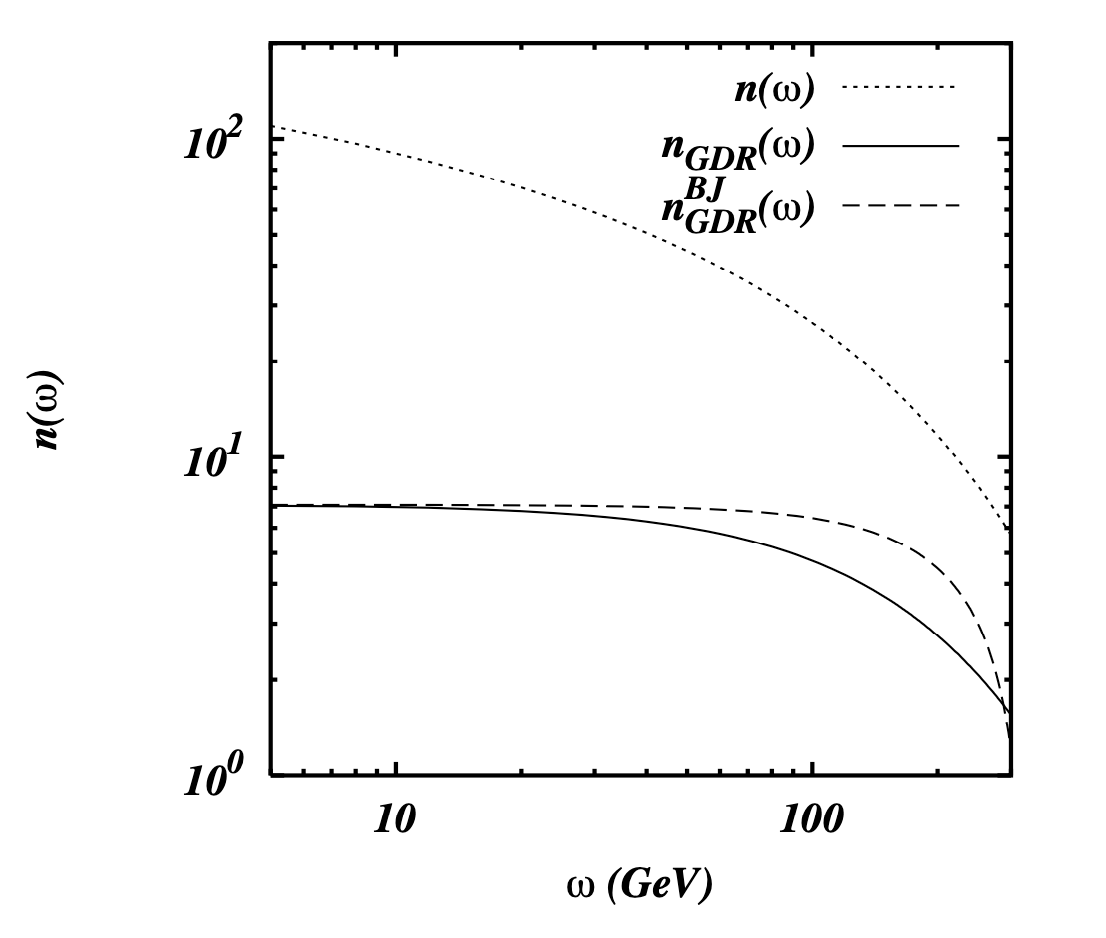}
\caption{(left) The electric (blue) and magnetic (red) fields of a relativistic ion are pancaked, and perpendicular to each other.  That field may be represented as a field of virtual photons parallel to the ion's path. $\Gamma$ is the Lorentz boost here. (right) This graph shows the photon flux from gold ions at RHIC - both unaccompanied photons (the dotted line), and those accompanied by a second photon which causes a GDR excitation.  From~\cite{Baur:2003ar}.}
\label{fig:EcrossB}
\end{figure}

As Fig.~\ref{fig:EcrossB} shows, relativistic charged particles are surrounded by a pancaked electric field, pointing radially outward, and a magnetic field that surrounds the particle. These electric and magnetic fields are always perpendicular, just as with a photon.  In 1924, Fermi expressed the fields as a flux of nearly-real photons , but did not account for relativistic contraction~\cite{Fermi:1924}.   A decade later, Weizs\"acker and Williams, included relativity, and calculated the virtual photon flux in a form that is still used~\cite{vonWeizsacker:1934nji,Williams:1934ad}.  

The electromagnetic radiation frequency spectrum is given by the Fourier transform of the electric field (assuming an equal-strength perpendicular magnetic field, as expected fro relativistic particles).  The differential photon flux ${\rm d}^3N/{\rm d}^2bdk$ at an energy $k$ and a distance $b$ from the center of a nucleus with charge $Z$, Lorentz boost $\gamma$ and velocity $v=\beta c$ is
\begin{equation}
\frac{d^3N}{d^2bdk} = \frac{Z^2\alpha k}{\pi^2\gamma^2\hbar^2c^2\beta^2} \bigg[K_1^2(x) + \frac{K_0^2(x)}{\gamma^2}\bigg],
\label{eq:photonflux}
\end{equation}
where $K_0$ and $K_1$ are Bessel functions, $\alpha\approx 1/137$ is the fine structure constant and $x=kb/\beta\gamma\hbar c$.  

The maximum photon energy $k_{\rm max}$ occurs when $x=1$,  
\begin{equation}
k_{\rm max} = \frac{\gamma \hbar c}{b}.
\end{equation}
At lower energies, when $x<1$, the flux scales as $1/k^2$, while for $x>1$, the photon flux is exponentially suppressed.  

In ultra-peripheral collisions, we require that the two nuclei physically miss each other (details below); this is roughly equivalent to requiring that the impact parameter (minimum ion-ion separation) $b> R_1 + R_2$ where $R_1$ and $R_2$ are the radii of the two ions. 

In a grazing collision with $b=R_1+R_2$, the photon may hit the target ion at a distance $R_1< b < R_1 + 2R_2$; the maximum photon energy varies across the face of the target nucleus.  Usually, this variation is ignored, and the maximum energy is taken at the center of the nucleus, where, for identical nuclei, $b=2R_A$.

With this, the total photon flux is the integral of Eq.
\ref{eq:photonflux} over $b>2R_A$:
\begin{equation}
\frac{{\rm d}N}{{\rm d}k} = \frac{4 Z^2\alpha R_A^2}{\pi^2\gamma^2\hbar^2\beta^2} 
\bigg[K_1^2(u) + \frac{K_0^2(u)}{\gamma^2}\bigg],
\label{eq:photonfluxk}
\end{equation}
where $u=2kR_A/\beta\gamma\hbar c$.  

This photon flux is calculated on the assumption that the nucleus acts as a point charge.  This works well for UPCs.  However, for photonuclear or two-photon interactions with $b<R_A$, such as photoproduction in peripheral collisions (with $b<2R_A$) or two-photon reactions within a nucleus, it is necessary to account for the distribution of charge within the nucleus, using a form factor.  
\subsection{Photon transverse momentum \label{sec:photonKt}}

To first order, the photon momentum is parallel to the nucleus direction of motion.  However, this is not exact, and the photon direction may be slightly different, with an opening angle roughly up to $\theta=1/\gamma$.   
This gives the photon a transverse momentum, $k_\perp$. Since transverse position and transverse momentum are conjugate variables, calculation of the $k_\perp$ spectrum is complex; it can only be calculated directly as long as there are no restrictions on $b$, {\it i. e.} integrating over $0 < b < \infty$.  Then, the $k_\perp$ spectrum is given by~\cite{Vidovic:1992ik}
\begin{equation}
    \frac{{\rm d}^3N}{{\rm d}^2k_\perp {\rm d}k} =
    \frac{Z^2\alpha F^2(k_\perp^2+k^2/\gamma^2)k_\perp^2}{\pi^2k(k_\perp^2+k^2/\gamma^2)^2}.
    \label{eq:kperp}
\end{equation}
This equation introduces $F(q^2)$, the form factor of the photon emitter. This is the Fourier transform of the nuclear charge distribution.  Different forms are used for different nuclei~\cite{Klein:2016yzr}.  Heavy nuclei are usually represented by a Woods-Saxon distribution, with the density profile:
\begin{equation}
\rho(r) = \frac{\rho_0}{1+\exp(r-R_{\rm WS}/d)},
\label{eq:WS}
\end{equation}
where is the hadronic Woods-Saxon radius and $d$ is its thickness.  Often, these values are obtained from electron scattering data, but, for less-known nuclei, they may be approximated $R_{\rm WS}\approx 1.2 A^{1/3}$ and
$d\approx 0.53$ fm.  The density $\rho_0$ may be calculated exactly using the normalization $\int d^3r \rho(r)=A$, where $A$ is the atomic number, or, for heavy-nuclei, approximated
$\rho_0 \approx 0.17/{\rm fm}^3$, with a slight $A-$dependent variation. 

Equation~\ref{eq:WS} gives a flat density distribution on the interior of the nucleus, with an exponential drop off near its surface.   The form factor of a Woods-Saxon nucleus may be nearly exactly represented by the convolution of a hard sphere with a Yukawa potential with range $a=0.7$ fm:
\begin{equation}
F(q) = \frac{4\pi\rho_0\hbar^3c^3}{Aq^3}
\bigg[\sin(qR_{WS}) - qR_A\cos(qR_{WS})\bigg]\bigg[\frac{1}{1+a^2q^2}\bigg]
\end{equation}
where the trigonometric terms are for the hard sphere, and the fraction accounts for the Yukawa potential with range $a$.  It should be noted that heavy nuclei like gold and lead are thought to have a neutron shell, so the hadronic radius $R_A > R_{\rm WS}$, by about 1/4 to 1/2 fm.

Protons are usually treated with a dipole form factor, 
\begin{equation}
    F(q^2) = \frac{1}{1+q^2/({\rm 0.71 \, GeV})^2},
    \label{eq:FFdipole}
\end{equation}
where $q^2=k_\perp^2+k^2/\gamma^2$.  The dipole form factor gives a good fit to electron elastic scattering data. For intermediate mass nuclei, roughly $1 < A < 14$, a Gaussian form factor works well.

As was noted, Eq.~\ref{eq:kperp} is derived assuming $0 < b < \infty$.  Since $b$ and $k_\perp$ are conjugate variables, when $b$ is constrained, such as $b>2R_A$, by the uncertainty principle, the average $k_\perp$ should rise. Unfortunately, there is no exact solution for restricted $b$. A solution does exist for two-photon interactions, as will be discussed later.  In the general case, it is possible to derive an approximate expression~\cite{Klein:2020jom}
\begin{equation}
\langle k_\perp^2 \rangle \propto \langle 1/b^2 \rangle.
\end{equation}

\subsection{Multiple photon exchange}

Because $Z^2\alpha >1$ for heavy nuclei, many UPC reactions involve multiple photon exchange: either two photons from one nucleus, or one from each.  As long as  $k\ll \gamma m_{\rm p}$, multiple photon emission is independent~\cite{Gupta:1955zz}, save for the common $\vec{b}$~\cite{Baur:2003ar}.     

The common $\vec{b}$ does introduce correlations, since the photons share the same polarization, and since photon energy spectrum depends on $b$.  
For multiple photon exchange, cross sections may be calculated in impact-parameter space:
\begin{equation}
\sigma = \int {\rm d}^2b P_1(b) P_2(b)...P_{\rm nohad} (b),
\label{eq:multisigma}
\end{equation}
where $P_1(b)$, $P_2(b)$ etc. are the probabilities for the given reactions to occur at a given impact parameter,
\begin{equation}
    P(b) = \frac{d^3N}{d^2bdk} \sigma(\gamma A\rightarrow X),
\end{equation}
where $\sigma(\gamma A\rightarrow X)$ is the reaction of interest; $P_{\rm nohad} (b)$ is the probability of not having a hadronic interaction.  

Per Eq.~\ref{eq:multisigma}, if $n$ photons are exchanged, then the interaction probability scales as $1/b^{2n}$; as $n$ increases, the impact parameter distribution is more and more peaked around $b=2R_A$~\cite{Baur:2003ar}.  Then, the mean impact parameter is
\begin{equation}
\langle b \rangle = 
\frac{\int_0^{b_{\rm max}} {\rm d}^2b \, b  P(b)}
{\int_0^{b_{\rm max}}   {\rm d}^2b P(b)}
\label{eq:bbar}
\end{equation}
where $b_{\rm max} = \gamma\hbar c/k$ is the maximum impact parameter for the photon energy. For $1-$ photon exchange 
\begin{equation}
\langle b \rangle =\frac{b_{\rm max}-2R_A}{\ln{b_{\rm max}/2R_A}},
\end{equation}
where $P(b)\propto 1/b^2$ up to $b_{\rm max}$.
For $n>1$, the equation simplifies, and
\begin{equation}
\langle b \rangle = \frac{2n-2}{2n-3},
\end{equation}
independent of the photon energy.  As $n$ increases, $\langle b \rangle$ decreases.
The changes in impact parameter distribution alter the photon energy spectra; a photon accompanied by one or more other photons has a smaller $\langle b \rangle$ and a harder energy spectrum than a single photon.  Figure~\ref{fig:EcrossB} compares the photon spectrum for single photons and those accompanied by a second photon which excites one of the nuclei. 

\subsection{Solving the two-fold photon energy ambiguity}
\label{sec:solvingambiguity}

The different photon spectra can solve an important problem in studying exclusive production, discussed in Sec.~\ref{sec:diffraction}: solving the two-fold ambiguity as to which nucleus emitted the photon, and which was the target (``pomeron emitter").   For a final state with mass $M_V$ and rapidity $y$, the photon energies for the two possibilities are
\begin{equation}
    k=\frac{M_V}{2} e^{\pm y}
    \label{eq:twofold}
\end{equation}
where the sign of the $\pm$ depends on the photon direction.  This complicates the problem of determining the photon-energy dependence of the cross section, as will be discussed in Section~\ref{ch:photonuclear}.  One can write equations with the form
\begin{equation}
\frac{{\rm d}\sigma(AA\rightarrow AAV)}{{\rm d}y} = 
\frac{{\rm d}N}{{\rm d}k_1}\sigma(\gamma(k_1) A\rightarrow VA) +
\frac{{\rm d}N}{{\rm d}k_2}\sigma(\gamma (k_2) A\rightarrow VA)
\label{eq:twofoldsolution}
\end{equation}
where $1$ and $2$ correspond to the two photon directions in Eq.~\ref{eq:twofold}.   One can write sets of equations for exclusive production without nuclear breakup, with nuclear breakup on one side, or nuclear breakup on both sides, corresponding to 1, 2 and 3-photon exchange.   The ratios of the ${\rm d}N/{\rm d}k$ for the two directions will be different, leading to a set of linear equations which can be solved to uniquely find $\sigma(\gamma(k_1) A\rightarrow VA)$ and 
$\sigma(\gamma(k_2) A\rightarrow VA)$.
One can also use photoproduction in peripheral collisions ($b<2R_A$)~\cite{Contreras:2016pkc}, albeit with some questions about the degree of coherence \cite{Zha:2017jch}.

There is one important caveat here: that each photon only does one thing.  This is sometimes known as 'factorization.'  Factorization holds for coherent photoproduction, in which the momentum transfer to the target is small.  However, it does not hold for incoherent photoproduction, whereby enough energy may be transferred to the target to excite it.  Comparisons between data on exclusive $\rho$ photoproduction from STAR~\cite{STAR:2007elq} and ALICE~\cite{ALICE:2020ugp} and theory~\cite{Baltz:2002pp} show that this factorization, Eq.~\ref{eq:multisigma}, does hold, although STAR sees some hints that it may fail at larger transverse momentum of the final state, where incoherent production is larger.    Both the STARlight~\cite{Klein:2016yzr} and n00n~\cite{Broz:2019kpl} generators can simulate these additional photons, and find good agreement with data.

\subsection{Low-energy photon interactions}

These equations rely on the fact that one can `tag' reactions of interest as being accompanied by additional photon exchange.   Because the photon spectrum is peaked at low photon energies, and because  $\sigma(\gamma A\rightarrow X)$ is also peaked at low energies, most of the excitations are soft, leading only to nuclear excitations, without the production of hadrons.   The rates for these excitations can be calculated by convoluting the photon flux with tabulations of the total photonuclear cross sections~\cite{Baltz:1996as, Baltz:2002pp,Baltz:2009jk}.

The lowest energy (and most common) excitation is the Giant Dipole Resonance (GDP)~\cite{Berman:1975tt}. A GDR is a broad collective oscillation of protons against neutrons in a nucleus; it usually decays by single neutron emission. For lead, the excitation is peaked at 13 MeV; the energy increases for lighter nuclei.  More energetic photons can cause more energetic excitations, which may decay by emitting multiple neutrons and/or protons~\cite{ALICE:2022iqi,ALICE:2024vpj}.  These data can generally be well described using nuclear fragmentation models~\cite{ALICE:2024vpj}.  These reactions are of some popular interest due to their ability to transform lead into gold.  

More energetic photons can excite individual nucleons.  Most commonly, a photon will excite a proton or neutron into a $\Delta$ resonance, which will then decay by emitting a pion.  These pions can be detected in forward detector elements.  If these forward detector elements are also used as veto detectors, it is necessary to correct the photon flux to account for these secondary interactions~\cite{ALICE:2020ugp}. Still more energetic interactions can result in the emission of multiple pions; the most energetic secondary photon exchanges can cause the primary reaction products to be obscured. 

\section{Photonuclear Interactions}
\label{ch:photonuclear}
Color charges are not found free in Nature, but they are bound inside hadrons---particles made of quarks and gluons---like nucleons and nuclei. The structure of a hadron in terms of QCD is dynamical, presenting a different configuration each time we take a snapshot of it, because quarks and gluons not only radiate, producing more quarks and gluons, but they can also annihilate reducing their number. Understanding the structure of hadrons when observed with a high-energy probe is one of the main research topics in QCD nowadays.

Even though photons lack color charge,
they are good probes of the QCD structure of nuclei. Differently than the field appearing in the Lagrangian, an actual photon is not only a photon, but also all other states to which it can couple. Well before the advent of QCD, it was recognized  that a physical photon has a hadronic structure~\cite{Bauer:1977iq}, so that it could fluctuate into vector mesons---hadrons with the same quantum numbers as the photon. Later on, deep-inelastic scattering~\cite{Devenish:2004pb} was described, in the infinite momentum frame, as the interaction of  a virtual photon with a proton made of partons---pointlike constituents of the nucleon, which in QCD are identified with quarks and gluons~\cite{Bjorken:1969ja}. In this picture, the photon can couple to the quarks in the proton because quarks, in addition to color, also carry electric charge. A third view, very useful when studying QCD at high energies, considers the fluctuation of a photon into a quark-antiquark color dipole in a reference frame where this dipole is long lived, such that it can interact with the hadron and probe its color structure~\cite{Mueller:1989st,Nikolaev:1990ja}.

To talk about nuclear structure is practical, but somehow imprecise. What is meant, is that a given probe and an observable are sensitive to specific characteristics of the nuclear structure, which are encoded in so called parton distributions. These 
are obtained from QCD through factorization theorems that separate the non-perturbative contributions from the hard part of the process which is in principle computable. As observables furnish a natural definition of a preferred direction---e. g. along or perpendicular to a particle incoming into a collision---parton distributions (besides depending on a normalization scale) are 
in principle dependent on ($i$) the longitudinal and transverse components of the four momentum of the incoming hadron, ($ii$) the corresponding four momenta of the  parton(s) participating in the hard process, and ($iii$)  on the four momentum exchanged in the interaction. Some observables are not sensitive to one or more of these variables, which means that an experimental measurement effectively integrates over them. For example, the well known parton distribution functions (PDFs) are extracted from inclusive deep-inelastic scattering experiments, which depend only on the fraction of the longitudinal momentum of the hadron carried by the parton participating in the hard interaction. This fraction is commonly denoted by $x$; to reach  small values of $x$ large interaction energies are required; i.e.  the high-energy limit of perturbative QCD corresponds to the small $x$ limit.  Other observables are sensitive for example to the transverse momentum of the partons in the hadrons, so their description requires transverse-momentum-dependent distributions (TMDs); or they may be sensitive to the location of the partons in the transverse plane, in which case generalized parton distributions (GPDs) are needed to describe the structure of hadrons. There are other possible distributions, for a review see e.g.~\cite{Diehl:2015uka} and references therein.

UPCs have been extensively used to study the structure of nuclei and their evolution---that is, how the structure is modified when one of the variables it depends upon changes. It was found at HERA~\cite{H1:2015ubc}, that the structure of protons at high energies is dominated by gluons whose density grows steeply as $x$ decreases. Specifically, at a fixed scale $Q^2$, the gluon density grows as a power law 
\begin{equation}
xg(x,Q^2)\propto x^\lambda,
\label{eq:gluonpowerlaw}
\end{equation}
 where $\lambda=\lambda(Q^2)$ varies from 0.1 to 0.4  when $Q^2$ changes from 1 GeV$^2$ to 150 GeV$^2$~\cite{H1:2001ert}. 
A similar power law growth of the gluon distribution is expected to occur for nuclei, 
with two phenomena  the main object of study at high energies:
\begin{itemize}
\item Shadowing  is the experimental fact that the structure of nuclei is different than  the sum of the structure of $A$ nucleons, with $A$ the atomic mass number~\cite{Armesto:2006ph}. In particular, at small $x$ the nuclei are measured to have less partons participating in interactions than $A$ protons. Within QCD this reduction is attributed to absorption corrections from multiple scattering of the probe with the different nucleons in the nucleus.
\item The other phenomenon is saturation: at low energies the partons in a hadron behave as quasi-free during a hard interaction and their density grows towards small $x$ due to gluon bremsstrahlung, but at some point the gluon density is so large that the rate of annihilation processes balances gluon splitting and the density reaches a dynamic equilibrium and ceases to grow.  Within QCD this is explained with an integro-differential equation for the dipole-target scattering amplitude. At low gluon densities, gluon splitting dominates and one obtains a linear equation. When the gluon density increases, the annihilation process gains in importance and the equation becomes non-linear~\cite{Kovchegov:2012mbw}. 
\end{itemize}
Saturation, even though proposed several decades ago~\cite{Gribov:1983ivg,Mueller:1989st}, has not been yet unambiguously determined experimentally. Finding saturation and studying its properties is one of the scientific pillars of the upcoming EIC (electron-ion collider), currently under construction~\cite{Accardi:2012qut,AbdulKhalek:2021gbh}. One of the main challenges to understand the high-energy structure of nuclei is to disentangle the effects due to shadowing, from those due to saturation. Photon-induced interactions offer a tool for exploring and understanding this issue.

Photonuclear strong interactions in UPCs can  proceed without the exchange of net color charge in diffractive processes, discussed in Sec.~\ref{sec:diffraction}, as well as through the exchange of color charges, that is, non-diffractive processes,  discussed in Sec.~\ref{sec:otherProbs}. The experimental and theoretical challenges are different and complementary in these two classes of processes, providing a wide spectrum of tools for exploring QCD at high energies.

\subsection{Probing nuclear structure at high energies with diffractive vector meson production \label{sec:diffraction}}
In diffractive interactions there is no net color transfer. Experimentally, this is reflected in the occurrence of large rapidity gaps, regions of the detector covering a range in rapidity with no activity above the noise level.  Theoretically, this is described by a pomeron, which is a complex system of at least two gluons with compensating color charges~\cite{Lipatov:1985uk,Forshaw:1997dc,Donnachie:2002en}. The fact that diffractive processes involve more than one gluon makes them very sensitive to the gluon structure of hadrons and provide  a prime tool to search for saturation and shadowing related phenomena in nuclei.

Diffractive vector meson photoproduction is probably the most studied  process in UPCs because it furnishes striking experimental signatures with large cross sections allowing for precise measurements and, being a diffractive process, is highly sensitive to the nuclear gluon distribution and its evolution.  Furthermore, the different masses of vector mesons provide a handle to study the scale dependence of the gluon density. In addition, within the Good-Walker approach (Sec.~\ref{sec:gw}) these processes offer a tool to study not only the average gluon density, but also are  sensitive to quantum fluctuations of the color charges in hadrons. 

\begin{figure}[t]
\centering 
\includegraphics[width=0.48\textwidth]{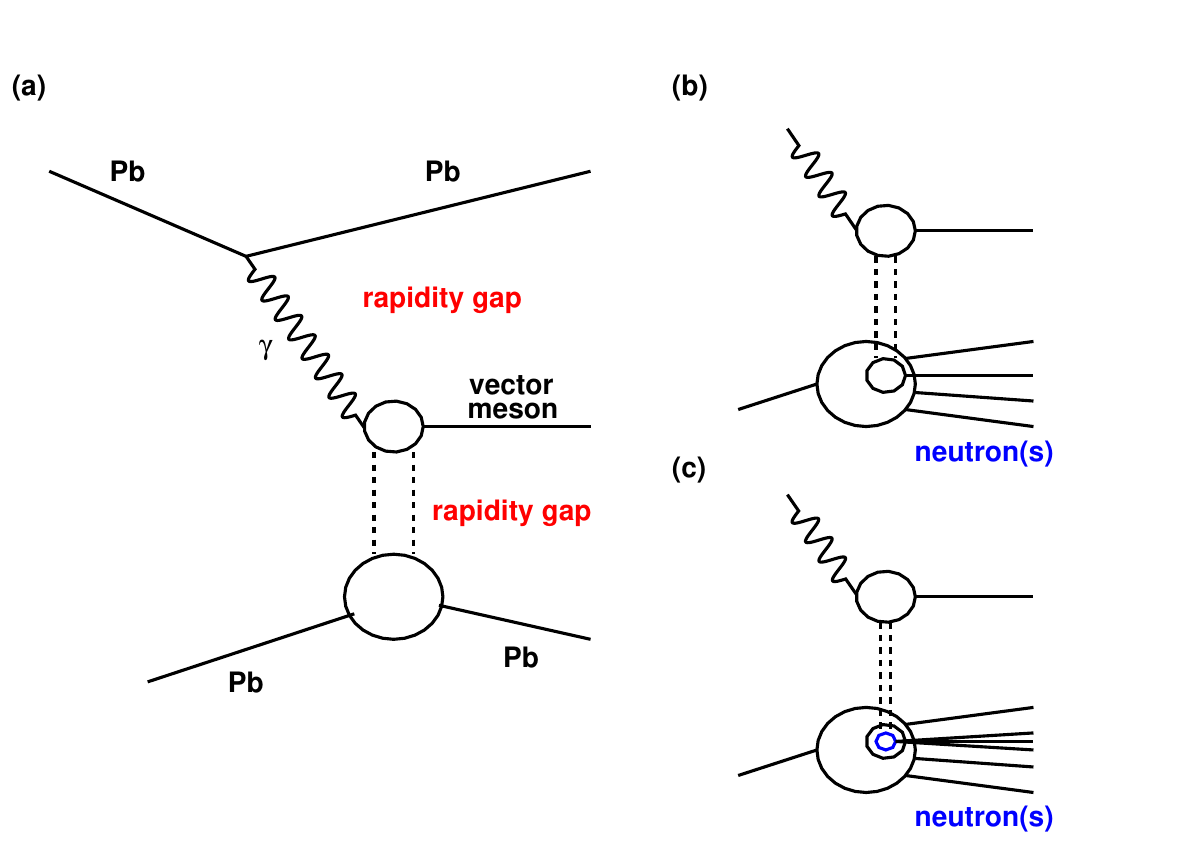}
\includegraphics[width=0.4\textwidth]{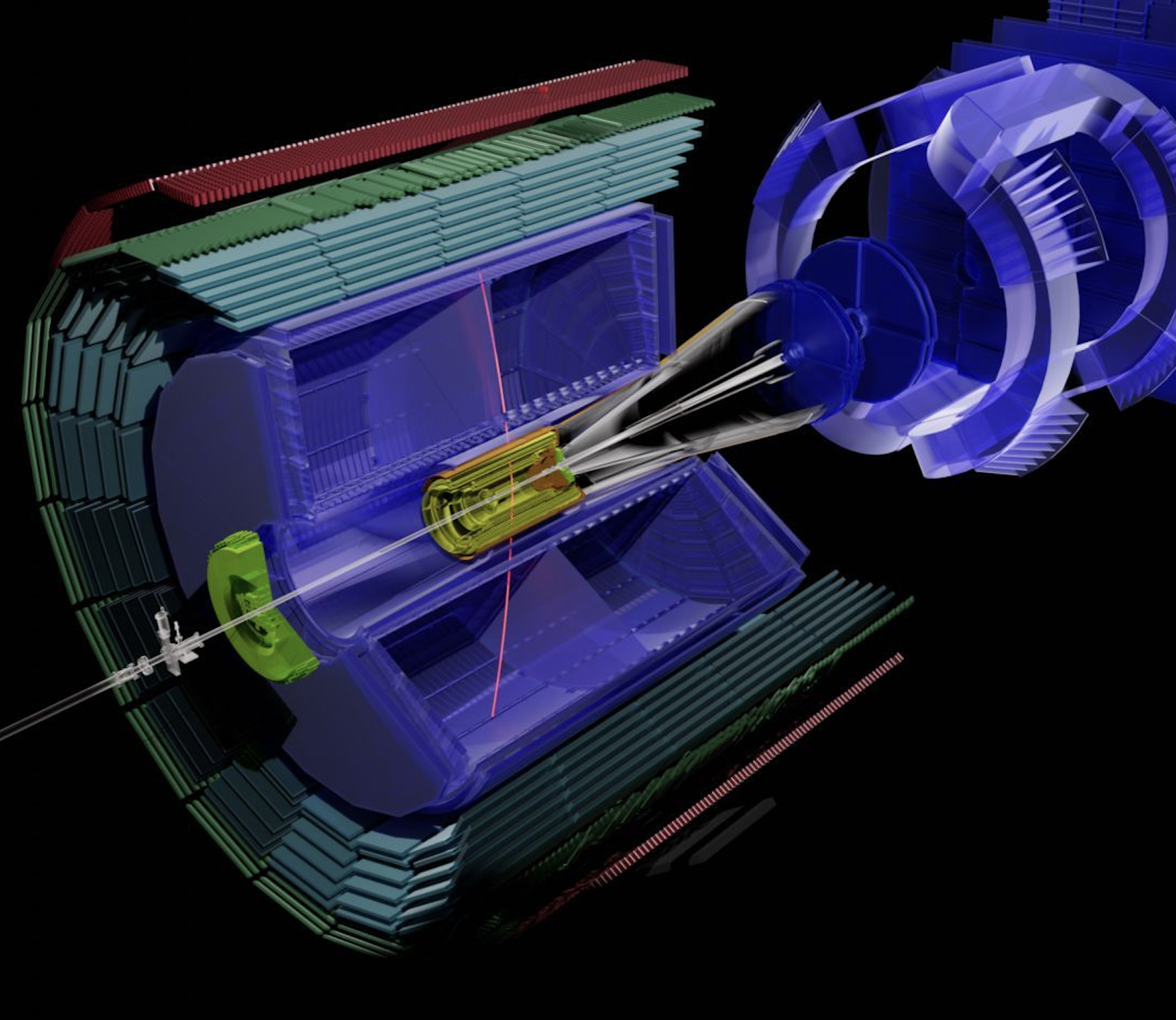}
\caption{
\label{fig:vmProd} 
(a) Diagram of diffractive coherent vector meson photoproduction in Pb--Pb collisions. One Pb nucleus  emits a quasi-real photon that fluctuates into a quark-antiquark color dipole which interacts with the full color field of the second Pb nucleus through the exchange of a pomeron depicted by the dash lines. A vector meson is produced and both nuclei remain intact.  In the case of incoherent production, the interaction happens with either (b) a nucleon which remains intact, but the nucleus dissociates or (c) with a subnucleon structure and both the nucleon and the nucleus dissociate. In both cases the dissociation products are produced at beam rapidities and normally include a neutron.  The right panel shows a real event, recorded by the ALICE experiment, of coherent vector meson production, with the two tracks from the vector meson decay shown in red.}
\end{figure}

In a reference frame where  photon fluctuations are  long-lived, this process occurs in three steps: ($i$) well before the interaction the photon fluctuates into a hadronic state---e. g. a vector meson or a color dipole---which ($ii$) exchanges a pomeron with the target, and ($iii$) lastly a vector meson is produced.  If the interaction takes place with the color field of the nucleus as a whole, one speaks of coherent production. If the interaction takes place with one part of the nucleus, the production is called incoherent. Incoherent interactions are subdivided into two classes: either one full nucleon participates in the collision or a subnucleon structure (commonly called a hotspot) is the one to interact; the latter class is known as incoherent dissociative production. Figure~\ref{fig:vmProd}  illustrates these processes whose experimental signature are large rapidity gaps, the decay products of the produced vector meson, and in the incoherent case the products of the nucleus dissociation which are emitted near beam rapidities.

A nice feature of these type of processes is that the full event kinematics can be reconstructed from  the four momentum of the vector meson and the energy of the incoming hadrons participating in the interaction. This is so, because the interaction of the pomeron with the target (either the full nucleus, a nucelon, or a hotspot) is elastic. The center-of-mass energy of the photon--hadron ($W_{\rm \gamma H} $) system is
\begin{equation}
W^2_{\rm \gamma H} = m \sqrt{s}\exp{(-y)},
\label{eq:W2}
\end{equation}
where $\sqrt{s}$ is the center-of-mass energy of the colliding hadrons, $m$ the mass of the vector meson, and $y$
its rapidity measured in the laboratory frame  with respect to the direction of the 
hadron participating in the photonuclear interaction. The fraction of the longitudinal momentum of the target carried by the pomeron, a proxy for $x$, is commonly defined as
\begin{equation}
x=\frac{m^2}{W^2_{\rm \gamma H} },
\label{eq:xProxy}
\end{equation}
where quasi-real photons are assumed. The momentum transferred in the interaction, given by Mandelstam-$t$, is obtained from 
 \begin{equation}
-t = p^2_\perp,
\label{eq:mant}
\end{equation}
with $p_\perp$ the magnitude of the transverse momentum of the vector meson. This last equation disregards the transverse momentum of the incoming photon in the laboratory frame, see Sec.~\ref{sec:photonKt}. This is  a good approximation except at very small values of $|t|$ where a correction must be applied. Finally, in the photoproduction regime, the scale of the process is related to the mass of the vector meson as 
\begin{equation}
    Q^2\propto (m/2)^2,
    \label{eq:q2m}
\end{equation}
where the proportionality constant is normally taken to be one.

\subsubsection{The Good-Walker approach to diffraction \label{sec:gw}}
Hadrons are quantum objects, so their color-field configuration is dynamical and differs from one interaction to another.  Well before the advent of the Standard Model, Good and Walker described diffraction by writing the incoming hadron as a superposition of basis
states, each of which is absorbed differently~\cite{Good:1960ba}. This approach was extended to diffraction in QCD~\cite{Miettinen:1978jb}.
It was then used to describe coherent vector meson production~\cite{Frankfurt:2011cs} and later on applied to the incoherent
production of vector mesons~\cite{Mantysaari:2016ykx}.
 
The main interest in this approach resides in the fact that coherent (coh) processes  depend in the average gluon density, while incoherent (inc) interactions  are sensitive to fluctuations of the gluon field~\cite{Mantysaari:2016ykx}:
 
 \begin{equation}
 \left. \frac{{\rm d}\sigma}{{\rm d} t}\right|_{\rm coh}  \propto \left| \left< A(x,Q^2,\vec{\Delta})\right>\right|^2,
 \ \ \
 \left. \frac{{\rm d}\sigma}{{\rm d} t} \right|_{\rm inc} \propto 
  \left(
  \left< \left| A(x,Q^2,\vec{\Delta})\right|^2\right> -  \left| \left< A(x,Q^2,\vec{\Delta})\right>\right|^2
\right).
 \end{equation}
Here, $A$ is the amplitude for the interaction of hadron $h$, i.e. the proton or nucleus, with the photon $\gamma$.  $\vec{\Delta}$ is the momentum transferred at the target vertex, the average ($\left<\right>$) is over all possible configurations  of the hadronic color field, $Q^2$ is the scale of the interaction which for the case of a quasi-real photon  is proportional to the mass squared of the vector meson, and  $x$ is the fraction of the longitudinal momentum of the hadron carried by the pomeron. In this approach there is a bound relating the total photonuclear cross section to those of the coherent and incoherent processes as
\begin{equation}
\sigma_{\rm coh}+\sigma_{\rm inc} \le \frac{1}{2}\sigma_{\rm tot}, 
\end{equation}
with the equality valid when all basis states are completely absorbed in the interaction. In this case, the incoherent cross section is zero. The rest of the cross section is comprised by non-diffractive inelastic scattering, with some processes in this category discussed in Sec.~\ref{sec:incJet} and Sec.~\ref{sec:openCharm}.

As mentioned before, diffractive processes involve the exchange of more than one gluon, so the parton distribution that is needed to compute the amplitude $A$ is the generalized parton distribution (GPD) or  even the generalized transverse momentum dependent parton distribution (GTMDs), see e.g.~\cite{Boer:2023mip}. Nonetheless, it has been shown that a good model for the GPDs entering vector meson diffractive photoproduction is the use of standard PDFs and a skewedness correction~\cite{Shuvaev:1999ce}. In this context, and taking into account that it is expected that one gluon takes most of the momentum and the rest of the gluons ensure the formation of a singlet color state, $x$ can be taken as a proxy of the variable entering the PDFs.
 
Incoherent diffractive vector meson production in the  Good-Walker formalism offers thus a tool to sample the  fluctuations of the color field of nuclei at femtometer (interaction with the full nucleus) and subfemtometer scales (dissociative). LHC provides for the first time the opportunity  to explore these  quantum phenomenon in nuclei over  large energy and Mandelstam-$t$ ranges.  First results will be discussed in Sec.~\ref{sec:fluct}.
 
Even though this formalism yields successful predictions, it is not clear how it works in quantum field theory.  
There are some conceptual issues with the Good-Walker approach~\cite{Klein:2023zlf}, connected with the definition of coherence.  In the Good-Walker formulation, coherence requires that the nucleus remain in its initial, ground, state.  However, as we have already discussed, the presence of additional photon exchange can lead to reactions like $AA\rightarrow A^*A^*V,$ where $A^*$ indicates that the nucleus has broken up.  Even in these cases, vector meson photoproduction exhibits a peak at low $p_\perp$, characteristic of addition of amplitudes in-phase---the signature of coherent production.  Similar low $p_\perp$ peaks have been seen in peripheral heavy-ion collisions (see Sec.~\ref{sec:rapdep}), where a coherently produced vector meson may be accompanied by hundreds of other particles~\cite{STAR:2019yox,ALICE:2022zso}.  In the Good-Walker paradigm, these reactions are classified as incoherent, despite the low $p_T$ peak that is a signature of coherent production.  It is unclear if they should actually be used as a measure of event-by-event fluctuations. 

Another issue comes from incoherent photoproduction at low $p_\perp$.  In terms of low$-x$ quarks and gluons, lead-208 and gold-197 look very similar.  However, in the nuclear shell model, they are very different.  Lead-208 is doubly magic, with a minimum excitation energies of 2.6 MeV, compared to 77 keV for gold-197.  This has important implications for incoherent photoproduction, since the 2.6 MeV first excited state is a lower limit on energy transfer, and so limits the phase space for production at low $p_\perp$, leading to significant differences between lead and gold.  The reasons for these apparent paradoxes are not understood, but they may be alleviated with higher-order formulations of Good-Walker. 

\subsubsection{Experimental challenges in diffractive vector meson photoproduction}
Hadron colliders like  LHC or RHIC are designed and operated to study physics processes, like Higgs production or the creation of the quark-gluon plasma, where many particles are produced and fill the full rapidity range covered by the detectors, which are also designed and operated accordingly. To study diffractive vector meson photoproduction where the detectors are basically empty except for the decay products of the vector meson is challenging, because the experimental requirements are very different to those of the main physics topics for which the facilities were built.

The fact that these measurements of diffractive vector meson photoproduction have been carried out, as described in the following sections, and that in some cases the achieved precision matches that obtained at specialized accelerators like HERA is a testament  to the creativity of the experimentalists at these facilities. Some of the principal experimental challenges to perform the analyzes described below are:
\begin{itemize}
\item Pileup. The LHC is operated at as high interaction rates as possible, which leads to  simultaneous collisions at the experiments; this effect is known as pileup and contaminates the rapidity gaps of diffractive collisions with particles originating in other interactions. This is particularly problematic in proton--proton (pp) collisions where the beam intensities are intentionally so high that tens and even hundreds of pp collisions happen in the same bunch crossing. For collisions of heavy ions at RHIC and  LHC the beam intensities are much less, so that the probability of having more than one hadronic collision in a single bunch crossing is very small. But, as the intensity of the electromagnetic field is so high, there is pileup from the production of soft electron-positron pairs in photon--photon interactions. Another form of pileup occurs when the latency of a subsystem of a large detector is longer than the time between consecutive interactions, for example the time-projection chambers (TPCs) of STAR~\cite{STAR:2002eio} or ALICE~\cite{ALICETPC:2020ann}, or when the full detector is continuously readout as is the case of ALICE during LHC Run 3~\cite{ALICE:2023udb}.
\item Rapidity-gap survival. Rapidity gaps can also be suppressed by secondary interactions of the same two hadrons participating in the diffractive process. This is particularly important in pp collisions where QCD inspired models are used to compute gap-survival factors which change the cross section  from around 10\% at low energies to more than 50\% at the highest energies reached at the LHC~\cite{Jones:2016icr}. 
\item Disentangling the contribution of the low and high energy photons. As previously explained, see Eq.~\ref{eq:twofoldsolution}, the measured cross section has two contributions, corresponding to photon emission from one or the other incoming hadron.  At midrapidity, both contributions are the same and there is no problem, but at rapidities different from zero one photon is more energetic than the other, see Eq.~\ref{eq:twofold}. There are several strategies to deal with this problem. In pp collisions the low energy photon covers a range of $W_{\rm \gamma p}$ previously measured at HERA. HERA measurements are then assumed for the low energy photonuclear cross section and the measurement, along with a computation of the photon fluxes, is used to extract the photonuclear cross section at high energies. In p--Pb collisions, owing to the $Z^2$ dependence of the photon flux (see Eq.~\ref{eq:photonfluxk}), the contribution where the proton emits the photon is negligible and there is no ambiguity. In Pb--Pb collisions, the strategy is to perform the measurement for different ranges of the impact parameter as explained in Sec.~\ref{sec:solvingambiguity}. 
\end{itemize}

\subsubsection{Longitudinal momentum dependence of proton structure\label{sec:protonxdep}}
The energy dependence of coherent production of J$/\psi$ was identified early on as an ideal process to search for gluon saturation, because the mass 
 of the J$/\psi$ furnishes a hard enough scale to justify the use of perturbative QCD, and the cross section, computed in the collinear approach and at leading logarithmic accuracy, depends on the {\em square} of the gluon distribution in the hadron~\cite{Ryskin:1992ui}. Equations~\ref{eq:W2} 
 and~\ref{eq:xProxy}  indicate that the measurement of the rapidity dependence of this process gives access to the $x$ evolution of the gluon distribution. 

\begin{figure}[t]
\centering 
\includegraphics[width=0.47\textwidth]{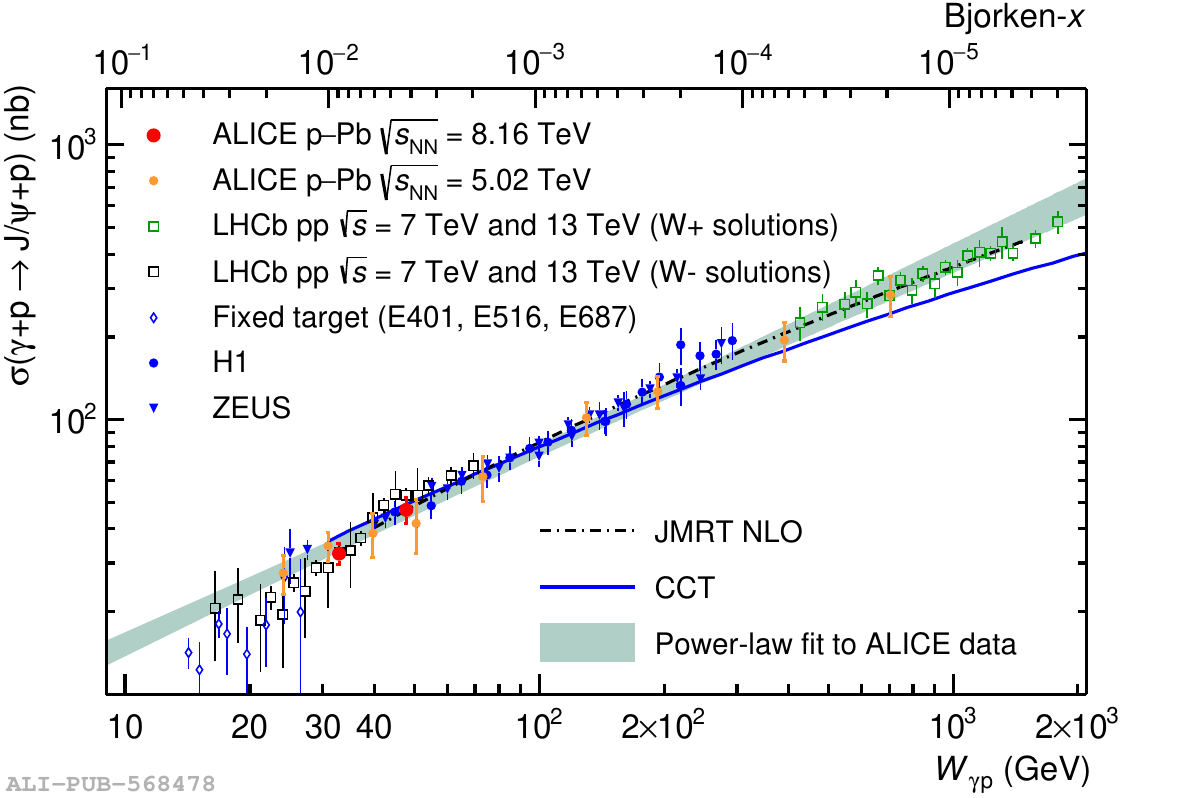}
\includegraphics[width=0.43\textwidth]{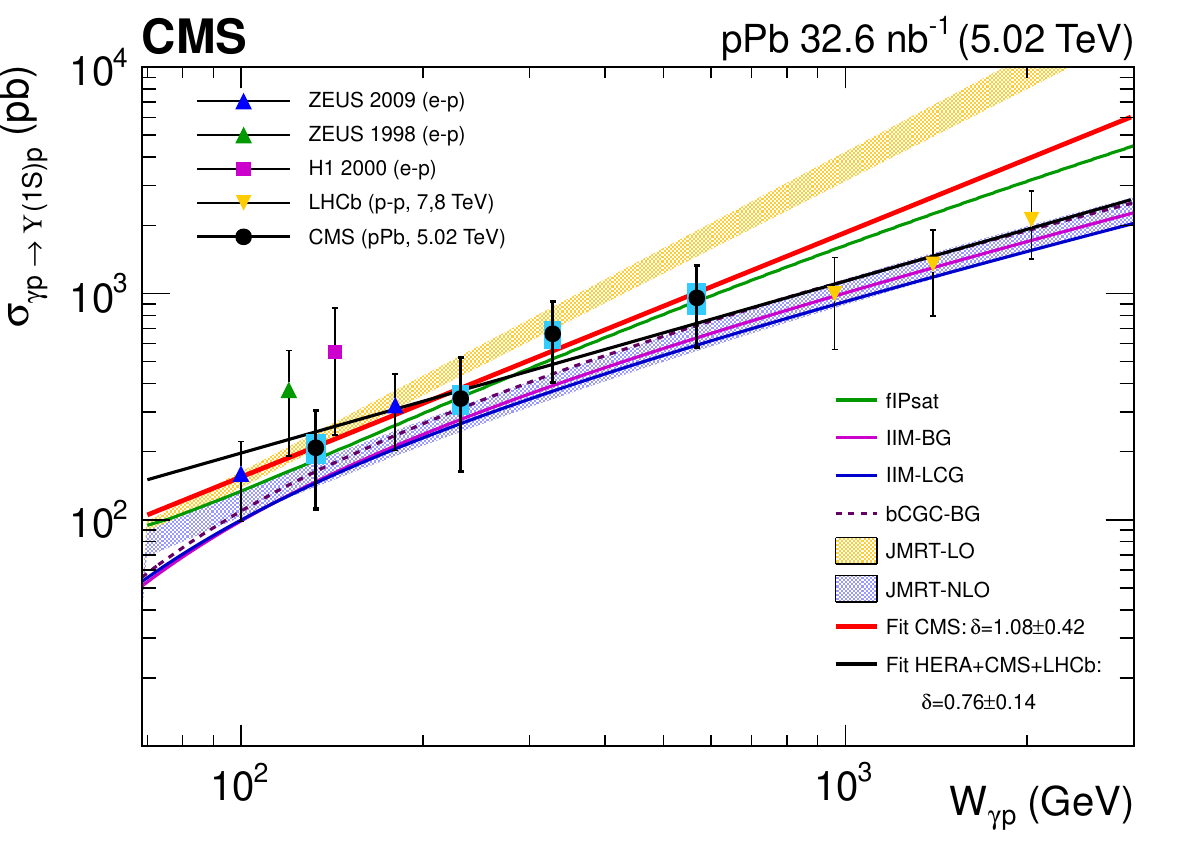}
\caption{
\label{fig:jpsiOffProtons} 
Summary of experimental data on coherent J$/\psi$ (left) and $\Upsilon(1S)$ (right) diffractive photoproduction off protons. Taken from Refs.~\cite{ALICE:2023mfc} and~\cite{CMS:2018bbk}, respectively.}
\end{figure}

Figure~\ref{fig:jpsiOffProtons}, taken from~\cite{ALICE:2023mfc}, summarizes data on the cross section of coherent J$/\psi$ diffractive photoproduction off protons. 
Several features are noteworthy. The measurements span more than three orders of magnitude in $x$ with a precision of around 10\% or less for most of the data points. This large continuous coverage of phase space is mandatory to really test models because QCD predictions stem from integro-differential equations depending on the logarithm of the evolved variable. The measurements carried out in different facilities and environments---fix target, HERA, LHC pp UPCs, LHC p-Pb UPCs---agree with each other. Finally, the measurements are consistent with a power-law growth with energy, correspondingly decreasing $x$. The exponent of the power law is found to be $\delta=0.70\pm0.04$~\cite{ALICE:2023mfc}, which follows  expectations based on the dependence of this process on the square of the gluon distribution  and the rise of the gluon distribution observed at HERA:
\begin{equation}
\sigma({\rm \gamma+ p\to J/\psi+p} ) \propto W^{\delta} \propto x^{-\frac{\delta}{2}} \approx x^{4\lambda},
\end{equation}
where the first equality uses Eq~\ref{eq:xProxy}, and the last expression Eq.~\ref{eq:gluonpowerlaw} is evaluated at the appropriate scale, see Eq.~\ref{eq:q2m}.

Figure~\ref{fig:jpsiOffProtons} also shows a summary of measurements for $\Upsilon$ production, where the same features as just discussed for the case of J$/\psi$ are present. These observations imply that saturation effects are not visible for this observable even at the highest  energies, which enables the use of these measurements as a baseline for nucleus-based searches of this phenomenon.

\subsubsection{Rapidity dependence of coherent photonuclear J$/\psi$ production\label{sec:rapdep}}
The coherent diffractive photoproduction of J$/\psi$ has also been studied utilizing nuclear targets in UPCs. The main motivation is that the larger number of partons in a nucleus, with respect to a single nucleon, increases the gluon densities  and saturation should set earlier, with $x$ larger by an $A^{1/3}$ factor~\cite{McLerran:1993ni}. 

A proof-of-principle measurement of $J/\psi$ photoproduction with a handful of events was  performed early at RHIC in Au--Au UPCs~\cite{PHENIX:2009xtn}. Since then, many measurements of this process have been carried at RHIC and LHC both in UPCs as well as in peripheral collisions. In some cases, the UPCs measurements have been performed in different classes using multi-photon exchange processes,  which allows to extract the photonuclear cross sections, see Sec.~\ref{sec:solvingambiguity}. 

Using UPCs at the LHC, the rapidity dependence of photonuclear J$/\psi$ production has been measured at large rapidities by the LHCb~\cite{LHCb:2025fzk} and ALICE~\cite{ALICE:2023jgu,ALICE:2026mlg} collaborations; at intermediate rapidities by the CMS collaboration~\cite{CMS:2023snh}; and at midrapidities by the ALICE~\cite{ALICE:2021gpt} and ATLAS~\cite{ATLAS:2025aav} collaborations. The measurements from the different collaborations agree reasonably with each other within the current experimental uncertainties, with one exception. The results from the ATLAS collaboration show an important disagreement with those from ALICE at midrapidity. ATLAS results are quite new and the tension with the previous ALICE results is  one of the main experimental open issues to be solved with the next generation of measurements.

\begin{figure}[t]
\centering 
\includegraphics[width=0.46\textwidth]{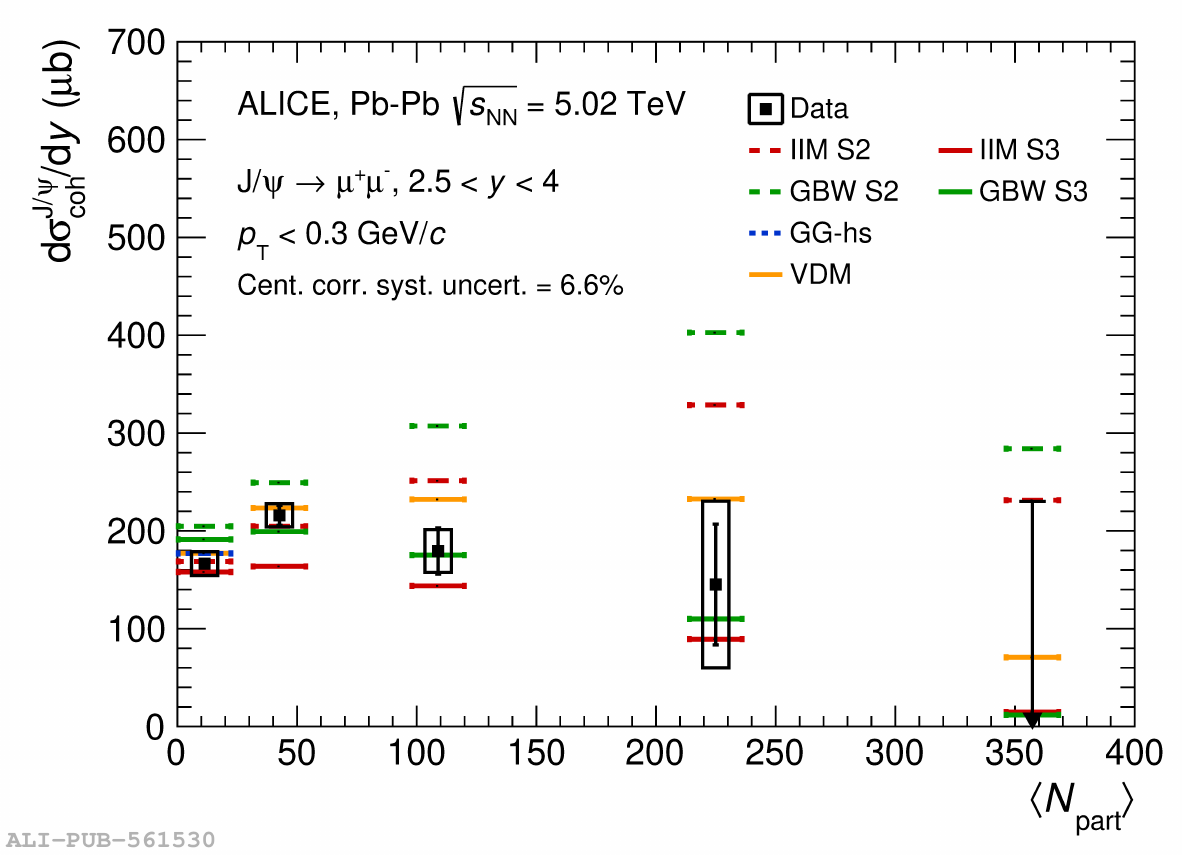}
\includegraphics[width=0.44\textwidth]{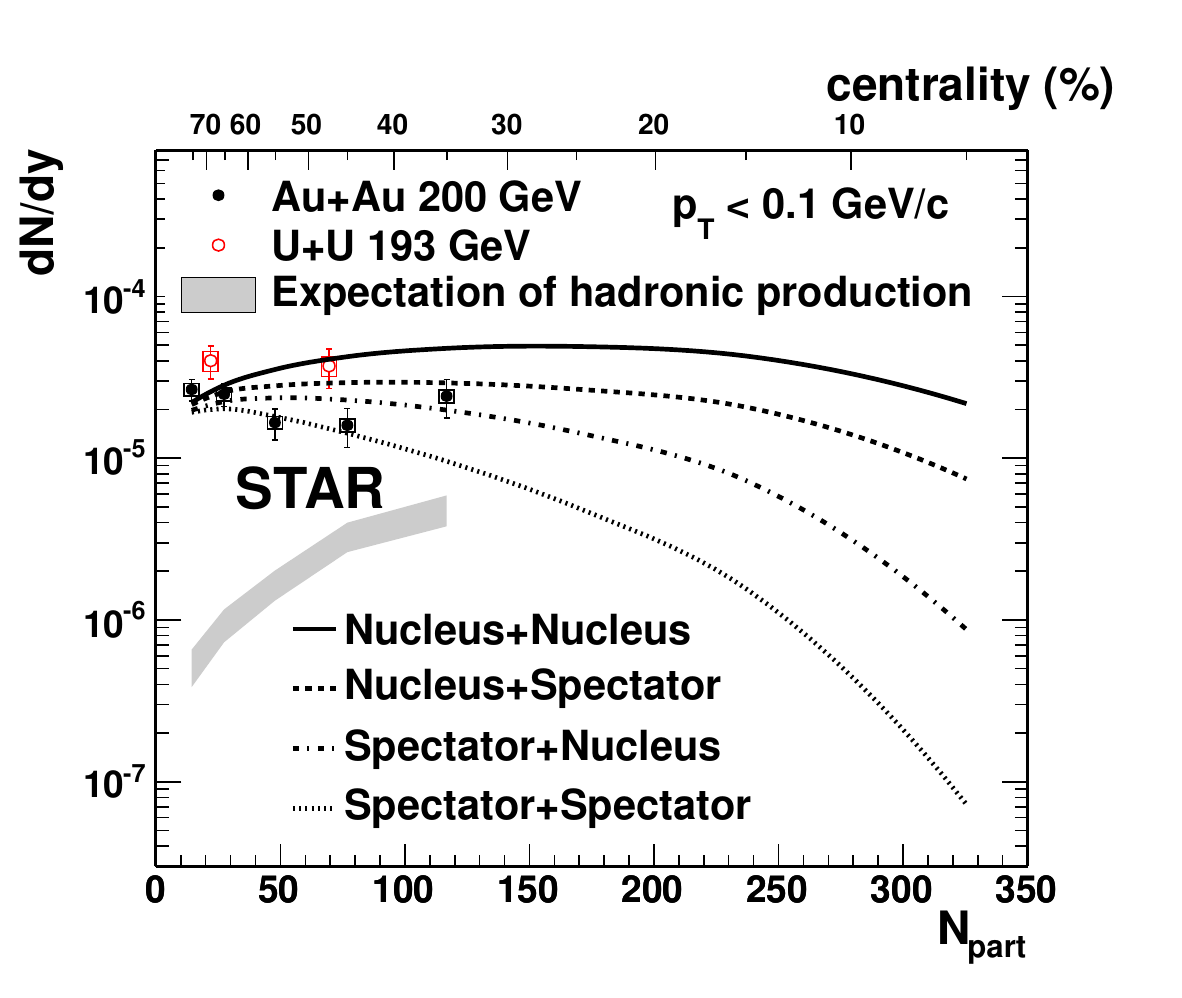}
\caption{
\label{fig:peri} 
Summary of experimental data on coherent J$/\psi$ diffractive photoproduction off Pb (left) and Au (right) in peripheral collisions of heavy nuclei. The horizontal scale is a measure of how many nucleons participated in the hadronic interaction. Taken from Refs.~\cite{ALICE:2022zso} and~\cite{STAR:2019yox}, respectively.}
\end{figure}

Photonuclear J$/\psi$ production has also been measured in peripheral heavy-ion hadronic collisions by the ALICE~\cite{ALICE:2015mzu,ALICE:2022zso} and LHCb~\cite{LHCb:2021hoq} collaborations at LHC as well as by the STAR collaboration at RHIC~\cite{STAR:2019yox}. Finding the photoproduced J$/\psi$ in the debris of the hadronic collision is possible due to the extremely small transverse momentum of the vector meson. Vector mesons produced by hadronic interactions have a larger average transverse momentum and can be rejected to obtain a clean photonuclear sample.  One motivation for these studies is the tantalizing possibility of using the photoproduced J$/\psi$ as a probe of the quark-gluon plasma formed in  hadronic  head-on collisions. The production of a vector meson in peripheral collisions also test our knowledge of the photon flux which in these cases occurs at impact parameters smaller than twice $R_A$ (see Chapter~\ref{ch:Photonflux}) as well as the size of the coherent region of the gluon field participating in the diffractive interaction~\cite{Zha:2017jch}. Recent results are shown in Fig.~\ref{fig:peri} where the lines represent model predictions using different coherent regions and photon fluxes. The spread of the curves around  data demonstrate that these measurements  constraint the model assumptions and improve our understanding of both, photon fluxes at small impact parameters and  the QCD structure of nuclei. 

\subsubsection{Energy dependence of nuclear structure from coherent photonuclear J$/\psi$ production} 
\begin{figure}[t]
\centering 
\includegraphics[width=0.37\textwidth]{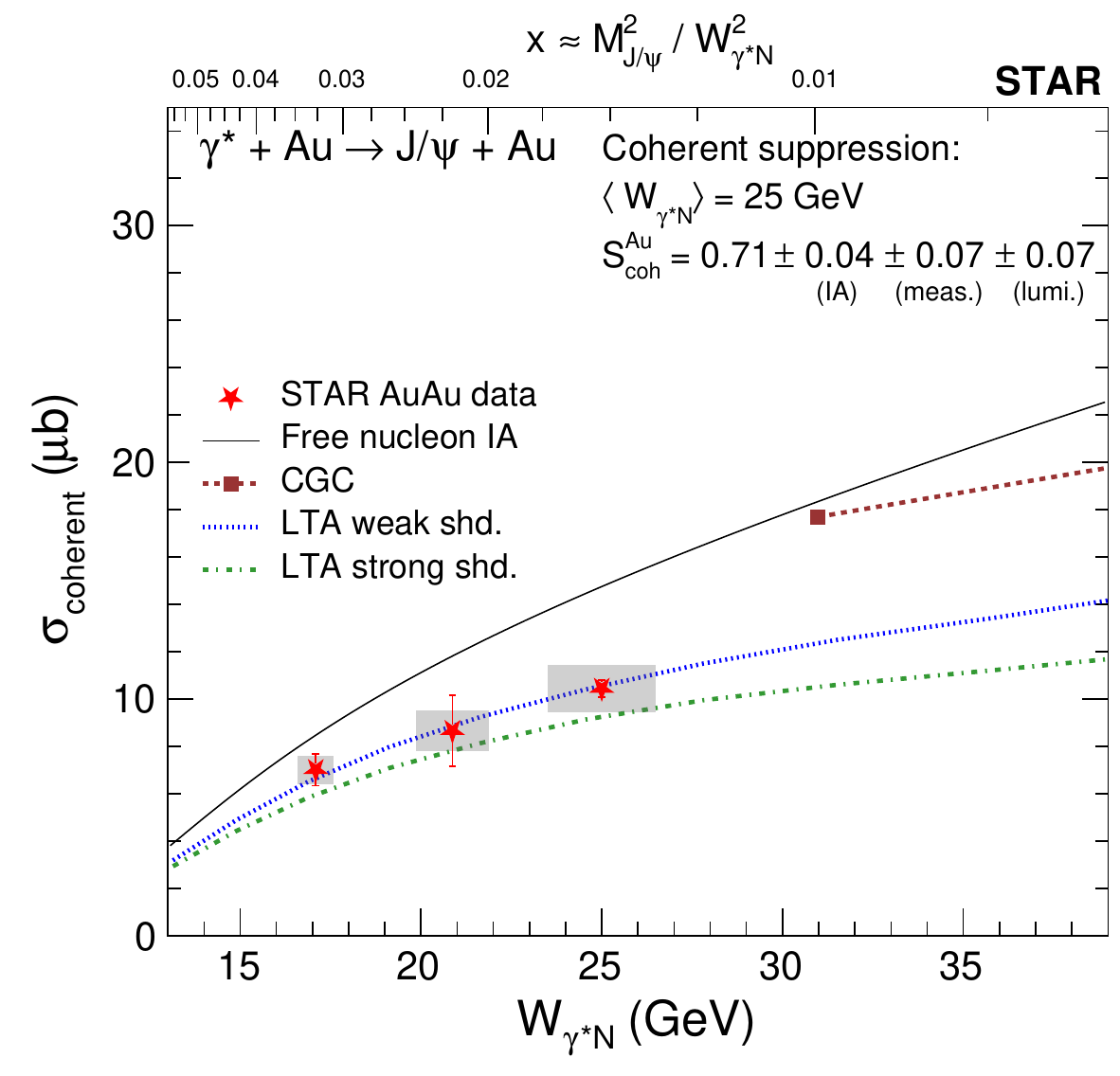}
\includegraphics[width=0.53\textwidth]{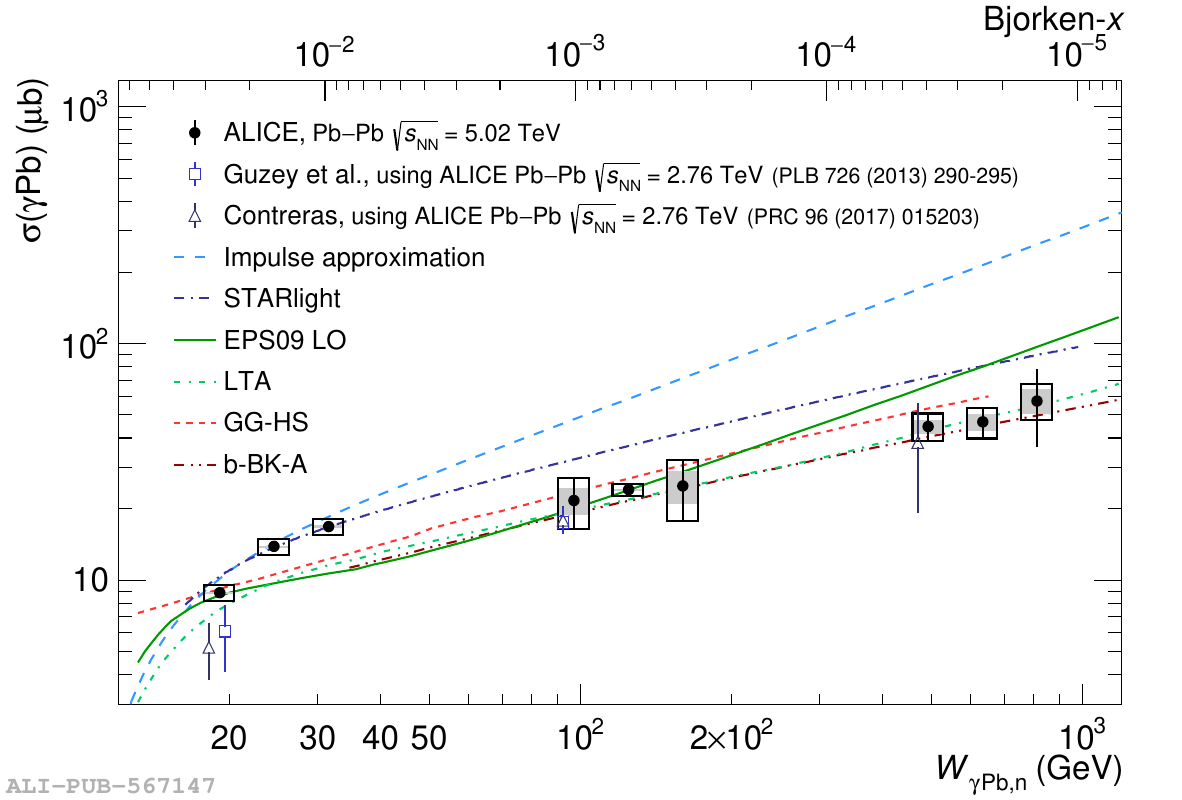}
\caption{
\label{fig:jpsiOffA} 
Summary of experimental data on coherent J$/\psi$ diffractive photoproduction off Au (left) and Pb (right). Taken from Refs.~\cite{STAR:2023vvb} and~\cite{ALICE:2023jgu}, respectively.}
\end{figure}

The previous section discussed the rapidity dependence of diffractive vector meson photonuclear production, that is the LHS of Eq.~\ref{eq:twofoldsolution}. These experiments are important to demonstrate their feasibility in hadron colliders and the precision that is achievable, but do not allow  for the extraction of the $x$ dependence of the hadronic structure. Experimentalists at RHIC and LHC have leveraged their zero-degree calorimeters to implement the strategy described in Sec.~\ref{sec:solvingambiguity} to access directly the energy evolution of the photonuclear cross section.

The current state-of-the-art measurement at RHIC is from the STAR collaboration and covers center-of-mass energies per nucleon corresponding to $15<W_{\rm \gamma Pb,n}<25$ GeV~\cite{STAR:2023vvb}.
Similar measurements have been performed at the LHC by the ALICE~\cite{ALICE:2023jgu} and CMS~\cite{CMS:2023snh} collaborations in Pb--Pb UPCs. In this case, the range covered by the measurements is  $20<W_{\rm \gamma Pb,n}<815$ GeV. Figure~\ref{fig:jpsiOffA} shows a summary of these data.

As for the case of $\gamma$p interactions discussed in Sec.~\ref{sec:protonxdep}, the data shown in Fig.~\ref{fig:jpsiOffA} covers  three orders of magnitude in $x$. Measurements from the STAR, ALICE, and CMS collaborations are consistent with each other.  In contrast to the results from $\gamma$p collisions, the cross section does not show a power-law increase with decreasing $x$. To explore the potential origins of this behavior, data are compared with different models of the nuclear structure. 
\begin{itemize}
\item The baseline is called the impulse approximation (IA) where the cross section is just $A$ times the cross section off protons, taking into account the different form factors of nucleons and nuclei~\cite{Guzey:2013jaa}. The prediction of this model agrees with the low energy data from Pb--Pb UPCs, but it is 10--30\% above the Au--Au measurements ~\cite{STAR:2023vvb}. At higher energies, this approach severely overestimates  the data. 
\item The STARlight model~\cite{Klein:1999qj} utilizes a parameterization of HERA data, like that shown in Fig.~\ref{fig:jpsiOffProtons}, the vector dominance model (VDM) and the optical theorem to obtain the total ${\rm J}/\psi+{\rm p} \to {\rm J}/\psi+{\rm p}$ cross section ($\sigma_{\rm J/\psi p}$). A classical Glauber model, where  the nuclear interaction is represented by independent nucleon interactions taking into account the nuclear shape, provides the corresponding nuclear cross section ${\rm J}/\psi+A \to {\rm J}/\psi+A$ ($\sigma_{{\rm J}/\psi A}$):
\begin{equation}
\sigma_{{\rm J}/\psi A} = \int{\rm d^2}\vec{b}\left(1-e^{-\sigma_{\rm J/\psi p} T_{AA}(b)}\right),
\end{equation}
where $T_{AA}(b)$ is the nuclear overlap function at an impact parameter $b=|\vec{b}|$. 
A second application of VDM and the optical theorem produces a prediction for the photonuclear cross section. This approach predicts correctly the low energy Pb--Pb measurements, and overestimates the cross sections at all larger energies. 
\item The leading twist approximation (LTA) describes shadowing based on the absorption corrections caused by multiple interactions. This model is formulated in the Good-Walker approach, where the interaction with each nucleon is the weighted average of the cross sections with each one of the states into which the nucleon can fluctuate~\cite{Frankfurt:2011cs}. The interaction with one nucleon corresponds to IA, the contribution of two nucleons is fixed by inclusive diffractive data from HERA, while the terms involving three or more nucleons are  unknown and have to be modeled. The authors considered two extreme scenarios corresponding to weak and strong shadowing. Both predictions are close at lower energies and describe the Au--Au data, with the weak shadowing option favored.  At large energies, and for Pb--Pb collisions, the weak and strong shadowing cases differ by a large energy-dependent factor which reaches values above two. Their average describes satisfactorily the data above $W_{\rm \gamma Pb,n}\approx100$ GeV. In this limit, given the size of the experimental uncertainties, both  weak and strong shadowing in LTA are strongly disfavored. At energies around 30 GeV LTA underestimates Pb--Pb data, in stark contrast with the nice description of Au--Au measurements. 
\item Another class of models, based on the dipole color approach, incorporate the concept of saturation. These models are only valid at small $x$ with their applicability restricted to $x\le 0.01$. In the model denoted b-BK-A~\cite{Bendova:2020hbb}, the physics is embodied in the solutions of the non-linear Balitsky-Kovchegov equation describing within QCD the evolution of the dipole-hadron scattering amplitude~\cite{Kovchegov:2012mbw}. The solution of this equation is very demanding. Up to now, it has not been solved over all the  phase space, with the consequence that solutions within the Good-Walker approach have not been implemented. To fill this gap, a QCD inspired calculation based on hotspots whose number increases with energy was implemented~\cite{Cepila:2017nef} and their predictions are shown as GG-HS in the figure. Both models predict correctly the high-energy behavior of the cross section, but do not describe the lower energy measurements.
Another dipole-based computation (CGC in the figure with Au--Au data), is obtained by implementing the Good-Walker approach in the Color Glass Condensate formalism, see e.g.~\cite{Mantysaari:2025ltq}. It is found that within this model the simultaneous description of data off protons and off nuclei is quite challenging.
 \end{itemize}
The current status is then that no model can describe all data at the same time. At high energies, the predictions of shadowing and saturation based models are quantitatively similar for this observable, so it is challenging to disentangle the effect of these different phenomena based only on coherent J$/\psi$ diffractive photoproduction off nucleons and nuclei.

\subsubsection{Coherent photonuclear production of heavier quarkonia} 
This successful program of measuring the  coherent photonuclear production of the J$/\psi$ vector meson could be repeated using other particles. 
Two other quarkonium states have been measured in UPCs of heavy ions: the $\Psi(2S)$ and the $\Upsilon(1S)$.

The photonuclear production of $\Psi(2S)$ has been measured at a Pb--Pb center-of-mass energy per nucleon pair of 5.02 TeV  by the ALICE collaboration at midrapidity~\cite{ALICE:2021gpt} and by the  LHCb~ collaboration at forward rapidity~\cite{LHCb:2022ahs}. This process has also been measured at lower energies in AuAu UPCs by the STAR collaboration~\cite{STAR:2023vvb}, although with roughly 25\% statistical uncertainties. The LHC results at midrapidity are consistent with both shadowing-based and saturation models, while those at forward rapidity, dominated by the low energy contributions, are described by shadowing-based models but saturation based models undershoot the data. This is the same pattern as observed for the case of J$/\psi$. 

The ratio of the $\Psi(2s)$  production cross sections to that of the $J/\psi$ is also of interest, as a potential signature of saturation~\cite{Peredo:2023oym}, although recent studies present a more nuanced view~\cite{Cepila:2025rkn}. The ratio as measured by all three collaborations is consistent with around 0.15, showing that the ratio does not have a strong dependence on $W_{\rm \gamma Pb,n}$. The measured ratio is also compatible to that found in photoproduction off protons in $e$p collisions at HERA~\cite{H1:2002yab} or in pp collisions at the LHC~\cite{LHCb:2024pcz}. This ratio can be described with shadowing-based and saturation-based models.

The LHCb~\cite{LHCb:2015wlx} and the CMS~\cite{CMS:2018bbk} collaborations have measured the coherent diffractive photoproduction of $\Upsilon(1S)$ in pp and p--Pb UPCs, respectively.  Figure~\ref{fig:jpsiOffProtons} summarizes these data. The cross section raises with energy as a power law whose exponent depends on which data set is considered. The potential source of this discrepancy is assigned to higher order corrections and not to saturation.  The cross section for $\Upsilon(1S)$ photoproduction in Pb--Pb UPCs is around three orders of magnitude less than that of J$/\psi$, so the number of events is small. Furthermore, the
measurement is complicated by the large background of lepton pairs from two-photon
collisions. Nonetheless, the CMS collaboration has submitted the
first results for the coherent production of this vector meson in
Pb--Pb UPCs at 5.02 TeV per nucleon pair~\cite{CMS:2026soy}. At midrapidity the cross section is around 5 $\mu$b and the $\gamma$Pb cross section at $W_{\rm \gamma Pb,n}=200$ GeV is around 60 nb.
These measurements of $\Upsilon(1S)$ production are important because no saturation effects are expected at these mass scales (around 22 GeV$^2$, see Eq.~\ref{eq:q2m}), so they can be used to improve our understanding of nuclear shadowing.

\begin{figure}[t]
\centering 
\includegraphics[width=0.485\textwidth]{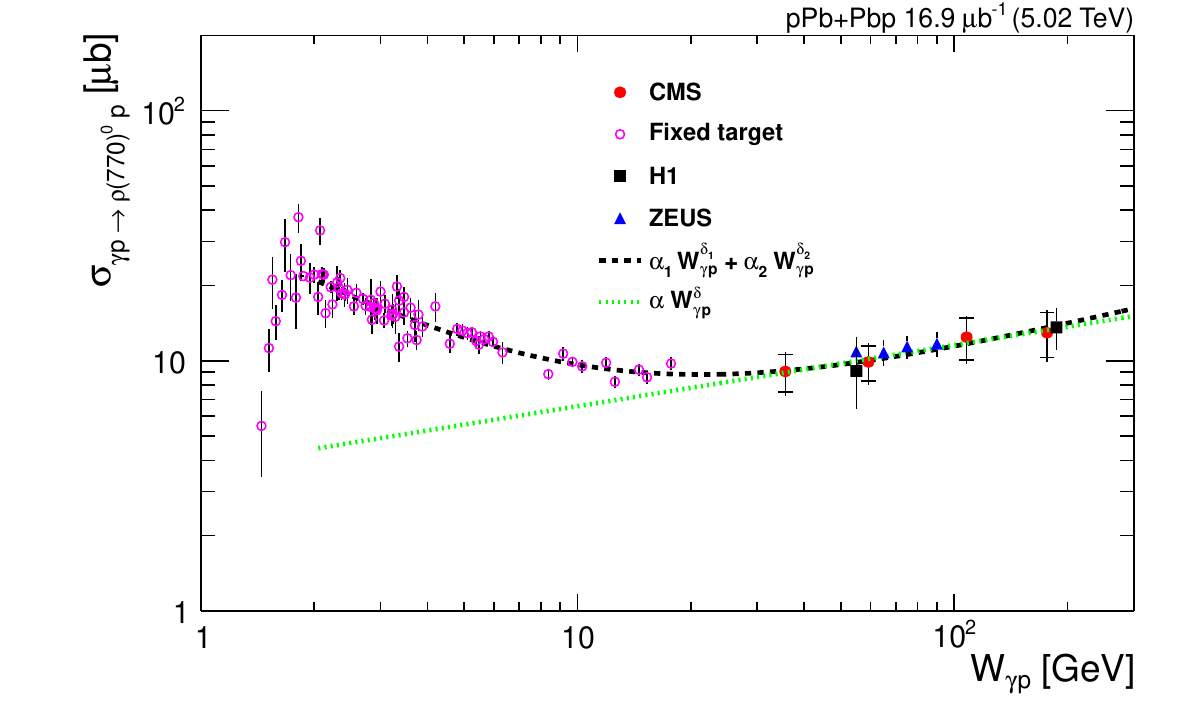}
\includegraphics[width=0.415\textwidth]{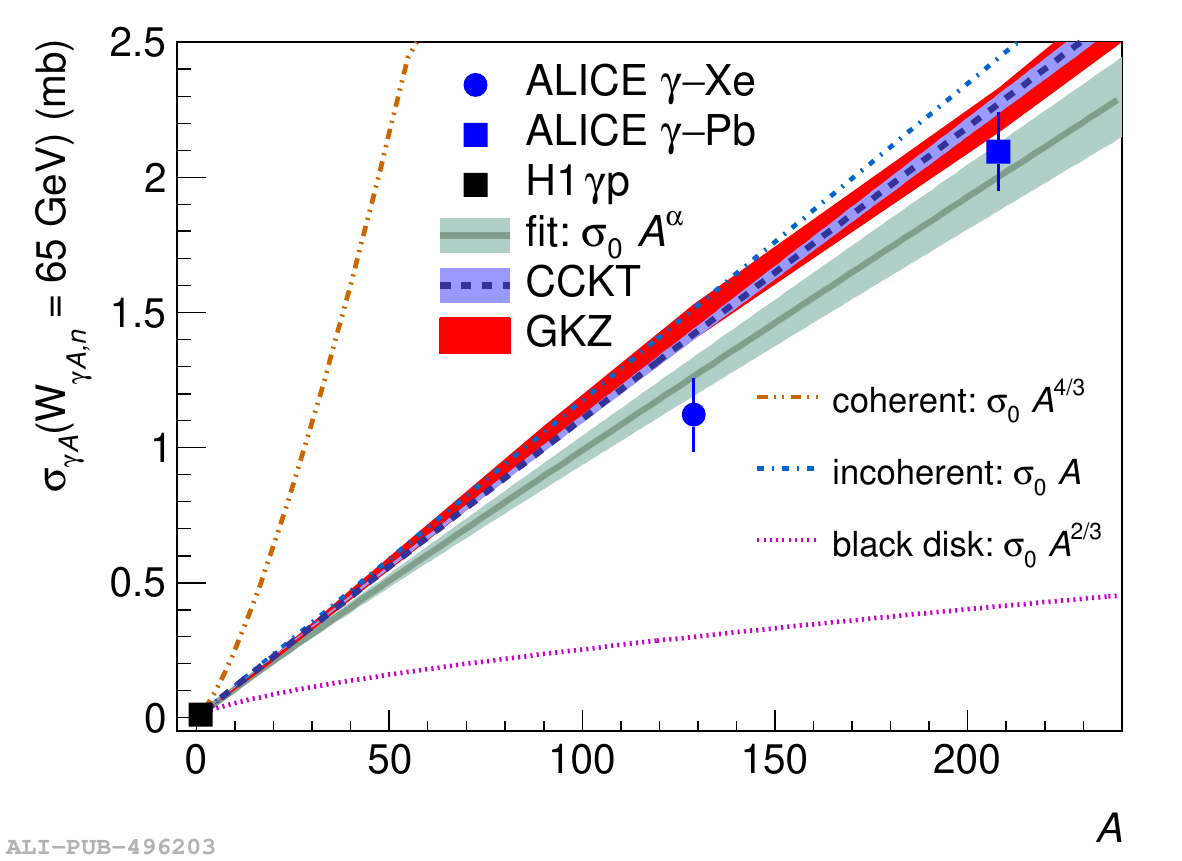}
\caption{
\label{fig:cohRho} 
Left: Summary of experimental data on coherent $\rho^0$ diffractive photoproduction off protons. Taken from~\cite{CMS:2019awk}.
Right: Atomic mass number dependence of coherent $\rho^0$ diffractive photoproduction  at a center-of-mass energy of the photonuclear system of 65 GeV.  Taken from~\cite{ALICE:2021jnv}}
\end{figure}

\subsubsection{Coherent photonuclear production of light mesons} 

\begin{figure}[t]
\centering 
\includegraphics[width=0.485\textwidth]{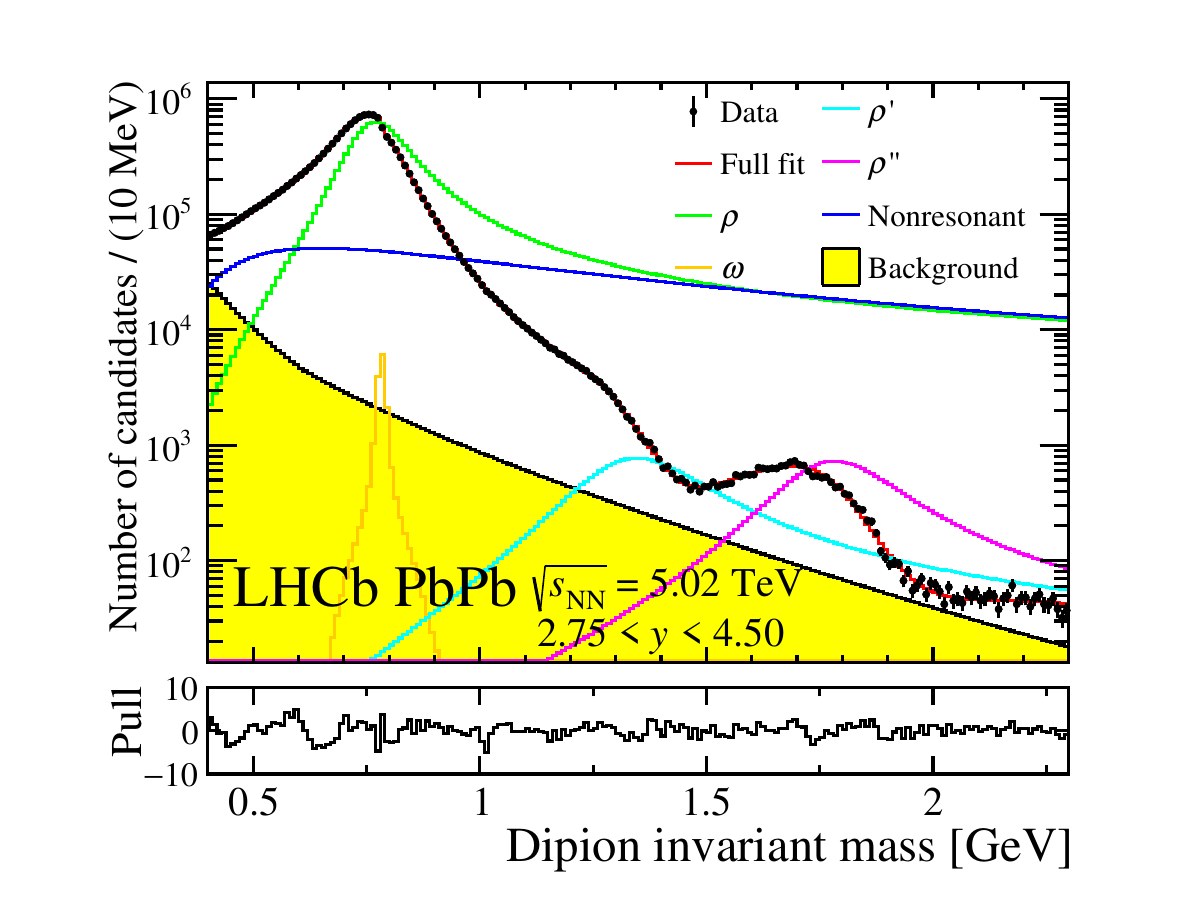}
\includegraphics[width=0.485\textwidth]{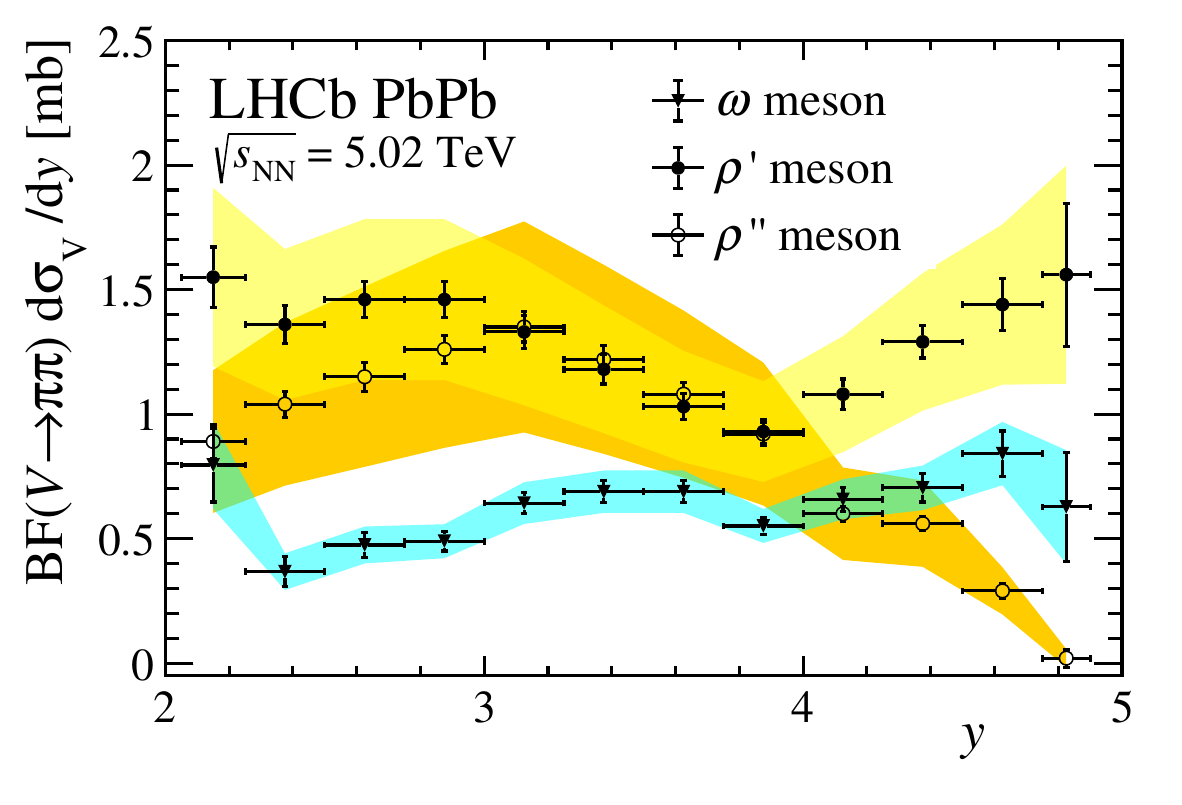}
\caption{
\label{fig:dipionmass} 
Left: Dipion mass distribution measured by the LHCb collaboration in Pb--Pb UPCs, and the contributions considered at the amplitude level to describe the spectrum. Right: Cross section times branching ration into dipions for three resonances as obtained with the model used to describe the mass spectrum. Taken from~\cite{LHCb:2025fzk}.}
\end{figure}

The lightest vector meson is the $\rho^0$ which has been extensively studied in UPCs both at RHIC and at LHC. The advantage of a lower mass is a larger cross section, but the $\rho^0$ is a wide resonance with a width of 150 MeV, which complicates the signal extraction. Furthermore, its low mass, 770 MeV, complicates the application of perturbative QCD techniques to understand the measured cross sections. Nonetheless there are several interesting results using this resonance.

The energy dependence of coherent $\rho^0$ diffractive photoproduction off protons has been measured at the LHC by the CMS collaboration~\cite{CMS:2019awk}, complementing earlier fixed-target experiments~\cite{Bauer:1977iq}. The cross section, shown in Fig.~\ref{fig:cohRho}, agrees with previous measurements from HERA, and also displays a power law growth with energy above some 20 GeV. Below this energy, the exchange off reggeons is dominant, causing the cross section to decrease with increasing energy. The slope of the power law is smaller than the slope observed for J$/\psi$ or $\Upsilon(1S)$ production, as expected from the behavior of the gluon distribution, see Eq.~\ref{eq:gluonpowerlaw}.

Coherent $\rho^0$ diffractive photoproduction has also been measured off a variety of nuclei, such as Au, Xe, and Pb. The STAR collaboration at RHIC has measured the cross section for Au-Au collisions at different per-nucleon-pair center-of-mass energies: 62.4 GeV~\cite{STAR:2011wtm}, 130 GeV~\cite{STAR:2002caw}, and 200 GeV~\cite{STAR:2017enh}. As these measurements were performed at midrapidity, the $\pm y$ ambiguity does not apply.  At $y=0$ these energies  correspond to $W_{\rm \gamma Au,n}$ of 7 GeV, 10 GeV, and 12.5 GeV, respectively, with the photonuclear cross section compatible with being constant around a value of 1.1 mb. The ALICE collaboration at the LHC has performed similar measurements in Pb--Pb UPCs at center-of-mass energies per nucleon pair of 2.76 TeV~\cite{ALICE:2015nbw} and 5.02 TeV~\cite{ALICE:2020ugp} which  at $y=0$ correspond to $W_{\rm \gamma Pb,n}$ of 46 GeV and 62 GeV, respectively. In both cases the photonuclear cross section is around 1.8 mb with an uncertainty of the order of 10\%. 

The cross section measured by ALICE were compared to a variety of models: STARlight predictions undershot data by some 2 standard deviations. While some saturation models describe data within one standard deviation.
An interesting case is the shadowing based model. As the mass of the $\rho^0$ meson does not provide a hard enough scale, this model is based on  VDM and a Glauber prescription.  The original model, including only photon fluctuations into the $\rho^0$ and intermediate elastic collisions with the nucleons forming the nucleus, did not describe data; this prompted the authors to include higher mass photon fluctuations and the contribution to shadowing from inelastic collisions, and approached called Gribov-Glauber, with the resulting model able to reproduce data~\cite{Frankfurt:2015cwa}. This is one example of how the availability of new data at high energies has helped us to understand better the measured phenomena and to improve  models. 

The ALICE collaboration also measured the cross section on xenon nuclei at a
center-of-mass energy per nucleon pair of 5.44 TeV~\cite{ALICE:2021jnv}. Since the energy dependence of the photonuclear cross sections for the coherent production of this vector meson is quite flat, several measurements at slightly different energies can be joined to study the dependence on the atomic mass number for this process. This is shown in Fig.~\ref{fig:cohRho}. These data are well described by a power law with an exponent close to one. Similar behavior is predicted in both, the  shadowing-based approach, and in the saturation-based hotspot model. This slope is in between the cases of pure coherence and the black-disk limit of QCD, and coincidentally agrees with expectations from a pure incoherent sum of nucleons making up the nucleus.

Recently, the LHCb collaboration has measured coherent dipion diffractive production in Pb--Pb UPCs at forward rapidities~\cite{LHCb:2025fzk}.  The measurement covers a large mass range of the dipion system. To describe the mass spectrum, the contributions of several terms have to be considered at the amplitude level.
It is an open problem how to treat all these contributions, how to deal with their potential different phases, and to decide on the inclusion or exclusion of other contributions.
The LHCb included the $\rho^0$ and $\omega$ resonances as well as continuum diffractive dipion production. It also included, at higher masses, two excited $\rho$ states. The resonances were described with a Breit-Wigner distribution with mass dependent widths. This fit function, shown in Fig.~\ref{fig:dipionmass} left, provides a good description of the dipion mass distribution. As mentioned above, to disentangle the energy dependence of the photonuclear process from this type of data is complicated without neutron detectors at beam rapidities. A  model dependent interpretation of the potential behavior of the low and high energy contributions to the cross section was attempted and reported in the article, but it  is too early to evaluate the feasibility of this approach. Nonetheless, the availability of these data is highly promising. 

The cross section of the $\omega$ meson and of excited states of the  $\rho$ meson, see Fig.~\ref{fig:dipionmass} right, was also measured. These results may help to bridge and explore the gap between the non- and perturbative regimes of QCD. These results join previous measurements at RHIC by the STAR collaboration~\cite{STAR:2009giy} and at  LHC by the ALICE collaboration~\cite{ALICE:2024kjy} of coherent production of excited $\rho$ states decaying into 4 pions. One of the interest of measuring multipion production in this mass region is to determine how many resonances are there, with data  compatible with both, the hypothesis of one broad resonance or of two resonances whose width overlaps~\cite{Klusek-Gawenda:2020gwa}. This process may also help to constraint the branching ratios of the resonance decay into four pions~\cite{Devi:2025ftf}.

Other recent measurements that promise  new insights  are those of coherent production of $K^+K^-$ pairs. There are two different results. The CMS collaboration measured the  photonuclear production of $\phi(1020)\to K^+K^-$ in UPCs~\cite{CMS:2025lsm}. The $\phi(1020)$ is a very narrow resonance with a mass at the boundary of applicability of perturbative QCD. Even though the measurement is very challenging because the kaons have very small momentum and do not cross the full detector, CMS demonstrated that the measurement is feasible at the LHC. In a related measurement, the ALICE collaboration studied coherent production of direct $K^+K^-$ pairs above the $\phi(1020)$ mass~\cite{ALICE:2023kgv}. This process tests simultaneously three different aspects: the modeling of resonances far away from its pole mass,  non-resonant production, and the interference between these two contributions.

\subsubsection{Nuclear structure in the transverse plane from coherent vector meson production}

\begin{figure}[t]
\centering 
\includegraphics[width=0.245\textwidth]{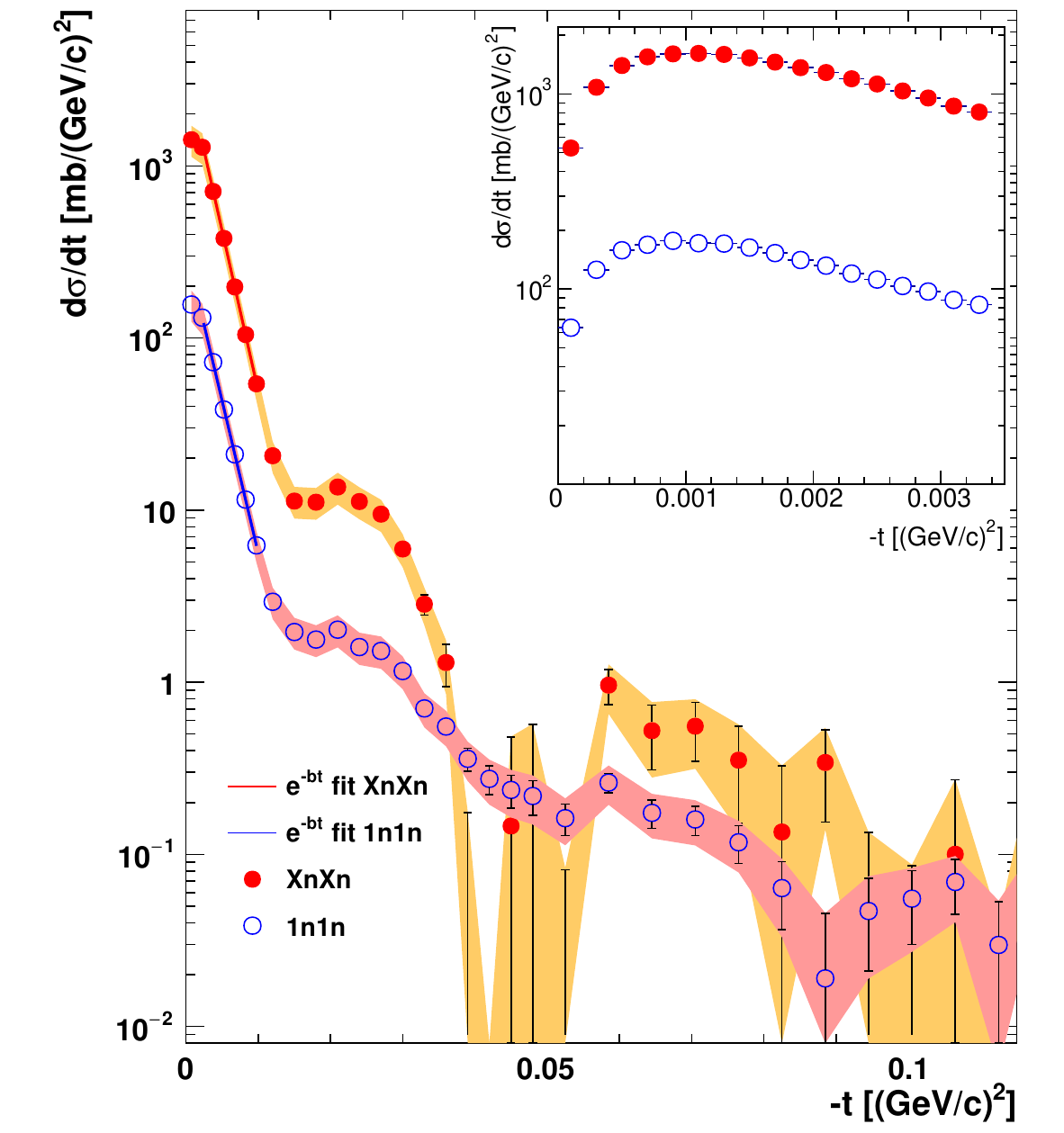}
\includegraphics[width=0.37\textwidth]{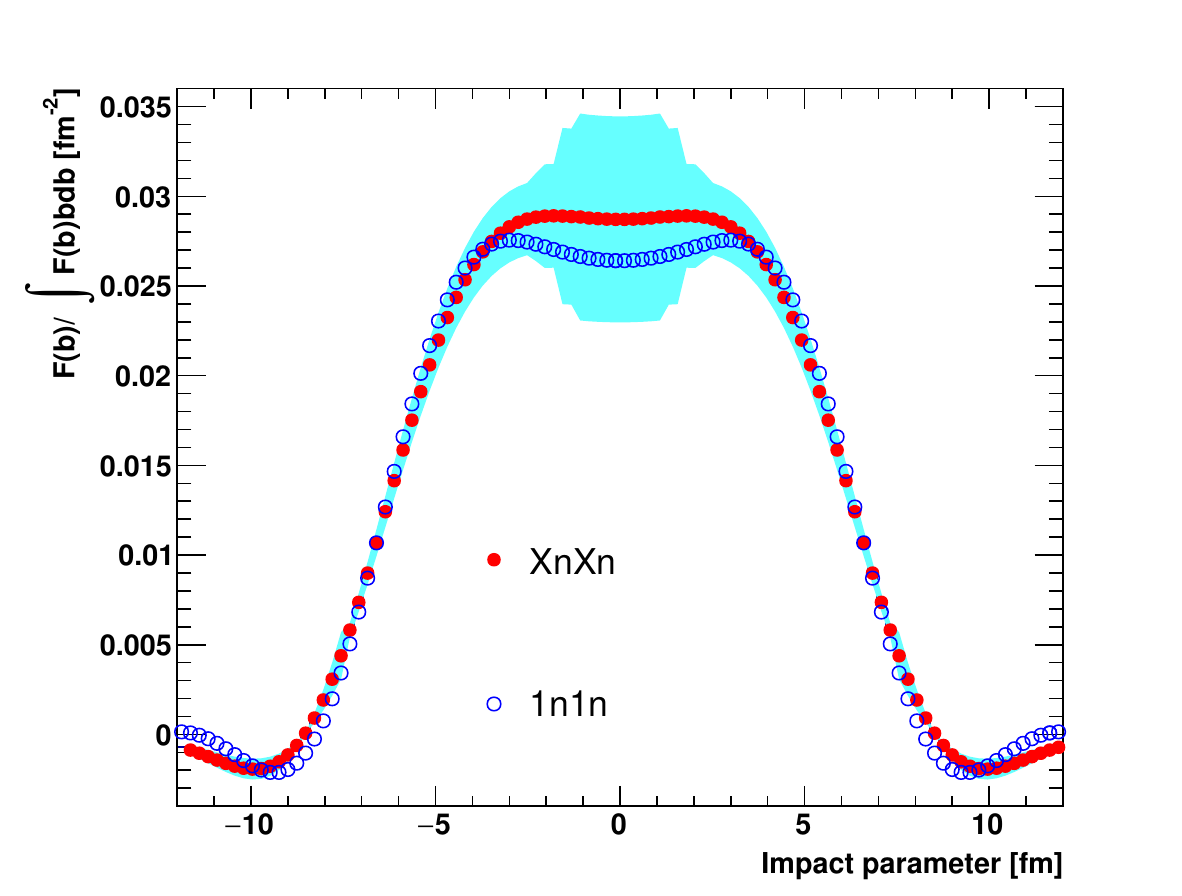}
\includegraphics[width=0.285\textwidth]{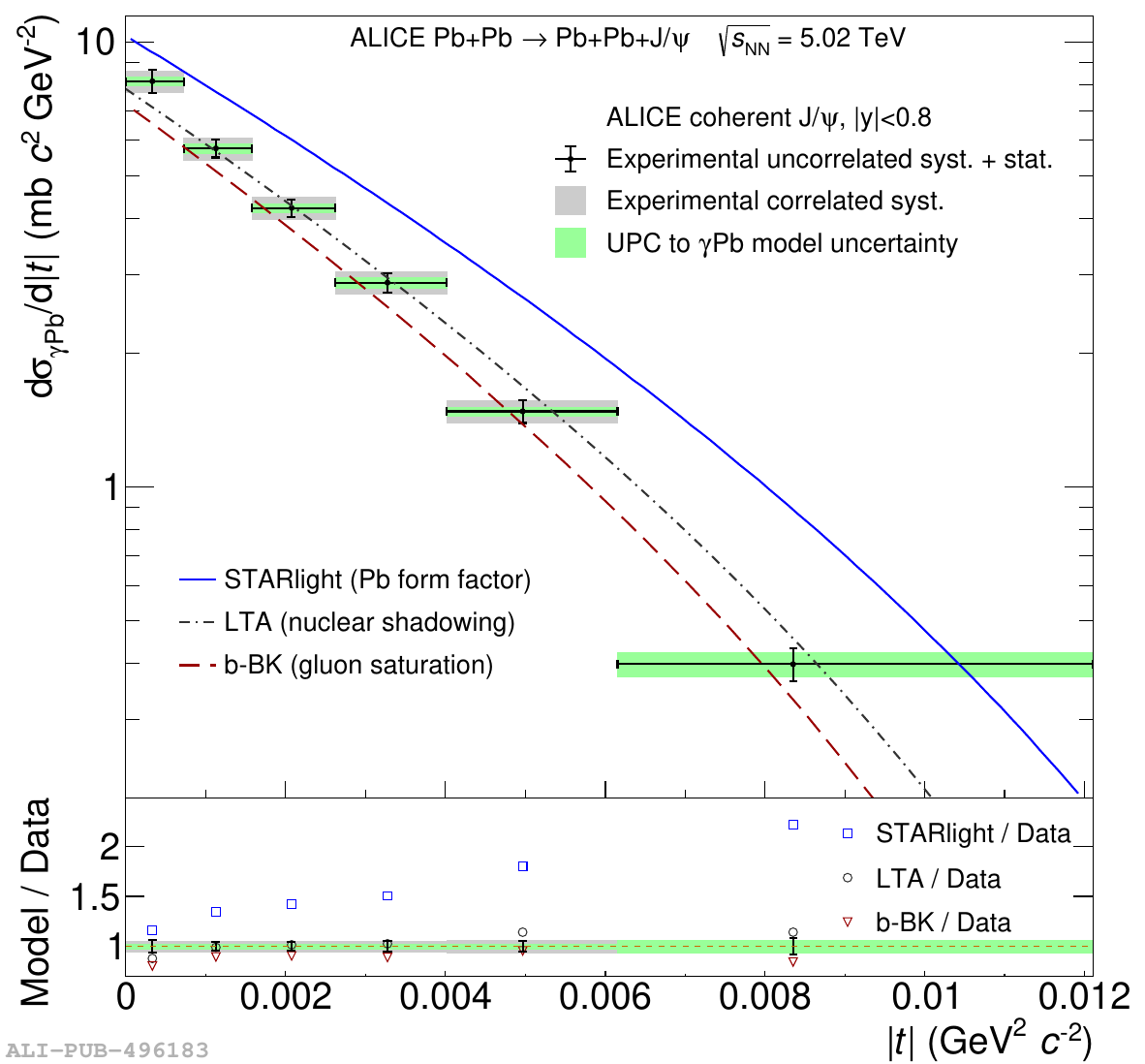}
\caption{
\label{fig:cohMant} 
Left: Mandelstam-$t$ dependence of coherent diffractive production of $\rho^0$ in Au--Au UPCs. Middle: Fourier transform of the of the data in the left panel. These two figures were taken from~\cite{STAR:2017enh}. Right: Mandelstam-$t$ dependence of coherent diffractive production of J$/\psi$ in Pb--Pb collisions. Taken from~\cite{ALICE:2021tyx}.}
\end{figure}

At modern ion colliders the Lorentz boost is so large that a nucleus becomes very thin in the direction of motion and its color field is contained in a plane transverse to its motion. This plane coincides with the impact-parameter plane. Regions of a given size in this plane are related through a Fourier transform to the transverse  momentum transferred in the interaction, which in this kinematics, is related to Mandelstam-$t$, see Eq.~\ref{eq:mant}. This means, that a measurement of the Mandelstam-$t$ dependence of the cross section for diffractive vector meson production provides information on the distribution of color charges inside the hadron in the transverse plane.

Such a measurement has been carried out by the STAR collaboration for the case of coherent production of $\rho^{0}$~\cite{STAR:2017enh} in Au--Au UPCs at  an energy per nucleon of $W_{\rm \gamma Au,n}=12.5$ GeV and it is shown in Fig.~\ref{fig:cohMant} (left).  Two samples selected with different number of beam-rapidity neutrons were studied, with 1n1n corresponding to one neutron on each side, mostly from giant dipole resonances---a smaller sample, but with lower background. In both cases, a steeply falling distribution with two diffractive minima is observed. The minima are at the same positions for both samples. 

The STAR collaboration also Fourier transformed these data to produce the density profile in Fig.~\ref{fig:cohMant} (middle).  The transform is~\cite{Boer:2011fh}
\begin{equation}
    F(b) \propto\int_0^\infty {\rm d}p_\perp p_\perp  J_0(bp_\perp) \sqrt{\frac{{\rm d}\sigma}{{\rm d}t}}
    \label{eq:fourier}
\end{equation}
where $J_0$ is a modified Bessel function and $-t\approx p_\perp^2$
There are several issues in this transform.  One is the square root; in going from cross section to amplitude, there is a sign ambiguity.  This ambiguity necessitates flipping the sign of the square root when integrating past each diffractive minimum.  As can be seen in the figure, it is not always easy to accurately determine the location of these minima. Also, the integral covers the range $0 < p_\perp < \infty$, but the data cut off at a finite $p_\perp$.  This introduces windowing artifacts~\cite{Toll:2012mb,Klein:2021mgd}, which might be partially alleviated by extrapolating the data~\cite{Klein:2026azm}. And, the measured $t$ includes contributions from the elastic scattering, the photon $p_\perp$ and detector resolution.  Detector resolution and the photon $p_\perp$ are subdominant, but non negligible.  Fortunately, at the LHC, the photon $p_\perp$ contribution is much reduced.  These components could also be removed by unfolding.

Alternately, one can directly compare the Mandelstam-$t$ distributions with different models.  The ALICE collaboration has done with for coherent $J/\psi$ diffractive photoproduction in Pb--Pb UPCs at midrapidity corresponding to $W_{\rm \gamma Pb,n}=125$ GeV~\cite{ALICE:2021tyx}, see Fig.~\ref{fig:cohMant} (right).  The cross section decays steeply with increasing $|t|$. The distribution is compared with three predictions. STARlight, based on the form factor for a Woods-Saxon distrubtion, with parameters measured using electron scattering, predicts a more gradual slope than data, indicating that the effective size of Pb seen by the diffractive interaction at LHC energies is larger than that measured with electromagnetic scattering at lower energies. Data are better described by the shadowing- and  saturation-based models, which both predict an increase, with respect to low-energy measurements, of the size of the nucleus when probed at LHC energies. Such a behavior was already observed at HERA when measuring the energy evolution of the Mandelstam-$t$ dependence for coherent J$/\psi$ diffractive photoproduction off protons~\cite{H1:2005dtp}, where it was found that the $|t|$ distribution at a fixed energy was well described in all cases by an exponential, with the slope growing logarithmically with increasing interaction energy.  This approach avoids some of the challenges presented by the Fourier-transform approach.

\subsubsection{Fluctuations of nuclear structure \label{sec:fluct}}

\begin{figure}[t]
\centering 
\includegraphics[width=0.95\textwidth]{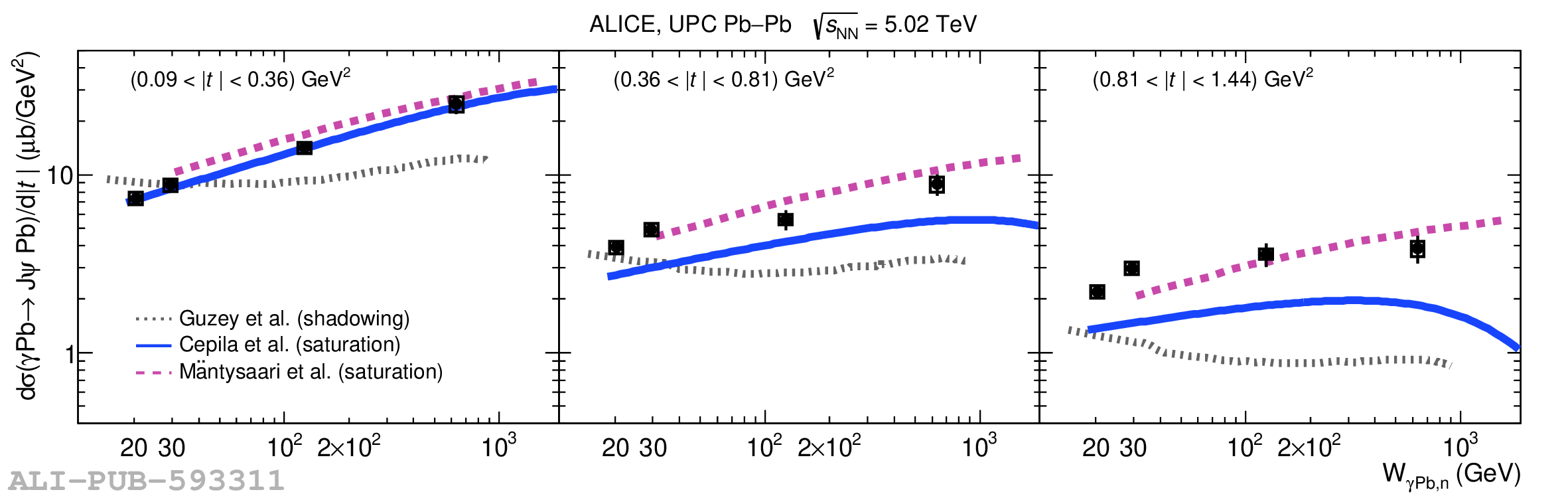}
\caption{
\label{fig:incJpsiPb} 
Energy and Mandelstam-$t$ dependence of the cross section for incoherent diffractive J$/\psi$ photonuclear production off Pb ions measured by the ALICE collaboration. Taken from~\cite{ALICE:2025cuw}.}
\end{figure}

As mentioned above, Sec.~\ref{sec:gw}, the incoherent diffractive photoproduction of J$/\psi$ is sensitive to quantum fluctuations of the gluon field in hadrons and thus offers a new window to study the high-energy limit of QCD.  
The first measurement of incoherent diffractive photoproduction of J$/\psi$ off protons was recently performed by the ALICE collaboration in p--Pb UPCs at $\sqrt{s}= 8.16$ TeV per nucleon pair~\cite{ALICE:2023mfc}. The cross section  was reported at $W_{\rm \gamma p}=33$ GeV and $W_{\rm \gamma p}=48$ GeV a ranged covered previously at HERA with more precision~\cite{H1:2013okq}. The measurements at the LHC agree with those from HERA, demonstrating the feasibility of measuring this process in hadron colliders. LHC Run 4 (2028-2033) should include a high luminosity period of p--Pb collisions that would allow to measure the incoherent diffractive photoproduction of J$/\psi$ off protons up to energies around 1 TeV.

The first measurement of incoherent diffractive photoproduction of J$/\psi$ off nuclei was performed by the ALICE collaboration in Pb--Pb UPCs at $\sqrt{s}= 2.76$ TeV per nucleon pair~\cite{ALICE:2013wjo}. The photonuclear cross section was found to be around 7 $\mu$b with an uncertainty of around 20\%.
This measurement is below the predictions of the shadowing-based model but still compatible with it within 1.5 $\sigma$ due to the large uncertainty. It is also below the saturation-based calculation using the CGC approach, in this case by around 2 $\sigma$.

New measurements with smaller uncertainties have been carried out recently by the STAR collaboration at RHIC in AU--AU UPCs at $\sqrt{s}= 200$ GeV per nucleon pair~\cite{STAR:2023vvb} and the CMS collaboration at LHC in Pb--Pb UPCs at $\sqrt{s}= 5.02$ TeV per nucleon pair~\cite{CMS:2025oxg}. At midrapidity, corresponding to $W_{\rm \gamma Au,n}=25$ GeV at RHIC, the cross section is about 5 $\mu$b. The measurements in the $\gamma$Pb system cover two ranges of energy $40<W_{\rm \gamma Pb,n}<50$ GeV and $300<W_{\rm \gamma Pb,n}<400$. The cross sections at lower energies are around 7 $\mu$b, while at larger energies are around 10 $\mu$b.  The weak shadowing LTA model describes the low energy data, but systematically overshoots the  high energy data by around 3 $\sigma$. The saturation-based CGC overshoots data and the weak shadowing model at all energies, showing that for this observable the predictions of saturation and shadowing based model are different, in contrast to predictions for the coherent process.

The ALICE collaboration went a step further and studied simultaneously the energy and Mandelstam-$t$ dependence of incoherent diffractive photoproduction of J$/\psi$ off Pb nuclei~\cite{ALICE:2023gcs,ALICE:2025cuw}. The measured cross sections are shown in Fig.~\ref{fig:incJpsiPb}. The panels, from left to right, correspond roughly to size scales in the transverse plan of around 0.6 fm, 0.3 fm, and 0.2 fm, respectively. The latter correspond to hotspot sizes where saturation effects may be first observed. At low values of $|t|$ the cross
section is seen to rise steeply, in the middle plane the rise is slower and the data in the last panel are
compatible with a still slower rise or with reaching a constant value. 
Data are also compared with three models discussed before. The LTA prediction shown in this figure is the average of the weak and strong shadowing cases, which described data on coherent production at high energies as shown in Fig.~\ref{fig:jpsiOffA}. Here, the LTA prediction does not describe the measurements and it is disfavored by data. The saturation models correspond to CGC (M\"antysaari et al) and the energy-dependent hotspot model (Cepila et al). The predictions of
these two models are qualitatively different at large energies and large Mandelstam-$t$, precisely where
one would like to look for saturation. The CGC model predicts a rising cross section, while the hotspot model predicts the cross section to reach a maximum and then decrease. The current precision of data does not allow to draw definitive conclusion at the moment. The new data recorded during LHC Run 3 (2022-2026) and to be acquired in LCH Run 4 would allow for substantially better measurements and should be able to determine experimentally the true energy dependence at large $W_{\rm \gamma Pb,n}$ and sub-femtometer transverse size scales.  

\subsection{Other QCD probes  with photon-induced interactions off nuclei \label{sec:otherProbs}}
In the last few years, the field of photonuclear interactions has expanded into other directions which are more challenging experimentally, but offer new perspectives to study the behavior of QCD in different areas. A few recent examples are discussed here.
\subsubsection{Diffractive dijet photonuclear production}
This process is similar to the one depicted in Fig.~\ref{fig:vmProd} except that the quarks, after interacting with the hadron via a pomeron do not form a bound state, but give rise to two jets. Angular correlations between the jets are sensitive to the polarization of the gluons within nuclei, a property of the QCD structure of hadrons for which there is very little data available and for which the corresponding predictions vary widely. The CMS collaboration measured the second Fourier coefficient ($\left<\cos(2\Phi \right>$) of  the azimuthal angle ($\Phi$) between the sum ($\vec{Q_\perp}$) and the difference ($\vec{P_\perp}$)
of the two jet transverse momentum vectors as a function of $Q_\perp$ in the region $P_\perp>Q_\perp$~\cite{CMS:2022lbi}. The measurements utilized data from Pb--Pb UPCs at $\sqrt{s}=5.02$ TeV.

From the experimental point of view, this measurement is more challenging because jets are large structures made of collimated hadrons which complicates the determination of rapidity gaps. On the other hand, the possibility of having variable $P_\perp$ and $Q_\perp$, which for the case of vector meson production are approximately fixed to be zero and the mass of the vector meson respectively, allows for an exploration of the gluon structure in a wider kinematic range.
 
Data shows that $\left<\cos(2\Phi \right>$ rises steadily with  $Q_\perp$ from about 0.07 at 2 GeV to 0.3 at 22 GeV, where the measurements still have sizable uncertainties above 20\%~\cite{CMS:2022lbi}.  A post prediction assigns a flat contribution to  $\left<\cos(2\Phi \right>$ of the order of 0.1 due to soft final state radiation of gluons from the quarks originating the jets, so it cannot explain the overall trend of data~\cite{Hatta:2020bgy}. This measurement demonstrate the feasibility of studying this type of observables in UPCs at LHC. The potential to access information about the polarization of gluons and their evolution opens a new window to probe the QCD structure of nuclei at small $x$.

\subsubsection{Inclusive jet photonuclear production\label{sec:incJet}}
\begin{figure}
\begin{subfigure}{.5\textwidth}
  \centering
  \includegraphics[width=.9\linewidth]{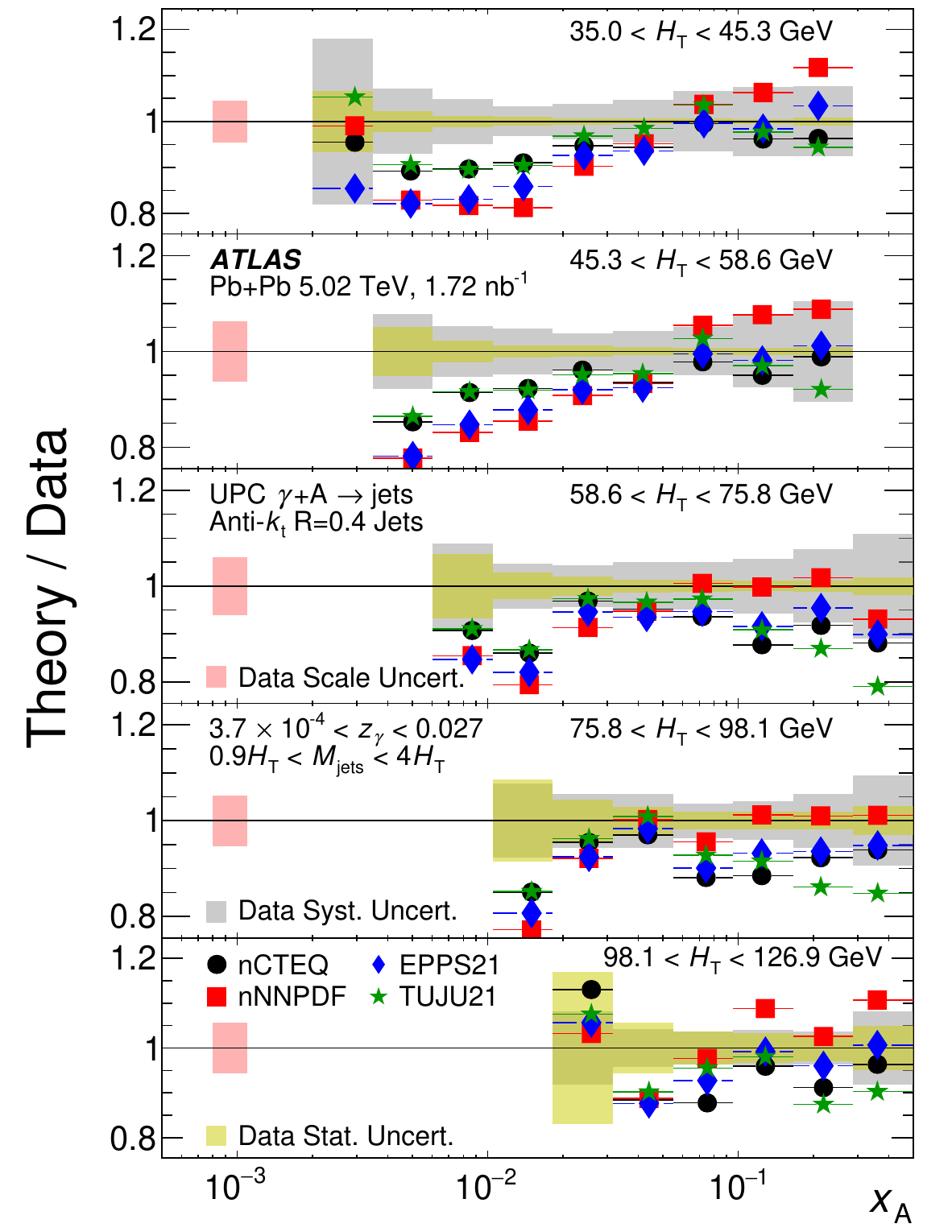}
\end{subfigure}%
\begin{subfigure}{.5\textwidth}
  \centering
  \includegraphics[width=.9\linewidth]{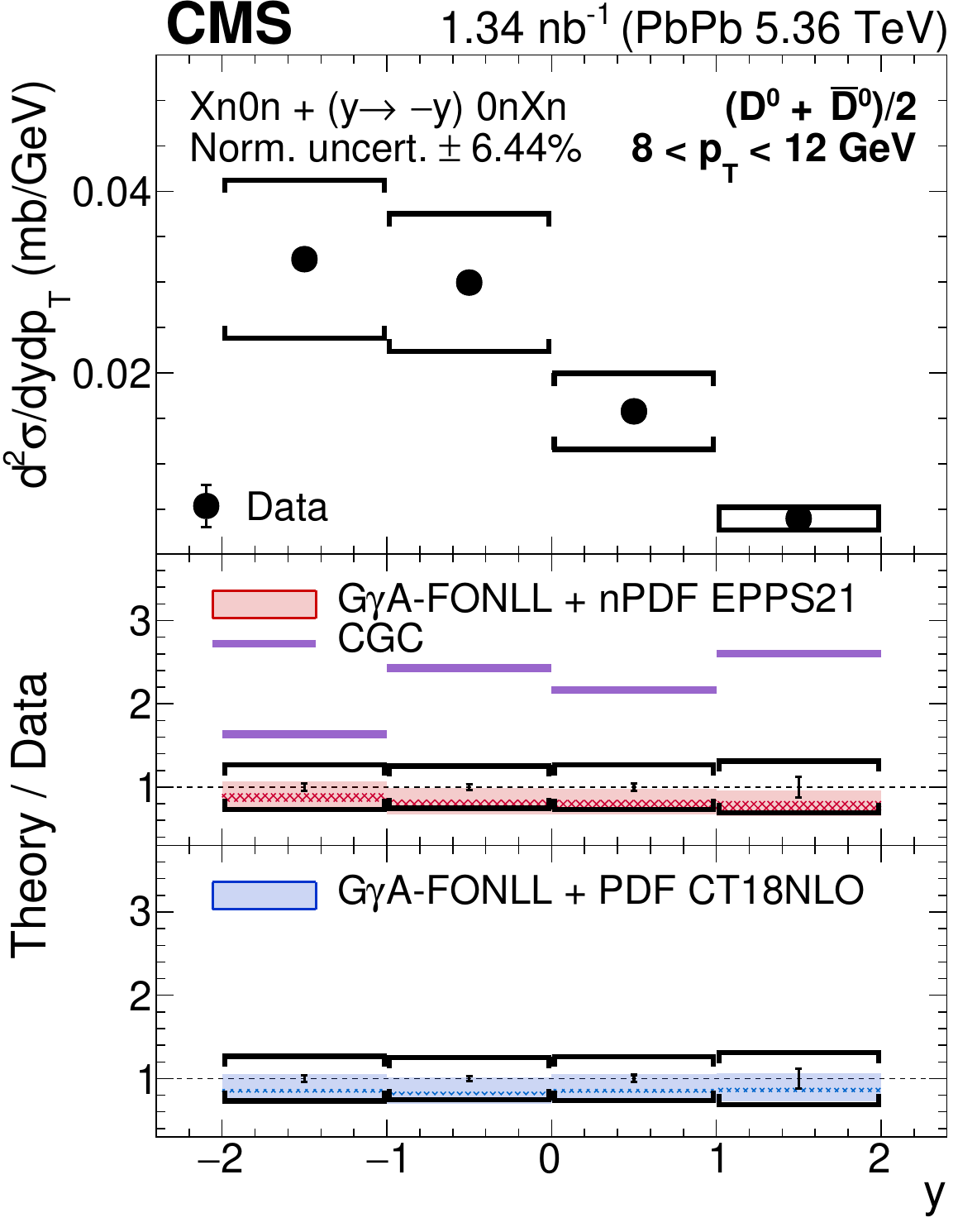}
\end{subfigure}
\caption{\label{fig:jetsD0} 
Left: Comparison of the measured triple-differential cross section for inclusive jet production in photonuclear interactions at the LHC with predictions at leading order based on different nuclear PDFs. Taken from~\cite{ATLAS:2024mvt}. Right: Comparison of $D^0$ production in photonuclear interactions at the LHC with models based on different nuclear PDFs as well as with a saturation-based (CGC) prediction. Taken from~\cite{CMS:2025jjx}.
}
\end{figure}

Jet production in photonuclear processes can also be studied for non-diffractive interactions, where instead of a pomeron, the interaction occurs with one gluon from the nucleus. At the LHC, this observable allows for the study of traditional nuclear PDFs over a large kinematic range. The main experimental challenge is that there is only one rapidity gap, in the side of the photon, which can be partially filled for the cases where the photon fluctuates into a complicated QCD state, e.g. a $\rho^0$ meson, and one parton of it participates in the hard scattering while the rest produce soft particles which potentially may fill, or reduce, the rapidity gap. 

Recently, the ATLAS collaboration made the first measurement of this process at the LHC~\cite{ATLAS:2024mvt}. The analysis utilizes data from Pb--Pb UPCs at $\sqrt{s}=5.02$ TeV and it covers a large kinematic range in three variables: the total transverse momentum $H_{\rm T}$ as well as the momentum fractions of the partons participating in the hard scattering from the photon $z_\gamma$ and the nucleus $x_{\rm A}$. An example of the impact of such measurements on our current knowledge of nuclear PDFs is shown in Fig.~\ref{fig:jetsD0}, left. First, one observes the large kinematic range covered by the measurements. Then, it is clear that none of the  predictions, which embody our current best knowledge of nuclear PDFs, is able to correctly describe data, with predictions undershooting  the measurements in large regions of phase space. Discrepancies between data and predictions increase towards lower scales ( $H_{\rm T}$ ) and low values of $x_{\rm A}$. It is also worth noting that the predictions vary across nuclear PDFs, which highlights the potential of this type of data to improve our knowledge of nuclear parton distributions.

In a very recent paper~\cite{ATLAS:2026sct}, the ATLAS collaboration measured inclusive jet production in two different classes: with and without neutrons produced at beam rapidity from the breaking up of the struck nucleus participating in the photonuclear interaction. They plot the ratio of the cross section without beam-rapidity neutrons to that with such neutrons as a function of $x_+$ the longitudinal momentum fraction of the struck quark. They find that the ratio is not unity, but increases with $x_+$. They interpret the process without neutrons as collisions involving nucleons at the periphery of the nucleus, implying an observation of impact-parameter dependence of the nuclear PDF.

\subsubsection{Photonuclear production of open charm\label{sec:openCharm}}

A similar measurement to the one just described is to look for charmed mesons instead of jets. Particles are better localized  and can be measured to lower transverse momenta than jets, offering experimental advantages and access to different kinematic regions. Such measurements have been advocated since long ago~\cite{Klein:2002wm} as an ideal tool to study the evolution of the gluon distribution at small $x$ because the photon interacts with a gluon to produce a charm-anticharm pair. Even measuring one of the two charm mesons puts constraints on our understanding of nuclear PDFs.

Very recently, the CMS collaboration measured the photonuclear production of $D^0$ mesons in UPCs of lead nuclei at 5.36 TeV per nuclear pair. The measurement was carried our in three ranges of transverse momentum from 2 GeV to 5 GeV, from 5 GeV to 8 GeV, and from 8 GeV to 12 GeV as well as three intervals of the $D^0$ rapidity $-2 < y <-1$, $-1 < y < 0$,  $0 < y < 1$, and $1 < y < 2$~\cite{CMS:2025jjx}. 
These data cover a range in $x$ form $3\times 10^{-4}$ to $3\times 10^{-2}$ with the smallest values of $x$ reached at largest  positive rapidities. 
The hard scales $Q^2$ probed by this measurement go from around 20 GeV$^2$ to some 600 GeV$^2$, with the smallest scales sampled by the lowest transverse momenta. An example of the measurements is shown in Fig.~\ref{fig:jetsD0}, right. The uncertainties are still large. The G$\gamma$A FONLL model based on the collinear factorization approach at next to leading order describes data reasonable well for both nuclear PDFs~\cite{Cacciari:2025tgr}, while the saturation-based calculation (CGC)~\cite{Gimeno-Estivill:2025rbw} which uses nonlinear  QCD evolution over predicts data by a factor of 3 at small $x$ (large $y$). The ALICE Collaboration has also observed open charm photoproduction~\cite{Nese:2025ohz}.

Profiting from the large luminosity available at the LHC during Runs 3 and 4, this program can be extended to the production of a bottom-antibottom quark pairs~\cite{Klein:2002wm,Goncalves:2003is}. The cross section for this process is suppressed by at least two orders of magnitude with respect to charm-anticharm pair production; nonetheless, the recent measurements of the CMS collaboration have motivated theorist to performed a detailed study of the production of B$^0$ mesons in UPCs at the LHC finding that such a measurement may be feasible~\cite{Goncalves:2026lea}. 

An even more ambitious possibility is to measure the photonuclear production of top-antitop quark pairs~\cite{Klein:2000dk,Goncalves:2013oga}. The cross section in Pb--Pb UPCs at the LHC is around 0.5 nb, which is more than six orders of magnitude lower than the cross section for the production of $b\bar{b}$-pairs, so it is outside the reach of the current UPC LHC program (although there may be a window of opportunity in pp UPCs~\cite{Goncalves:2020saa}), but it is a potential topic of study at  FCC-ee (in $\gamma\gamma$ interactions) and at FCC-hh.

\subsubsection{Probing for quark-gluon plasma formation in photon-nuclear interactions \label{sec:upcqgp}}
\begin{figure}[t]
\centering 
\includegraphics[width=0.45\textwidth]{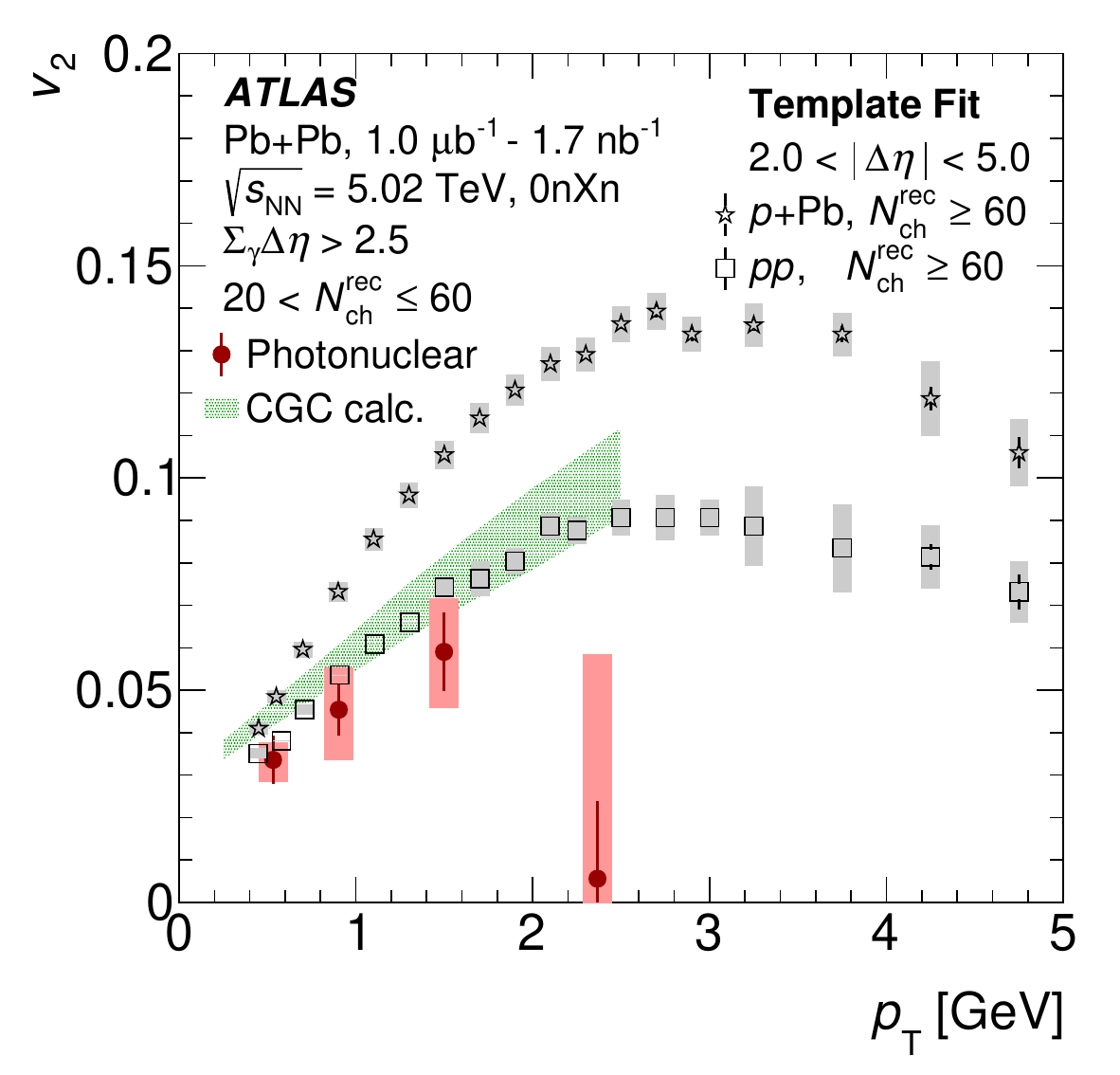}
\includegraphics[width=0.45\textwidth]{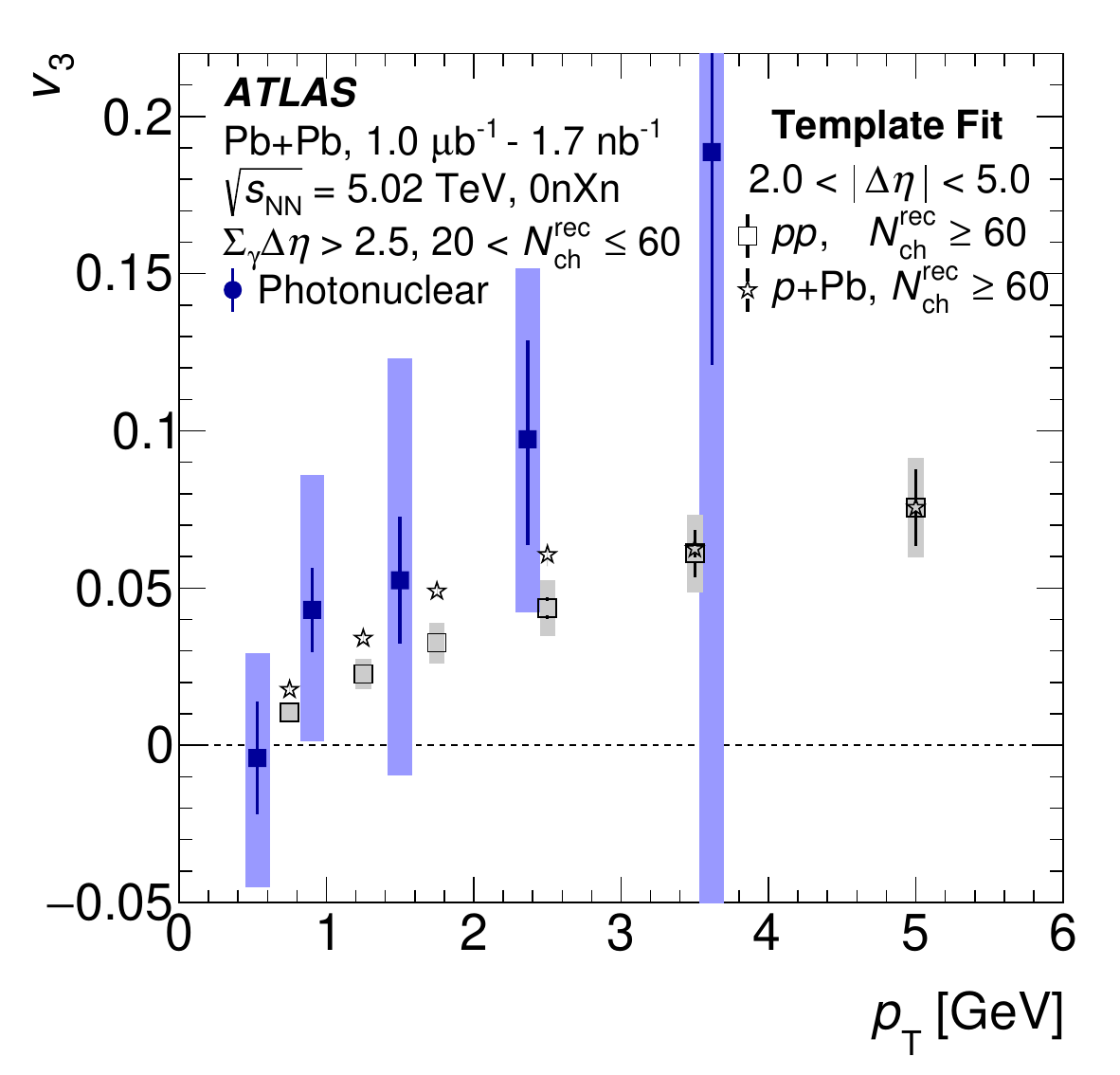}
\caption{
\label{fig:v2} 
Comparison of the measured triple-differential cross section for inclusive jet production in photonuclear interactions at the LHC with predictions at leading order based on different nuclear PDFs. Taken from~\cite{ATLAS:2024mvt}.}
\end{figure}

When the universe reached the age of $1 \mu s$ after the Big Bang  it was in a state called the quark-gluon plasma (QGP), where the temperature is so high that quarks and gluons are not bound into hadrons, but move freely forming a low viscosity medium that behaves as a perfect liquid. Such a state is nowadays routinely created in head-on collisions of heavy-ions at high energies, where enormous energies densities and temperatures are reached. The experimental discovery of the QGP was based on several different observables each with specific characteristics and all together forming a convincing portrait of the QGP~\cite{STAR:2005gfr,ALICE:2022wpn}. A natural question is to ask for the minimum requirements to form such a state. Surprisingly, several of the phenomena originally though to be a smoking gun of the QGP have been observed in so called small systems, like proton--proton, or proton--lead collisions among others~\cite{Grosse-Oetringhaus:2024bwr}.
Photonuclear interactions in UPCs have also been recently used to explore observables traditionally associated to the formation of QGP. In this case, the view is that the photon fluctuates into a hadronic state, like the $\rho^0$ vector meson, which then scatters head-on with the other incoming nucleus. 

The ATLAS collaboration has studied the so-called $v_2$ and $v_3$ flow coefficients that characterize potential collective behavior among the particles produced in the interaction~\cite{ATLAS:2021jhn}. These coefficients are obtained by selecting one particle and correlating it with all other particles produced in a collisions via their distance in the rapidity ($\Delta\eta$) and azimuthal angle ($\Delta\phi$) plane. The range $2<\Delta\eta<5$ of the corrected distribution is projected in $\Delta\phi$ and parametrized with help of a truncated Fourier series whose second and third coefficients are precisely $v_2$ and $v_3$. Figure~\ref{fig:v2} shows the dependence of  $v_2$ and $v_3$ as a function of the transverse momentum of the selected particle. The coefficients are compared to similar measurements performed in pp and p--Pb collisions as well as to the predictions of a model which generates non-zero $v_2$ values from fluctuations in the initial state of the gluon field of the  hadronic state from the photon fluctuation as well as of the Pb ion and does not contain a QGP component. Photonuclear data are seen to agree with those data from pp collisions and for $v_2$ also with the predictions of the model. But it is important to note that the model uses quite extreme values for some of the parameters, suggesting that the agreement with data is not necessarily a property of the model, but of the improbable values used for the parameters. The fact that the observed values of  $v_2$ and $v_3$ are not zero demonstrate that the particles produced in $\gamma$Pb collisions participate in long-range collective motion. The origin of this behavior, whether QGP or initial-state fluctuations, is not known, but these measurements provide new information from a system naively not associated with the  creation of QGP.

Another paper from the ATLAS collaboration tests in photonuclear interactions in UPCs three observables commonly used to search for QGP~\cite{ATLAS:2025tof}.
\begin{itemize}
\item The multiplicity dependence of the average transverse momentum of identified particles is sensitive to radial flow, the average velocity of the collective expansion of the collision products in the transverse plane.  If all particles expand collectively, they will share a common average velocity and the average transverse momentum of heavier particles will be higher than that of lighter ones. The average transverse momentum increases slowly with the multiplicity of the collision. Interestingly, one can compare the measurements in the Pb-going and $\gamma$ going direction and even though the mass ordering is present in both cases, the value of the average transverse momentum is reduced in the $\gamma$ going direction for each of the $\Lambda$, $K^0_{\rm S}$ and pion cases. In the  Pb-going  the results are similar in magnitude and shape to those obtained in p--Pb collisions. 
\item The transverse momentum dependence of the ratio of baryons to mesons is studied in the Pb-going and $\gamma$ going directions for the cases $\Lambda/K^0_{\rm S}$ and $\Xi^-/K^0_{\rm S}$. The $\Lambda/K^0_{\rm S}$ ratio in the Pb-going direction grows with transverse momentum up to around 2.5 GeV where it reaches a value of 0.6 and then decreases. In the $\gamma$ going direction the shape is the same but the ratio values are systematically below than in the  Pb-going direction. In studies of QGP this behavior is associated with an enhanced recombination probability of quarks because in a collective expansion they are close in phase space. The $\Xi^-/K^0_{\rm S}$ ratio has a similar shape, but reaches only up to around 0.15 and the current data uncertainties do not allow for a firm statement regarding a difference between the Pb- and $\gamma$-going sides.
\item The last observable studied in this paper is the ratio of yields of strange hadrons to charged hadrons as a function of the multiplicity of the event. The ratios in the $\gamma$-going direction are systematically below those in the Pb-going direction, which is consistent with an enhancement of strangeness production but can also be explained with other mechanisms.
\end{itemize}
In summary, these observations confirm that photonuclear interactions behave for these observables as hadronic collisions, probably dominated by $\rho^0$--Pb like interactions. A model based on QGP formation describes some of the characteristics of the measurements but not all. Together with the measurement of  $v_2$ and $v_3$ these studies confirm the importance of photonuclear interactions to explore the possible formation of QGP in small systems.

\subsubsection{Rapidity transport of baryon number}

Another area where photonuclear interactions may contribute to our
understanding of QCD is to explore how baryon number is transported in
rapidity. That is, the initial colliding hadrons have a baryon number
traveling at beam rapidities. Experiments have found that it is
possible to measure a baryon number different from zero at midrapidity, but it is not know how this baryon number was transported by several units of rapidity.

The STAR collaboration has addressed this question by measuring ${\rm
d}N/{\rm d}y$ the net
difference of protons and anti-protons as a function of rapidity near  midrapidity in $\gamma$--Au
collisions at a center-of-mass energy of the Au--Au system of 54.4 GeV
per nucleon pair~\cite{STAR:2024lvy}. This measurement revealed that
\begin{equation}
\frac{{\rm d}N}{{\rm d}y}\propto e^{-\alpha_{\rm B} y}, 
\end{equation}
with $\alpha_{\rm B} =1.04 \pm 0.22$.

The photon  has zero baryon number, so the transport
has to be from the beam rapidity of the ion participating in the
photonuclear collision to midrapidity, some 4 rapidity units away.
There are two competing explanations of baryon transport. The baryon
number can be  carried by the the valence quarks of the nucleons in the
nucleus, with models based on this assumption predicting $\alpha_{\rm
B}$ around 2.5. An alternative explanation is that there is a
particular configuration of the gluon field, called a junction that
has a center (the junction) to which all three valence quarks are
connected \cite{Kharzeev:1996sq}. As the junction is mainly formed by low momentum gluons and these can also connect to quarks from quark-antiquark pairs, it is
easier for them to move in rapidity. Models based on this idea predict  $\alpha_{\rm
B}$ around 1 for a junction-junction scenario and around 0.5 for a
junction-pomeron case. This measurement  favors 
the explanation that baryon number is carried in junctions that can
transport it across large distances in rapidity.

The simplest case of baryon transport in electromagnetic interaction is in `backward production' exclusive photoproduction reactions, like $\gamma + {\rm p}\rightarrow V + {\rm p}$, but where the vector meson $V$ takes most of the proton momentum, and the proton takes most of the photon momentum; in the center of mass frame, the photon/meson and the proton essentially trade places~\cite{Sweger:2023bmx}. Although most studies of backward production have been done at fixed-target accelerators,  some reactions can also be studied using UPCs~\cite{Gayoso:2021rzj}, enabling measurements at higher photon energies.

\section{Two-photon interactions}

Two-photon interactions encompass a wide variety of physics, from purely quantum electrodynamic processes to studies of hadron structure, to searches for beyond-standard-model physics.   In fact, the first widely recognized `application' of UPCs was to look for two-photon production of the Higgs boson
\cite{Papageorgiu:1988yg,Cahn:1990jk}.  

\subsection{Two-photon luminosity}

The cross section for two-photon reactions may be factorized into a two-photon luminosity and cross section:
\begin{equation}
    \sigma (AA\rightarrow AA\gamma\gamma \rightarrow AAX) = 
    \int {\rm d}k_1 {\rm d}k_2 
    \frac{dN(k_1)}{dk_1} \frac{dN(k_2)}{dk_2}
  =    \sigma(\gamma\gamma\rightarrow X)
    \frac{{\rm d}^2\ell_{\gamma\gamma}}{{\rm d}W{\rm d}y}    \sigma(\gamma\gamma\rightarrow X)
    \label{eq:twophotonmaster}
\end{equation}
where $k_1$ and $k_2$ are the two photon energies, $ \frac{n(k_1)}{k_1}$ are the photon fluxes, and ${\rm d}^2\ell_{\gamma\gamma}/{\rm d}W{\rm d}y$ is the two-photon luminosity.  Here, we have applied a change of variables from photon energies to pair invariant mass $W$ and rapidity $y$: $W^2=4k_1k_2$ and $y=1/2\ln(k_1/k_2)$.  

For UPCs, as usual, we require $b>2R_A$.  This breaks the factorization into two separate photon fluxes.  Instead the two-photon luminosity can be calculated in impact-parameter space:
\begin{equation}
  \frac{{\rm d}^2\ell_{\gamma\gamma}}{{\rm d}W{\rm d}y} = \ell_{AA} \frac{W}{2} \int {\rm d}^2b_1 \int {\rm d}^2b_2 n(k_1,b_1) n(k_2,b_2)
  P_{\rm nohad}(|\vec{b_1}-\vec{b_2}|)
  \label{eq:twophotonluminosity}
\end{equation}
where $\ell_{AA}$ is the ion-ion luminosity, and $b_1$ and $b_2$ are two-vectors used to integrate over transverse space.  This four-dimensional integral may be reduced to a three-dimensional integral using a trick: the 4-dimensional integral over ${\rm d}^2b_1 {\rm d}^2b_2$ reduces to the 3-dimensional ${\rm d}|b_1|{\rm d}|b_2|\cos{\theta}$, where $\theta$ is the angle between $\vec{b_1}$ and $\vec{b_2}$~\cite{Baur:1990fx}.  This approach works well for the hard-sphere model, where, for a given $|b_1|$ and $|b_2|$ the $b>2R_A$ requirement maps to a hard cut on $\theta$, but it is less advantageous when one includes more realistic nuclear density profiles. 

The range of integration for $b_1$ and $b_2$ merit discussion.  The use of $ P_{\rm nohad}$ eliminates interactions where the nuclei overlap.  However, there is another possibility; even when $b>2R_A$, it is possible to have the two-photon interaction point within one of the nuclei, so $b_1<R_A$ or $b_2<R_A$.  In most early calculations, it was implicitly assumed that the final state would be hadronic, so any production within a nucleus would lead to nuclear breakup and the destruction/alteration of that hadronic state.  So, luminosity calculations would explicitly avoid this region.     In this approach, it is easy to add conditions about nuclear breakup $P_{\rm \gamma breakup}(b)$ to Eq.~\ref{eq:twophotonluminosity}~\cite{Baltz:2009jk}. This scheme is used in the STARlight Monte Carlo generator for photon-induced interactions in UPCs~\cite{Klein:2016yzr}. Recent Monte Carlo generators like Upcgen~\cite{Burmasov:2021phy,Burmasov:2026ytn} and SuperChic~\cite{Harland-Lang:2020veo}
already include the contribution of the photon flux for the cases where the $\gamma\gamma$ interaction point is within one of the nucleus.

Most of the current experimental interest in two-photon reactions involves the production of lepton pairs, which may emerge from within a nucleus unscathed.  Two photon production of $W^+W^-$ has also been observed~\cite{ATLAS:2020iwi}; this process has some sensitivity to BSM physics. 

\subsection{Lepton Pair Production}

Pair production is one of the oldest particle reactions known, with the first theoretical discussions dating back to 1934~\cite{Landau:1934zj}.  Pair production from colliding relativistic heavy ions can have extremely large cross sections~\cite{Baur:2007zz}.  The cross section to produce an $e^+e^-$ pair in a lead--lead collision at the LHC is about 200,000 barns!  The cross section is large enough so that in collisions with $b\gtrsim 2R_A$, an average of 2 1/2 pairs are produced.  Of course, most of these pairs have pair mass $M_{ee} \gtrsim 2m_e$, and lepton $p_\perp\approx m_e$, so they are invisible in typical RHIC or LHC detectors. 

With heavy nuclei, the couplings are large ($Z^2\alpha >1$) and perturbative cross-section calculations were considered suspect~\cite{Klein:2004is} and references therein.  Before the Relativistic Heavy Ion Collider (RHIC) began operation, there was concern that the cross sections for pair production and pair production with capture (discussed below) would be large enough to quickly degrade the circulating beams. This led to a series of coupled-channel cross-section calculations, some of which found very large cross sections~\cite{Rumrich:1991xs}.  Some of the calculations used very large sets of basis states, and so were run on supercomputers.  These calculations were considered inconclusive because the number of basis states may have been too small. 

Later theorists developed non-perturbative techniques to solve the Dirac equation directly, and so to calculate cross sections to all orders.   The first found cross sections  were consistent with the lowest-order calculations, without the expected Coulomb corrections~\cite{Baltz:1998zb}. This was a surprise; later calculations changed the order of integration, and found results that were consistent with a moderate Coulomb correction~\cite{Ivanov:1998ru,Lee:1999eya}, of order 25\% comparable in size to the Bethe-Maximon correction seen for pair production with real photons on heavy targets.  However, a more recent next-to-leading order calculation found much smaller  Coulomb corrections, albeit mostly for relatively near-threshold pairs~\cite{Lee:2009vu}. 

Because the total cross section is so large, the possibility of multiple $e^+e^-$ pair production involving a single ion pair becomes large as $M_{ee}\to 2m_e$, and $P(b)$ can grow larger than one.  These equations can continue to be used, but $P(b)$ can no longer be treated as a probability.  Instead, it becomes the mean number of pairs, with the event-by-event number of pairs following a Poisson distribution, with the given mean~\cite{Alscher:1996mja}. 

For higher-energy pairs, with $M_{ee}$ and lepton $p_\perp$ large enough to be visible in RHIC and LHC detectors, the cross section has always been expected to be close to the lowest order Breit-Wheeler cross section
\cite{Brodsky:1971ud}:
\begin{equation}
    \sigma{\gamma\gamma\rightarrow e^+e^-} =
    \frac{4\pi\alpha^2}{W^2}
    \bigg( \big[2 + 2D^2 -D^4\big] 
    \ln{([\frac{W+\sqrt{W^2-4m^2}}{2m})}
    - \sqrt{1-D^2}\big[1+D^2\big]\bigg),
    \label{eq:pairangle}
\end{equation}
{\bf check equation}
where $W$ is the pair mass and $D^2=4m^2/W^2$.
As previously noted, the cross section is peaked near threshold.  
The rapidity depends on the ratio of the photon energies, and its distribution is very broad.  

UPC pair production can be modeled using either the equivalent photon approximation or as a full lowest-order quantum electrodynamics (QED) calculation.   The pair $p_\perp$ is the vector sum of the two photon $k_T$s, so is typically very small. These two approaches lead to very similar results, except that the QED calculation has a higher average pair $p_\perp$~\cite{STAR:2004bzo,Hencken:2004td}, since it accounts for photon azimuthal orientations.

The angular distribution of these pairs is 
\begin{equation}
G(\theta)=2+4(1-Z^2)\frac{(1-Z^2)\sin^2(\theta)\cos^2(\theta)+Z^2}
{[1-(1-Z^2)\cos^2(\theta)]^2},
\end{equation}
where $\theta$ is the angle between the beam direction and lepton direction, in the pair center-of-mass frame. $G(\theta)$ is peaked at small angles, so most of the leptons have large absolute pseudorapidities.  The lepton $p_\perp$ depends on the pair $p_\perp$ and on $\theta$. 

The CHALMNP Collaboration at the CERN Intersecting Storage Rings~\cite{CERN-Harvard-AnnecyLAPP-MIT-Naples-Pisa:1980fls}, the STAR experiment at RHIC~\cite{STAR:2004bzo,STAR:2019wlg}, the CDF experiment at the Fermilab Tevatron (for $p\overline p$ collisions)~\cite{CDF:2006apx},  and ATLAS~\cite{ATLAS:2015wnx, ATLAS:2020epq,ATLAS:2022srr}, ALICE~\cite{ALICE:2013wjo,ALICE:2026mlg} and CMS~\cite{CMS:2011vma} at the LHC have studied pair production.  The cross sections are compatible with the lowest-order cross section, usually in between the cross sections calculated allowing for two-photon reactions inside the nucleus and the cross sections that exclude this region.  

\begin{figure}[t]
\centering 
\includegraphics[width=0.6\textwidth]{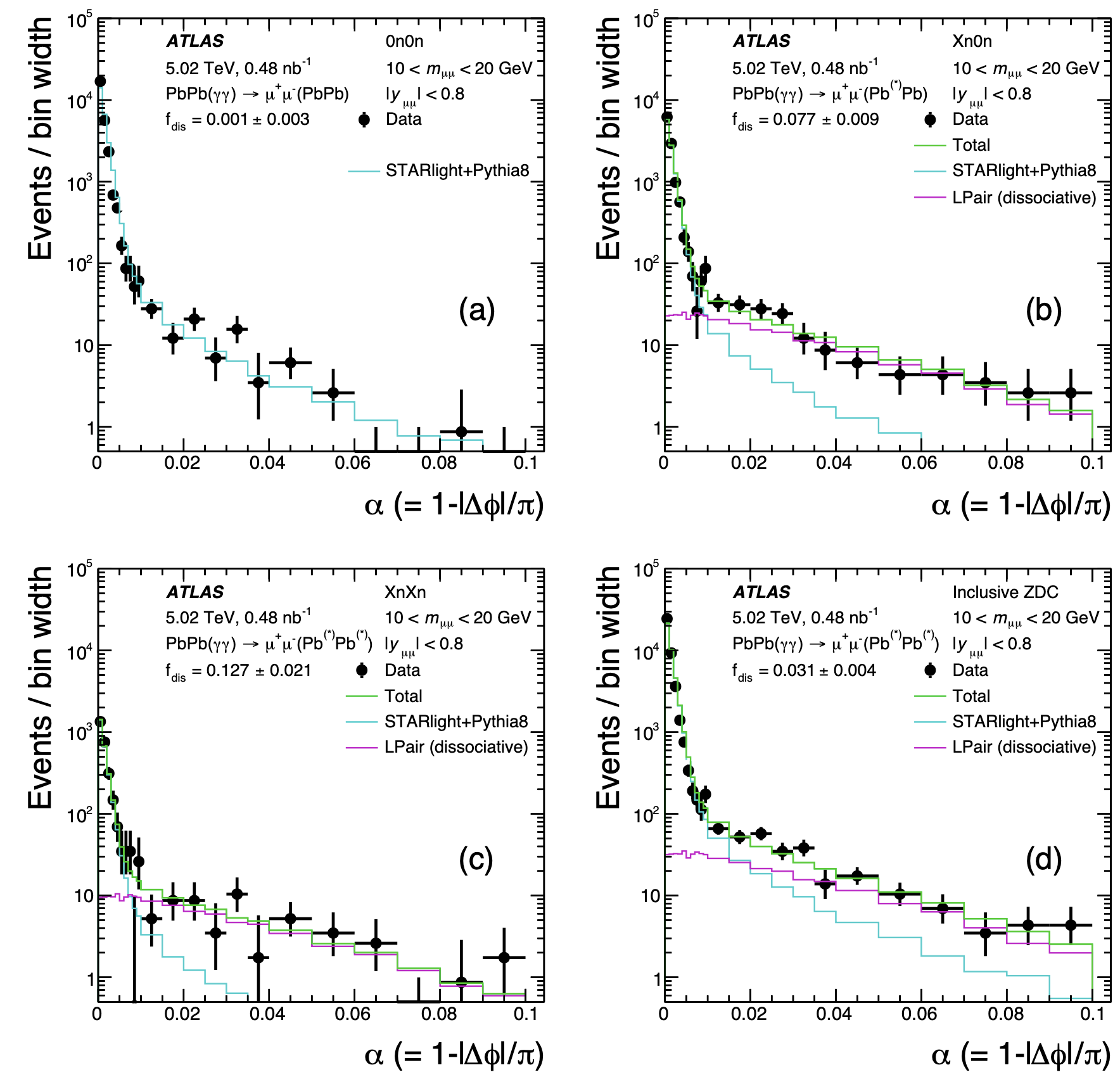}
\includegraphics[width=0.6\textwidth]{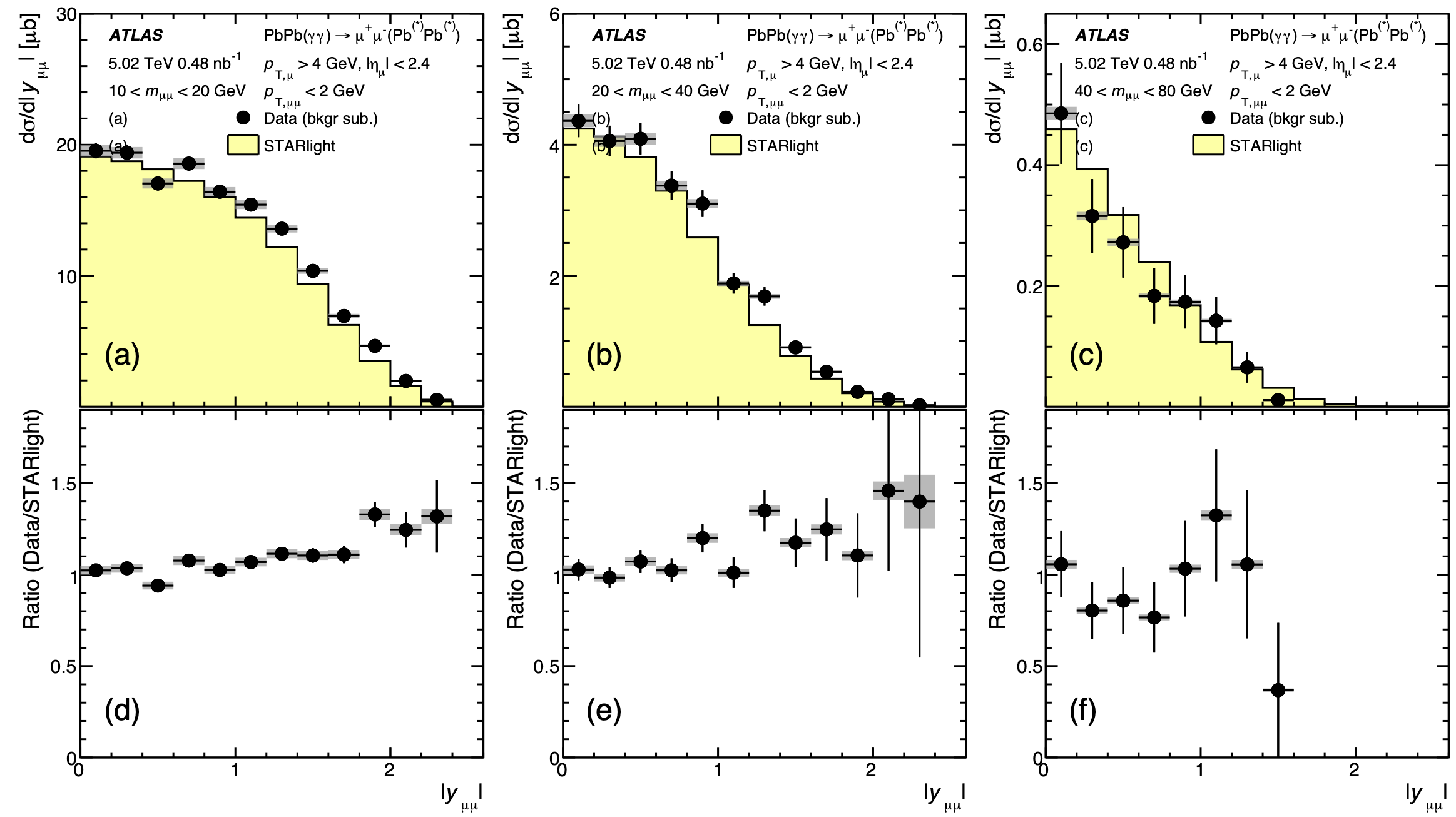}
\caption{
\label{fig:leptonpairs} 
ATLAS data on $\gamma\gamma\rightarrow \mu^+\mu^-$.  The top panel shows the azimuthal acoplanarity $\alpha=1-|\delta\phi|/\pi$, for four different combinations of neutrons in their ZDCs,  while the bottom panel shows the pair rapidity distribution for three different pair mass ranges.
From~\cite{ATLAS:2020epq}.
}
\end{figure}

Figure~\ref{fig:leptonpairs} shows recent ATLAS results on muon pair production.  The collaboration presented their results in terms of azimuthal acoplanarity $\alpha=1-|\delta\phi|/\pi$ instead of pair $p_\perp$ because the detector resolution for $\alpha$ is better than for $p_\perp$.  The top panel shows  $\alpha$ for four different selections of neutrons in their ZDCs; neutrons can come either from additional photon exchange or from a relatively rare process whereby a nucleus breaks up as it emits a photon.  

Events without neutrons should mostly be pure $\gamma\gamma\rightarrow\mu^+\mu^-$.  The ATLAS data are peaked at small $\alpha$, but with a high$-\alpha$ tail that is suppressed by a few orders of magnitude.  The peak and tail are well described by a STARlight~\cite{Klein:2016yzr} plus PYTHIA~\cite{Sjostrand:2014zea} simulation. PYTHIA is needed to account for the high$-\alpha$ tail, which is due to final-state radiation (FSR) from the leptons.  It may also be well described analytically using a Sudakov resummation to account for the FSR~\cite{Klein:2018fmp,Klein:2020jom}.  For the event classes that include neutrons, the high$-\alpha$ tail is higher, in accord with calculations that account for the $p_\perp$ kick from nuclear breakup~\cite{Vermaseren:1982cz}. 

The rapidity distributions (bottom panel) are in good agreement with predictions based on Eqs.~\ref{eq:twophotonluminosity} and~\ref{eq:pairangle}. 

\begin{figure}[t]
\centering 
\includegraphics[width=0.6\textwidth]{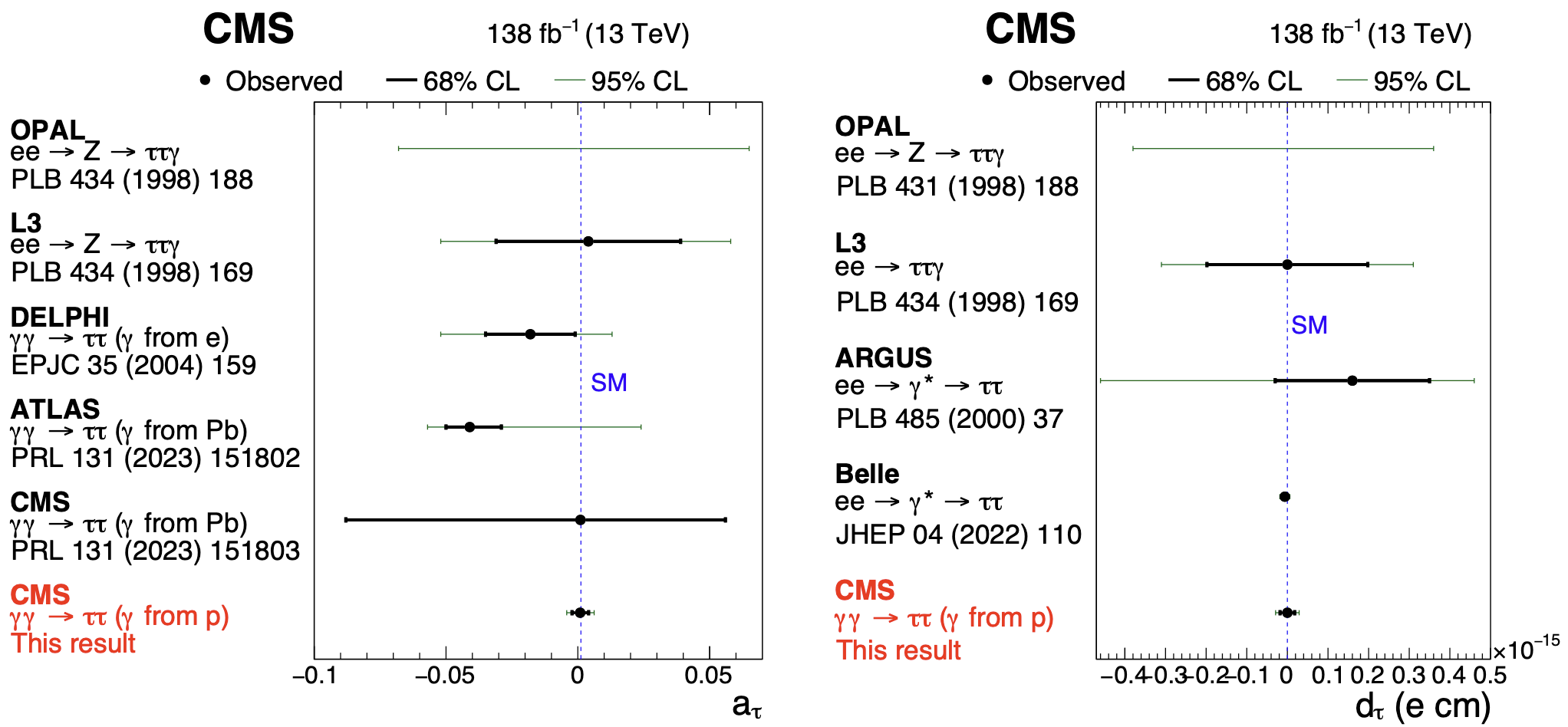}
\caption{
\label{fig:taulimits} 
Limits on the anomalous magnetic moment  and electric dipole moment of the $\tau^\pm$ lepton from UPC two-photon production of $\tau^+\tau^-$, compared to limits from other sources.
From~\cite{ATLAS:2022ryk}.
}
\end{figure}

One particularly interesting two-photon channel is the reaction $\gamma\gamma\rightarrow\tau^+\tau^-$.  The kinematic distributions are sensitive to BSM physics, including a possible $\tau^\pm$ anomalous magnetic moment, $a_\tau$ and electric dipole moment $d_\tau$.  This reaction  has been studied for both pp and Pb--Pb UPCs. Tau pair production has been studied by a number of experiments~\cite{ATLAS:2022ryk,CMS:2022arf,CMS:2024qjo,ATLAS:2026wrz}. As Fig.~\ref{fig:taulimits} shows, the higher luminosity with pp collisions leads to far tighter limits than in ion collisions.  

\subsection{$e^+e^-$ production with capture}

Pair production with capture is similar to open pair production, except that the electron is produced bound to one of the incident nuclei, lowering its charge by one.  This process was first observed at LBNL's Bevalac, in collisions of $^{92+}U$ on a lead target~\cite{Belkacem:1993wq}.  The fixed target geometry simplifies the observation of the altered ions. 

In $\overline{\rm p} {\rm p}$ collisions, as at the CERN's LEAR (Low Energy Antiproton Ring) or at Fermilab's accelerator complex, the positron can be produced bound to the $\overline{\rm p}$, producing an antihydrogen atom.  The first antihydrogen was produced at LEAR using this reaction~\cite{Baur:1995ck}, followed by production at the Fermilab antiproton accelerator~\cite{Blanford:1997up}.

At RHIC and the LHC, pair production with capture leads to single-electron gold and lead, respectively, changing the ions charge:mass ratio. The resulting ions will be lost to the beam.  This is a primary source of beam loss at the LHC.    Because the electron carries little momentum, the resulting beam of single-electron ions remains reasonably well collimated, depositing significant energy at a fairly specific point on the beampipe~\cite{Klein:2000ba}.   This secondary beam was first observed at RHIC in a retrospective analysis of Cu$^{+29}$ beams, using data from radiation-monitoring photodiodes about 140 m downstream of the PHENIX interaction point~\cite{Bruce:2007zza}.  At the LHC, this beam of single-electron lead will hit the beam pipe about 440 m downstream from the interaction regions, depositing significant power into the superconducting magnets~\cite{Schaumann:2020lsl}.  At a  peak (instantaneous) luminosity of $6\times 10^{27}$ cm$^{-2}$ s$^{-1}$, these beams can carry up to 140 watts of power.  It has been demonstrated that this is enough energy to quench the LHC magnets~\cite{Jowett:2016yfa}.  Absent alleviation methods, such as defocusing or adding a collimator, this secondary beam sets one limit on the LHC ion--ion luminosity.  For higher energy accelerators, such as FCC-hh, the problem worsens, due to the combined effects of the higher luminosity, higher cross section, and higher energy carried by each ion. 

\subsection{Light-by-Light scattering}

\begin{figure}[t]
\centering 
\includegraphics[width=0.3\textwidth]{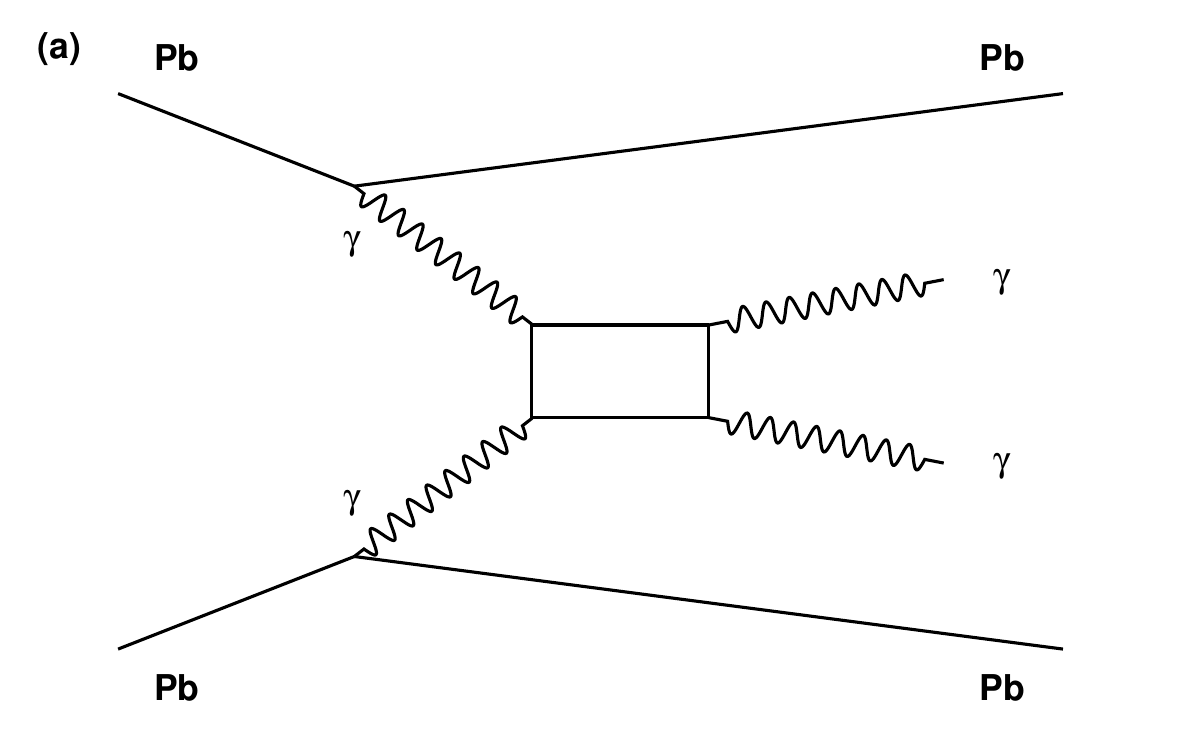}
\includegraphics[width=0.3\textwidth]{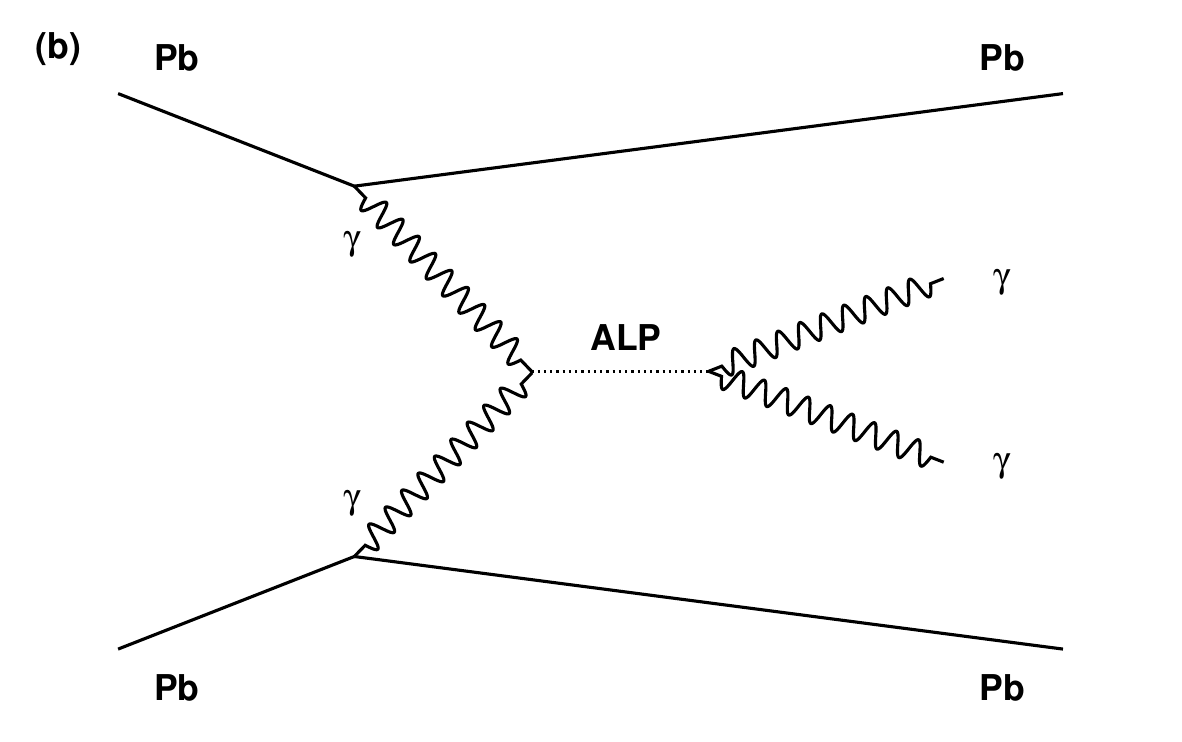}
\includegraphics[width=0.35\textwidth]{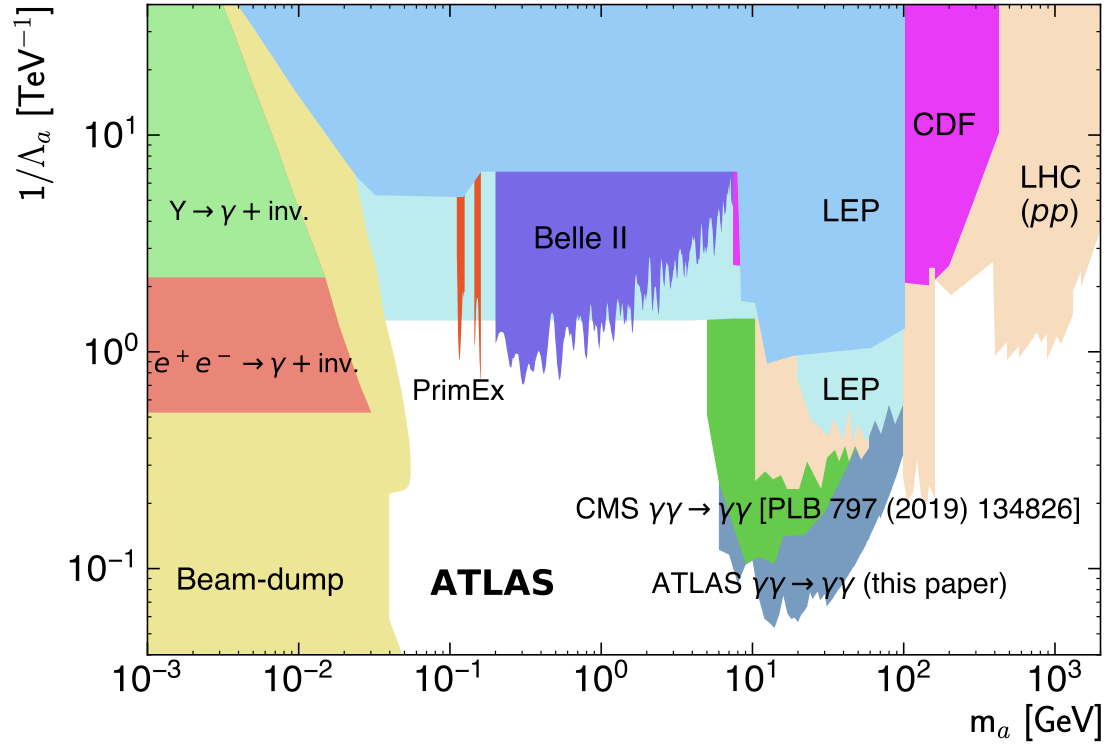}
\caption{
\label{fig:lightbylight} 
Two diagrams that can give rise to light-by-light scattering.  The box diagram (left) has a loop that includes all electrically charged particles, including hypothetical BSM particles.  Light-by-light scattering can also be mediated by an axion-like-particle (center).  The latter channel produces $\gamma\gamma$ events concentrated at the mediator mass.  Axion limits (right) from UPC light-by-light scattering and other sources. UPCs are the most sensitive probe for axions with masses between about 5 GeV and 90 GeV; from~\cite{CMS:2018erd}.
}
\end{figure}

Although the cross section is small, light-by-light scattering, $\gamma\gamma\rightarrow\gamma\gamma$ is of interest for a couple of reasons.  It can occur via either of the diagrams in Fig.~\ref{fig:lightbylight}.  

In the first diagram, Fig.~\ref{fig:lightbylight} (left) scattering occurs via a box diagram which includes amplitudes from all electrically charged particles, whether standard model or beyond.  So, the total cross section is sensitive to BSM physics.  Although this process has been observed, the measurement is not yet precise enough to constrain BSM physics.

Alternately, Fig.~\ref{fig:lightbylight} (center), the process can proceed via an axion-like particle (ALP) intermediary.  This channel has a more distinctive signature: a peak in the $\gamma\gamma$ invariant mass at the axion mass.  Both ATLAS~\cite{ATLAS:2020hii} and CMS~\cite{CMS:2018erd}  have searched for this process and set limits on possible ALPs. These limits are set in a 2d plane of axion mass and coupling, see Fig.~\ref{fig:lightbylight} (right).  UPCs are the most sensitive probe for axion masses between about 5 GeV and 90 GeV.  Future searches by the proposed ALICE 3 detector should probe down to lower masses~\cite{ALICE:2022wwr,Jucha:2023hjg}.

One difficulty for both channels is the presence of background from hadronic resonances decaying to two photons~\cite{Klusek-Gawenda:2025ffh}.  This has been considered for hadronic resonances that are produced by two photons, but there is a larger background from much more copious photoproduction of vector mesons, which then decay radiatively to particles like the $\eta_c$ or $\eta_b$, which then decay to two photons~\cite{Klein:2018ypk}. If the soft photon from the vector meson decay is missed, then the observed final state will be two photons. 

\subsection{Two-photon production of mesons}

Two-photon production of mesons was one of the first topics proposed for investigation using UPCs~\cite{Baur:1990fx,Greiner:1992fz,Nystrand:1998hw,Baur:2001jj}. Compared to two-photon reactions at $e^+e^-$ colliders like LEP, UPCs offer somewhat higher luminosities, and, at the LHC, the potential of higher $\gamma\gamma$ collision energies.  The partial width (coupling) of a meson to two photons is sensitive to the charge content of mesons, so glueballs should have a smaller coupling than conventional $q\overline q$ mesons.

The cross sections for two-photon production can be calculated from Eq.~\ref{eq:twophotonmaster}, with the cross section to produce a narrow meson with spin $J$ and mass $m$:
\begin{equation}
\sigma(\gamma\gamma\rightarrow X) = 8\pi^2 (2J+1)
\frac{\Gamma_{\gamma\gamma}}{2W^2 \delta (W-m)}
\end{equation}
where $\Gamma_{\gamma\gamma}$ is the partial width of the meson to two photons.  For wider resonances, it is necessary to integrate over Eq.~\ref{eq:twophotonmaster}, including the mass-dependent coupling.

Unfortunately, the cross sections for two-photon production of mesons is small.  Because of the extra photon, the cross section is reduced by a factor of $\alpha_{\rm EM}\approx 1/137$ compared to photoproduction (not counting factors of a few due to the pomeron coupling, $Z$ vs. $A$, etc.).  For example, with lead beams at the LHC, the $\rho^0$ production cross section is 5200 mb~\cite{Klein:1999qj}, compared with 23 mb for the $f_2(1270)$~\cite{Baltz:2009jk}---a factor of 225 difference.  The rate difference is large enough so that coherent photoproduction of vector mesons, followed by radiative decay (with the emission of a soft, unobserved photon) can be a larger source of pseudoscalar mesons like the $\eta_c$ than direct production from two photons~\cite{Klein:2018ypk}. Beyond that, two-photon interactions at UPCs have quite similar kinematics to coherent photoproduction, so the two reactions are not easy to separate. 

Two-photon production of mesons has not yet definitively been observed in UPCs.  However, with the larger LHC Run 3 and Run 4 samples now being collected, observation should not be too far off.  
The likely first-sighting will be the decay $f_2(1270)\rightarrow\pi^+\pi^-$. Already, in Fig. \ref{fig:dipionmass} (left), there is an apparent enhancement at the $f_2(1270)$ mass, with a width that appears comparable with the $f_2(1270)$.  With lead beams at the LHC, the predicted cross section is about 23 mb \cite{Baltz:2009jk}, about 0.4\% of that for the $\rho^0$ - roughly consistent with Fig. \ref{fig:dipionmass}.  

\subsection{Monopole searches in two-photon interactions}

In 1931, Dirac proposed a model of a monopole, a particle that carries an isolated magnetic charge, as an infinitely thin and long solenoid. Using this scheme, he discovered that the monopole charge is quantized and related to the inversely proportional to the electric charge, which implies that the existence of a monopole would require charge quantization~\cite{Dirac:1931kp}. Other forms of monopoles have been found in quantum field theories attempting to extend the Standard Model, for a review see~\cite{Mavromatos:2020gwk}.

Recently, it was realized that the Schwinger mechanism~\cite{Schwinger:1951nm} could be used to search for monopoles in UPCs of heavy ions~\cite{Gould:2019myj}. The idea is that the magnetic field, see Fig.~\ref{fig:EcrossB}, is so intense that it could suffice to extract a monopole-antimonopole pair out of the vacuum. This idea has been tested, using Pb--Pb UPC collisions at the LHC, by the Moedal~\cite{MoEDAL:2021vix} and ATLAS~\cite{ATLAS:2024nzp} collaborations at the LHC. No monopoles have been found. These measurements have been used to put strong limits on monopole production at small masses. Moedal uses a trapping technique that allows them  to set constraints, at 95\% CL, for monopoles with Dirac charges 1, 2, and 3 up to masses of 75 GeV. ATLAS uses the high ionizing power, and consequent energy loss, as well as exploiting that the traces in the detector are  different for monopoles and  standard electrically charged particles. They exclude  monopoles with a single Dirac magnetic charge and
mass below 80 GeV to 120 GeV with the range corresponding to different models.

\section{Quantum interferometry, polarization and nuclear structure \label{sec:interference}}
Both coherent photoproduction and two-photon interactions exhibit some unique quantum effects: the two ions act like a two-slit interferometer, albeit on in which the two sources have no prior overlap.  These interference effects have been recently reviewed~\cite{Brandenburg:2025one}.   For coherent photoproduction, our inability to tell which nucleus is the photon emitter and which is the target leads to interference between the two possibilities.   Because of this in-distinguishability, the two amplitudes are added~\cite{Klein:1999gv}:
\begin{equation}
\sigma(b,k_\perp,b) = \big| A(k_\perp,k_1,b)e^{i\phi_1}  - A (k_\perp,k_2,b) e^{-\phi_2+\vec{k_\perp}\cdot\vec{b}/\hbar} \big|^2
\label{eq:interfere}
\end{equation}
where $k_1$ and $k_2$ are the two photon energies from Eq.~\ref{eq:twofold}, $A$ is the production amplitude at a given photon energy, $p_\perp$ and $b$,  $\phi_1$ and $\phi_2$ are the two phases associated with  vector meson production.  For light mesons, these phases are believed to vary only slowly with photon energy~\cite{Bauer:1977iq} so we will take $\phi_1=\phi_2$.

The minus sign between the two amplitudes is because one can swap scattering from nucleus 1 with scattering from nucleus 2 by a parity exchange.  Vector mesons are negative parity, so the interference is destructive as $p_T\rightarrow 0$.  For ${\rm p}\overline {\rm p}$ collisions, such as at the Fermilab Tevatron, swapping requires both charge and parity flips, so the sign becomes positive~\cite{Klein:2003vd}.  

To find the total cross section, it is necessary to integrate over $\vec{b}$.  The behavior of the cross section depends on $y$ and $p_\perp$.  When $|y|$ is large, then one amplitude or the other will dominate, and there will be no interference.  However, when $|y|$ and $p_\perp$ are both small, then the amplitudes are similar and $\vec{k_\perp}\cdot\vec{b} < \hbar$, so the exponential is 1 and the cross section is reduced.  This interference is  visible when $p_\perp < \hbar/\langle b\rangle$; this is easiest to see when additional photons are exchanged, so $\langle b \rangle$ is smaller.  The STAR Collaboration has observed this interference in $\rho^0$ photoproduction at low $p_\perp$ and midrapidity~\cite{STAR:2008llz}.   
%The impact parameter distribution depends on the number of accompanying photons; this interference is most visible in reactions involving multiple photons. 

This interference is of considerable interest as a quantum mechanical system---the two ions act like a double-slit interferometer, albeit one in which the two sources share no common history.  Moreover, the $\rho^0$ decay almost immediately, before the $\rho^0$ amplitudes from the two sources can overlap; any interference must collectively involve the vector meson decay products, despite the fact that they are traveling in opposite directions, in many cases at nearly the speed of light.  So, the system is an example of the Einstein-Podolsky-Rosen paradox~\cite{Klein:2002gc}.

\begin{figure}[t]
\centering 
\includegraphics[width=0.39\textwidth]{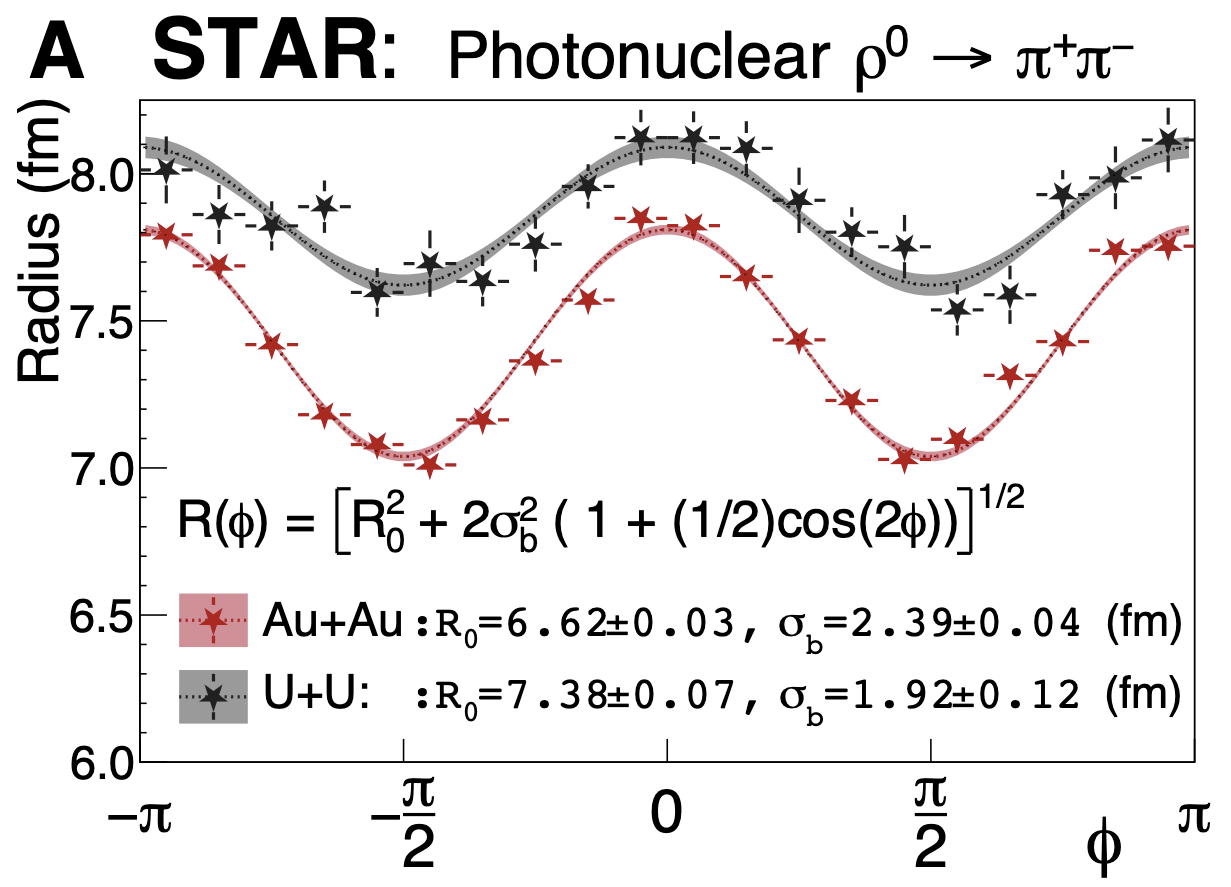}
\includegraphics[width=0.46\textwidth]{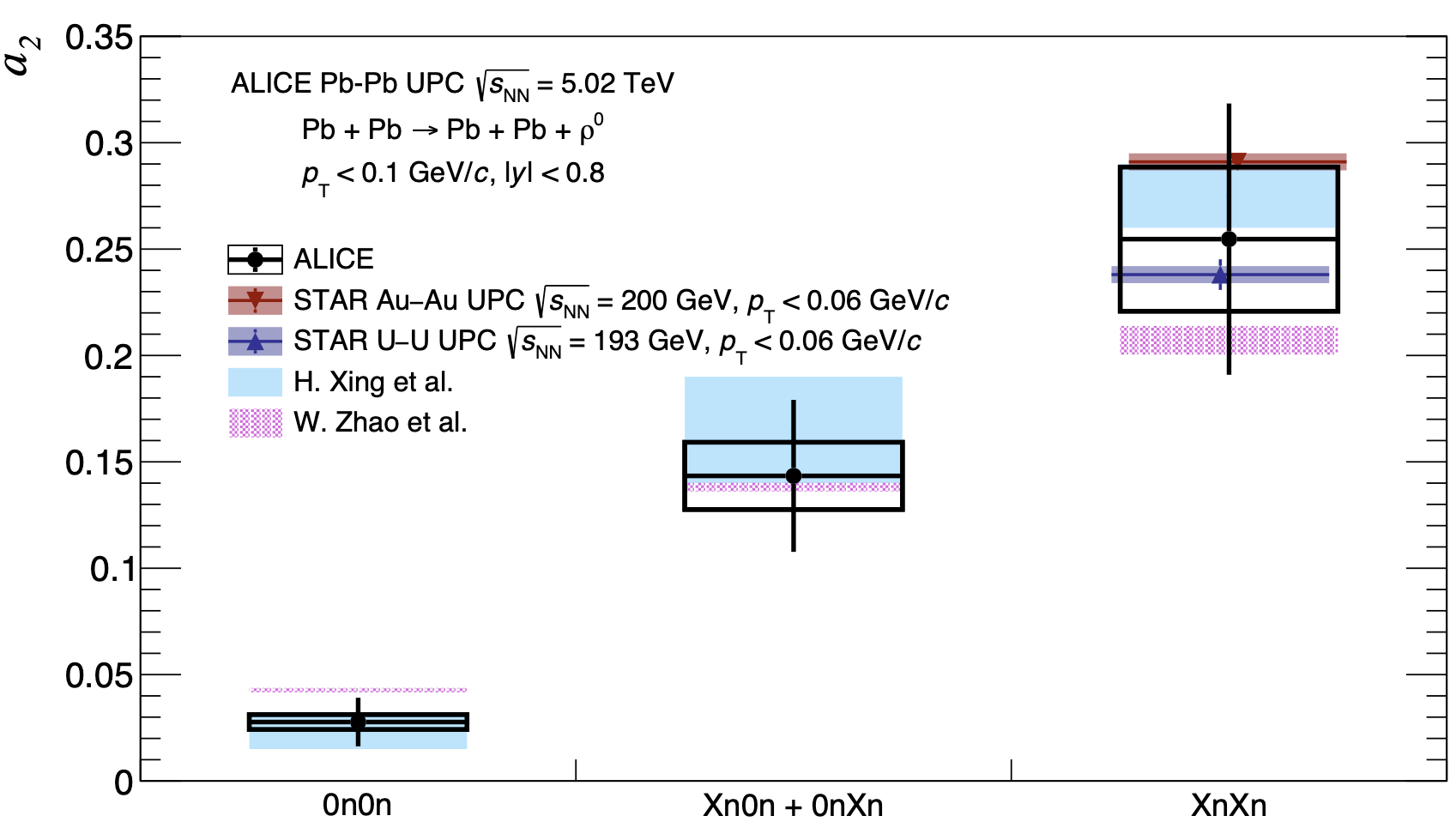}
\caption{
\label{fig:interfere}
(Left) The azimuthal asymmetries measured by STAR, for gold and uranium targets).  The data is a good match to the theoretical predictions.  From~\cite{STAR:2022wfe}.  (Right) Magnitude of the dipole assymmetries measured by ALICE for different combinations of accompanying neutrons, which map to different impact parameter distributions.   From~\cite{ALICE:2024ife}.
}
\end{figure}

The interference has an angular dependence, due to the $\vec{k_\perp}\cdot\vec{b}$ term in Eq.~\ref{eq:interfere}~\cite{Li:2019yzy}.  The interference is largest when $\vec{k_\perp}$ and $\vec{b}$ are parallel, and disappears when they are perpendicular. Figure~\ref{fig:interfere} (left) shows the angular dependence measured by the STAR Collaboration. 

The impact-parameter vector $\vec{b}$ has additional significance because, per s-channel helicity conservation, the vector meson polarization follows the photon polarization, which itself follows the electric field. This field extends radially outward from the photon-emitting nucleus, following $\vec{b}$.  And, when $\rho^0$ decay, the $\pi^+\pi^-$ plane follows the $\rho$ polarization; as long as the $\rho^0$ transverse momentum is small, the azimuthal angle (transverse to the beam direction) $\theta$ between the $\pi^+$ or $\pi^-$ and the $\rho$ $p_\perp$ follows a $\cos^2(\theta)$ distribution.  

After integration over $\vec{b}$, this leads to an azimuthal asymmetry in $\phi$, the azimuthal angle between the $\pi^+$ and $\rho^0$ $p_\perp$,  with both $\cos(2\theta)$ and $\cos(4\theta)$ components; the strengths of these asymmetries depend on the $|\vec{b}|$ distribution, {\it i. e.} on the number of accompanying photons~\cite{Xing:2020hwh}.  
The STAR Collaboration has studied these interference phenomena in $\rho^0$ photoproduction~\cite{STAR:2022wfe}.  By including the angle $\phi$ in their fits, they were able to make good fits to $d\sigma/dt$.  This advance allowed them to make the first UPC-based precision measurements of nuclear size for lead and gold nuclei.  the ALICE Collaboration has also studied this interference, using the presence of neutrons to define event classes with different $\langle b\rangle$~\cite{ALICE:2024ife}.  As is shown in Fig.~\ref{fig:interfere} (right), the asymmetry was largest for the smallest impact parameter distributions, and largely disappeared for events without accompanying neutrons.   These interference phenomena are sometimes described using the language of Entanglement Enabled Intensity Interferometry~\cite{Brandenburg:2024ksp,Brandenburg:2025one}.  

Because of the strong electromagnetic fields, UPCs can also be used to study quantum correlations between multiple vector particles produced in a single ion-ion encounter~\cite{Klein:2026jgj}.  Here, `vector particle' can be either a vector meson, or a vector excitation of a nucleus, most notably a Giant Dipole Resonance.  All of these particle share a common polarization (due to the common $\vec{b}$), so, in a quantum calculation,  their decay products will tend to end up pointing in the same direction.  The decay products are more tightly clustered than in a classical calculation, where $\vec{b}$ is well defined.

One also expects to see other correlations, such as pairs that are close in rapidity~\cite{Klein:2024wos}.  Phenomena such as stimulated production and even stimulated decay are expected, although observation of the latter may be a challenge. 

Some similar angular modulations are also expected in two-photon reactions, such as $\gamma\gamma\rightarrow l^+l^-$. This comes about because the collisions involve two linearly-polarized photons, polarized along the vector from the photon emitting ion to the two-photon interaction point.  The $\gamma\gamma\rightarrow l^+l^-$ cross section depends on if the photon polarizations are parallel or perpendicular, introducing an azimuthal modulation to the cross section~\cite{Li:2019yzy}. Data from STAR confirms these asymmetries~\cite{STAR:2019wlg}.

\section{Conclusions, and looking ahead}

UPCs at RHIC and the LHC are the energy frontier for photon physics.  Over the past few years, UPC analyses have greatly advanced our understanding of physics in multiple areas.

Data on vector meson photoproduction provided measurements of nuclear shadowing---changes in low Bjorken-$x$ quark and gluon distributions when protons are inserted into heavy nuclei.
Measurements of jet photoproduction have led to precise measurements of these gluon densities.  Moreover, $J/\psi$ photoproduction data show that not only is the gluon density reduced at low $x$, but that the effective shape of the nucleus changes, consistent with the reduction in gluon density. These data reach down to $x\approx10^{-6}$ with protons and $x\approx10^{-5}$ with nuclei.   Data on incoherent photoproduction probes event-by-event partonic fluctuations at subnucleonic scales, providing evidence for the existing of gluonic hotspots. 

Data on the photoproduction of lighter mesons has been used to precisely measure the hadronic radius of gold and uranium, providing new measurements of the thickness of their neutron skins.  It has also contributed to the studies of these mesons themselves, with new results on $\rho$-$\omega$ mixing, higher mass resonances ($\rho'$ states) and even photon fluctuations directly into $K^+K^-$ pairs.  Measurements of $\rho$ production on different nuclei has shown a roughly linear scaling in atomic number, midway between full coherence and the black disk limit. 

Photoproduction has also become another laboratory to study QGP in small systems, including for the study of baryon transport across rapidity. 

Two-photon interactions have also greatly extended our understanding of lepton-pair production, allowing us, for example, to set new limits on the anomalous magnetic moment of $\tau$. Light-by-light scattering, a topic of great interest intrinsically, can also  be leveraged for BSM studies, specifically to search for axion-like particles.   Pair production, with electron capture, has also attracted much interest, as the method used to first produce antihydrogen - a positron bound to an antiproton. 

Another burgeoning field is the study of quantum interference phenomena that utilizes the linear polarization nature of the photons in UPCs to enable, for example, precise measurements of nuclear radii or the study of entanglement in processes that act as a double-slit interferometer.

All these data have been accompanied by corresponding advances on the theory side, where we are entering a new precision era, where theorist are attempting the simultaneous description of several observables including novel ways of assessing the uncertainty of their predictions.

In the short term, a multi-fold increase of data is expected at the
high-luminosity LHC (HL-LHC). All the detectors systems are being upgraded and some of the improvements will have a great positive impact for UPCs. For example, the ALICE collaboration will collect data in  LHC Run 3 and 4  (2023-2026, 2030-2033) in continuous readout mode~\cite{ALICE:2023udb}, which will increase the amount of collected data with the central barrel detector by several orders of magnitude. For signals at higher scales, where the cross sections are small, the new HL-LHC conditions and the upgrades of CMS and ATLAS detectors will open new kinematic ranges for exploration ~\cite{Citron:2018lsq}.  For the period from 2035 to 2041 the main new addition will be the ALICE 3 detector~\cite{ALICE:2022wwr} being designed to have a much larger rapidity coverage than the current ALICE detector.

The UPC community is eagerly following  the construction of the Electron-Ion Collider~\cite{AbdulKhalek:2021gbh} and looking forward to the start of operations, expected for early 2030s. Here, the various ions species that can be accelerated and the availability of polarized beams promise to open new horizons for the study of photon-induced interactions.  Because the EIC is a dedicated accelerator, and because the ePIC detector is optimized for this physics, the EIC will be able to make very precise measurements. 

On a longer time scale, the global community has chosen the Future Circular Collider (FCC) as the new accelerator to be built at CERN~\cite{FCC:2025lpp}. The project envisions a first stage of electron-positron collisions (FCC-ee, from around late 2040 and lasting some 15 years) followed by a hadron collider (FCC-hh, from 2070 onward) with the capacity of pp as well as of heavy-ion collisions. The FCC-ee will open new horizons for two-photon interactions, while the FCC-hh will allow us to study the evolution of the color structure of hadrons to extremely small Bjorken--$x$, ensuring that UPCs will remain at the energy frontier for the foreseeable future.

\section{Symbols}
\hspace{2em}In this review, we adopt natural units ($\hbar c$) with $\rm{GeV}$ as the default energy  scale. Our conventions for relativity follow all recent field theory texts. We use the metric tensor with Greek indices running over 0, 1, 2, 3 or $t$, $x$, $y$, $z$. Roman indices denote only the three spatial components.  Einstein's summation convention is employed to handle repeated indices. Four vectors are denoted by light italic type; three vectors are denoted by boldface type; unit three vectors are denoted by a light italic label with a hat over it. 

The selection of symbols is motivated by the need for precision and uniformity in conveying concepts throughout the review. Each symbol introduced serves a specific purpose within the context of our discussion. For clarity, we define the key symbols and provide illustrative examples to demonstrate their meaning. Readers are encouraged to refer to the symbol index at the end of this section for a comprehensive overview of the symbols used and their corresponding meanings.

\begin{itemize}
        \item $\omega$: energy of a quasi-real photon
        \item $k$: four momentum of a quasi-real photon
        \item $\mathbf{k}$: three momentum of a quasi-real photon 
         \item $\mathbf{k_{\perp}}$: transverse momentum of a quasi-real photon  
        \item $\Delta$: the recoil four momentum from a target nucleus
        \item $\mathbf{\Delta}$: the recoil momentum from a target nucleus
        \item $\mathbf{\Delta_{\perp}}$: the recoil transverse momentum from a target nucleus
        \item $y$: the rapidity of the photoproduced vector meson or dilepton pair
        \item $\mathbf{p_{\perp}}$: the transverse momentum of the photoproduced vector meson or dilepton pair
        \item $\mathbf{\beta}$: the velocity of the relativistic nucleus, in the lab frame.
        \item $\gamma$: the Lorentz contraction factor of the relativistic nucleus in the lab frame.
        \item $\alpha$: the fine-structure constant
        \item $Z$: the charge carried by the nucleus
        \item $A$: the atomic number of the nucleus
        \item $W_{\rm \gamma p}$: photon-nucleon center of mass energy
        \item $d\sigma_{\rm coherent}/dt$: the differential cross section for coherent production
        \item $R_A$: nuclear radius
        \item $m_{\rm p}$: proton mass (or this could be changed to $m_N$, the nucleon mass)
     
\end{itemize}

\begin{ack}[Acknowledgments]

This work is supported in part by the U.S. Department of Energy, Office of Science, Office of Nuclear Physics, under contract number DE-AC-76SF00098, an the FORTE project, reg. no. CZ.02.01.01/00/22\_008/0004632, co-funded by the European Union.
\end{ack}

\bibliographystyle{Harvard}
\bibliography{reference}
\end{document}